\documentclass[12pt,reqno]{amsart}
\usepackage{color}
\usepackage{amssymb,amsthm}
\usepackage{amsmath}
\usepackage{hyperref}
\usepackage{mathrsfs}
\usepackage[all]{xy}
\usepackage{amsmath, amscd}
\usepackage{bbold}

\usepackage{comment}
\usepackage[draft]{fixme}

\begin{document}

\theoremstyle{plain}
\newtheorem{proposition}{Proposition}[section]
\newtheorem{theorem}[proposition]{Theorem}
\newtheorem{lemma}[proposition]{Lemma}
\newtheorem{corollary}[proposition]{Corollary}

\theoremstyle{definition} 
\newtheorem{definition}[proposition]{Definition}

\theoremstyle{remark}
\newtheorem{remark}[proposition]{Remark}

\renewcommand{\thesection}{\Roman{section}}
\renewcommand{\thesubsection}{\thesection.\arabic{subsection}}
\renewcommand{\theequation}{\thesection.\arabic{equation}}

\title[Complex  Classical Fields 
and Partial Wick Rotations]{Complex  Classical Fields\\ 
and Partial Wick Rotations}

\author[A. M. Jaffe]{Arthur Jaffe}
\address{Harvard University\\
Cambridge, MA 02138}
\email{arthur\_jaffe@harvard.edu}

\author[C. D. J\"{a}kel]{Christian D.\ J\"{a}kel}
\address{School of Mathematics\\ 
Cardiff University, Wales}
\email{christian.jaekel@mac.com} 

\author[R. E. Martinez II]{Roberto E. Martinez II}
\address{Harvard University\\
Cambridge, MA 02138}
\email{remartin@fas.harvard.edu}

\begin{abstract}
We study some examples of complex, classical, scalar fields within the new framework that we introduced in a previous work. In these particular examples, we replace the usual functional integral by a complex functional arising from  {\em partial Wick rotation} of a quantum field.  We generalize the Feynman-Kac relation to this setting, and use it to establish the spectral condition on a cylinder. We also consider positive-temperature states. 
 \end{abstract}

\maketitle

\begin{center}
{\small \bf Dedicated to Arthur Strong Wightman}\footnote{A. S. Wightman was an inspiration to all of the authors.  We learned of his passing while finishing this work, which relates to topics that fascinated him.}
\end{center}

\tableofcontents

\section{Introduction} 
In \cite{JJM Framework} we introduced a framework for using complex classical fields to describe neutral, scalar quantum fields.  In that work we replace the real functional integral by a complex functional.  In this work we study in detail one particular family of  examples that provide a classical interpretation for  {\em partial Wick rotation} of a quantum field. 

Complex fields arise naturally when the heat kernel of a Hamiltonian is complex, as in the case when an interaction breaks time-reversal symmetry. A simple family of  examples arises when one adds a multiple of the momentum to the Hamiltonian---the case that we study in this paper. 

In the usual situation for scalar bosons, the Euclidean action $\mathfrak A = \mathfrak A(\Phi)$ is real, and the Feynman-Kac density ${\rm e}^{-\mathfrak A}$ is positive.  In the case that $\mathfrak A$ has some other nice properties, the density ${\rm e}^{-\mathfrak A}$ can be normalized to define a probability measure. However, when the classical fields $\Phi$ are complex, the action $\mathfrak A$ may also be complex. Consequently, the problem of integrating ${\rm e}^{-\mathfrak A}$ is more subtle.   (See also \cite{Gu1, Gu2}.)   

The mathematics of complex measures on finite-dimensional spaces poses no difficulty provided the absolute value of the measure can be integrated. The situation is more complicated for measures on function spaces, such as the measures in functional integrals. Not only can the density grow in certain complex directions, but also oscillations may lead to other difficulties with normalization. Even the case of Gaussian measures is not  straightforward, so one can imagine more difficulty in the study of interactions with non-quadratic actions.

In this paper we consider perturbations of a Hamiltonian $H$ with zero ground-state energy and with a positive heat kernel.  We study perturbations of the form 
	\begin{equation}
		H_{\vec {\it v}} = H + \vec P\, \cdot {\vec {\it v}}\;.
	\label{Boosted Hamiltonian}
	\end{equation}
The momentum $\vec P$ commutes with $H$ and generates spatial translations. As we require that  $\vert {\vec {\it v}}\vert<1$, the operator  $H_{\vec v}$ is a multiple of the Hamiltonian in a Lorentz frame boosted by velocity $\vec v$, in units for which the speed of light $c=1$.

We study free fields in arbitrary dimension; in spacetime dimension two we also treat ${\mathscr P}(\varphi)_{2}$-interactions on the spatial circle. 
In \S\ref{Sect:FK} we introduce a generalized Feynman-Kac relation to deal with the non-linear interaction in the absence of a measure. We define a functional on a sufficiently large sub-algebra of functions of the classical fields to analyze the corresponding quantum fields.  

As discussed in \cite{JJM Framework}, the property of \emph{reflection positivity} plays a key role in our setting. 
Further insight arises from having two reflection-positive planes in the classical framework,  for one then has a 
symmetry relating two different quantum theories.   

In \S\ref{Section:Spectral_Condition} we use this symmetry to give a  proof of 
	\begin{equation}
		0  \leqslant H_{{\vec {\it v}}} \;,
	\label{Spectral Condition Pphi}
	\end{equation}
for ${\mathscr P}(\varphi)_{2}$-interactions on a spatial circle. This spectrum condition was conjectured in \cite{Glimm-Jaffe-Spectrum} and proved in \cite{Osipov}; see \S\ref{Section:Spectral_Condition} for further discussion. Without appealing to Lorentz symmetry, the estimate \eqref{Spectral Condition Pphi} results in analyticity of the imaginary-time field $\varphi(t,\vec x)$ in the spatial variable $\vec x$. In particular, the spectrum condition holds  when the spacetime manifold is compact in the spatial directions.

In \S\ref{Sect:Positive Temperature-SF} and \S\ref{Sect:The Fully Compactified Case} we analyze quantization for positive temperatures. We hope that the methods we develop here can be useful in a wider context.  We are currently  studying a second application to charged fields.  

\setcounter{equation}{0} 
\section{Quantization\label{Sect:Quantization}}
We adopt the notations and conventions of our earlier work. We analyze classical fields on space-times of the general form
\[
\boldsymbol{X} =X_1\times \cdots \times X_d  \; , 
\]
where each factor $X_i$ either equals $\mathbb{R}$ (the real line) or $S^{1}$ (a circle) of length $\ell_i$. The classical Gaussian, neutral, scalar field $\Phi$ is an operator-valued distributions, and all classical fields commute.   The classical field acts on the Fock space 
\begin{equation}
	\label{Euclidean Space}
        \mathcal{E} 
        		= \mathbb{C} \oplus
        			\bigoplus_{n=1}^\infty \,\mathcal{E} _n\;,
        			\quad \text{where} \quad 
        		\mathcal{E} _n
        			= \underbrace{\mathcal{E} _1 \otimes_s
                			\cdots\otimes_s \mathcal{E} _1}_{n\ \rm factors}\;,
\end{equation}
over the Hilbert space $\mathcal{E} _{1}=L_{2}({\boldsymbol X} ; dx)$.

\subsection{Quantization of Vectors}
Let $\Phi$ be a field on $  \mathcal{E}  $ for which a unitary reflection~$\Theta$ is doubly reflection-positive. 
While $\Theta$ usually denotes time reflection, our quantization method applies to any reflection satisfying the 
following list of properties:

\begin{itemize}
\item [$ i.)$] $\Theta^{-1}=\Theta^{*}=\Theta$ on $  \mathcal{E}  $; 
\item [$ii.)$] $\Theta  \mathcal{E}  _{\pm}= \mathcal{E}  _{\mp} \, $, where $\mathcal{E}  _{\pm}$ are subspaces of $\mathcal{E} $; and,
\item [$iii.)$] $0\leqslant \Theta$ on $  \mathcal{E}  _{\pm} \, $.  
\end{itemize}
The sesquilinear form 
	\[ 
		(A, B) \mapsto \langle A,\Theta B \rangle_{  \mathcal{E}  }
	\]
on $  \mathcal{E}_{+}\times  \mathcal{E}_{+}$ (or on $  \mathcal{E}_{-}\times  \mathcal{E}_{-}$) defines  
pre-Hilbert spaces $\mathcal{H}_{\pm,0}\, $, which are the quantizations of $  \mathcal{E}  _{\pm}$ with respect 
to the reflection $\Theta$.  The vectors in $\mathcal{H}_{\pm,0}$ are equivalence classes  
	\[
		\widehat A=A+N\in  \mathcal{E}  _{\pm} / {\mathcal N}_{\pm}\; , 
	\]
where $A\in  \mathcal{E}  _{\pm}$, and where $N\in {\mathcal N} \cap  \mathcal{E}  _{\pm}$ is an element 
of the null space~${\mathcal N}$ of the form \eqref{Quantization Form}. The inner products  
\begin{equation}
	\label{Quantization Form}
		\bigl\langle \widehat A,\widehat B \bigr\rangle_{\mathcal{H}_{\pm, 0}}
		= \bigl\langle  A ,\Theta B \bigr\rangle_{  \mathcal{E}  }\;,
		\quad
		A,B\in  \mathcal{E}  _{\pm}\; , 
\end{equation}
defined initially on $\mathcal{H}_{\pm, 0}\,$, extend to inner products on the Hilbert space $\mathcal{H}_{\pm}\,$, the completion of $\mathcal{H}_{\pm, 0}\,$.  As $\Theta $ is unitary, property~$ii.)$ in the list above~ensures that the Hilbert spaces $\mathcal{H}_{\pm}$ are 
isomorphic, so for simplicity we denote both spaces as $\mathcal{H}$. The quantization map $\ \widehat{}\ $ is a contraction on vectors,  namely 
	\begin{equation*}
\Vert \widehat A \Vert_{\mathcal{H}}
		\leqslant   \Vert A \Vert_{  \mathcal{E}  }\;. 
	\end{equation*}

\subsection{Quantization Domains\label{Quantization Domains}}
To simplify notation in this section, we only consider reflections in the first coordinate. Consider an open subset $\mathcal{O}$ of the product space
\[
\boldsymbol{X}_{+} = X_{1, +} \times X_{2} \times \cdots \times X_{d} \; ,
\]
where $X_{1,+}$ equals the half-circle $S^{1}_{+} $ or the half-line $\mathbb{R}_{+}$ and $X_{j} = S^{1}$  or $ \mathbb{R}$, $ j= 2, \ldots, d$.  Let ${\mathscr P}(\mathcal{O})$ denote the algebra of formal\footnote{The formal product is replaced 
by the operator product as the formal expression is applied to the vacuum vector $\Omega_{0}^{E}$.}
polynomials in field operators averaged with $C^{\infty}$-functions supported in $\mathcal{O}$.  

\begin{definition}
An open set $\mathcal{O}\subset \boldsymbol{X}_{+}$ is 
a {\em quantization domain} if the 
quantization map $A\mapsto \widehat A$ takes the linear subspace $ \mathcal{D}(\mathcal{O})=
{\mathcal P}(\mathcal{O})\Omega_{0}^{E}$ onto a subspace $\widehat{ \mathcal{D}(\mathcal{O})}$ that 
is dense in $\mathcal{H}$.  
\end{definition}

\begin{remark}
The Reeh-Schlieder theorem of Wightman quantum field theory says that products of Minkowski space fields, 
smeared with test functions supported in an arbitrary open bounded spacetime region and applied to the vacuum vector, 
form a total set of vectors in $\mathcal{H}$. One can think of a quantization domain $\mathcal{O}\subset \boldsymbol{X}_{+}$ as a \emph{classical} version of this property. 
\end{remark}


\begin{proposition}[\bf Non-trivial Quantization Domains \cite{Hoole}]\label{Theorem:Quantization Domain}  
Consider a covariant classical scalar field $\Phi(f)$ on~$  \mathcal{E}  ( \boldsymbol{X})$ with ${X}_{1,+}= \mathbb{R}_{+}$, which satisfies 
\[
 \Theta\,\Phi(x)\,\Theta
        		= \Phi(\vartheta x)^* \; . 
\]
Assume that:
\begin{itemize}
\item[$i.)$] The characteristic function $S(f)=\langle\Omega_{0}^{E} , {\rm e}^{i\Phi(f)}\Omega_{0}^{E}\rangle_{\mathcal E}$ 
is invariant under the action of the spacetime translation group and the time reflection on the test  functions; and,
\item[$ii.)$]  There is a constant $M<\infty$ and a Schwartz-space norm $\| \cdot \|_{\alpha}$ on time-zero test  functions such that the following estimates hold:
\[
		0\leqslant H\;,
		\quad
		\pm | \vec P | \leqslant M(H+ \mathbb{1}) \; , 
\]
and
\begin{equation*}
		\pm \varphi(h) 
		\leqslant M \| h \|_{\alpha} \,(H+\mathbb{1}) \;.
\end{equation*}
\end{itemize}
Then any open set $\mathcal{O}\subset \boldsymbol{X}_{+}$  is a quantization domain. 
\end{proposition}

\subsection{Quantization of Operators}
Consider a linear transformation~$T$ whose domain is a quantization domain $\mathcal{D}(T)\subset  \mathcal{E}$.  If $T$ 
maps~$\mathcal{E}  _{+}\cap \mathcal{D}(T)$ into~$\mathcal{E}  _{+}$  and $T$  maps ${\mathcal N}_{+}$ into ${\mathcal N}_{+}$, then $T$ has a quantization~$\widehat T_{+}$ on $\mathcal{H}_{+}$ with domain $ \mathcal{D}(\widehat T_{+})= (\mathcal{E}  _{+}\cap \mathcal{D}(T))^{\wedge}$. 
Explicitly,   
\begin{equation*}
		\widehat T_{+} \,\widehat A
		= \widehat {TA} \quad \text{for} \quad
		A \in \mathcal{D}(T)\cap   \mathcal{E}  _{+}\;.
\end{equation*}
If in addition, $T$ extends to a densely-defined operator on ${\mathcal E}$ with 
adjoint~$T^{*}$, let  
	\[
		T^{+}=\Theta T^{*} \Theta \; .  
	\]
Assume that $T^{+}$ leaves $\mathcal E_{+}$ invariant, that is, $T^{+} \colon \mathcal{E}  _{+}\cap  \mathcal{D}(T^{+})\to  \mathcal{E}  _{+}$. In this case, a Schwarz inequality in $\mathcal E$ shows that $T$  maps~${\mathcal N}_{+}$ into~${\mathcal N}_{+}$.  In addition the adjoint of~$\widehat T_{+}$ on $\mathcal{H}_{+}$ extends~$\widehat{T^{+}}$.   The latter denotes the quantizations of~$T^{+}$ on $\mathcal{H}_{+}$. Similarly, one has a quantization $T^{-}$ of $T$ in case that $T$ 
maps $  \mathcal{E}  _{-}\cap \mathcal{D}(T)$ into~$  \mathcal{E}  _{-}$  and $T$  maps ${\mathcal N}_{-}$ into ${\mathcal N}_{-}$.

\subsection{The Heat Kernel Semigroups\label{Example A}}
In case $X_{1}= \mathbb{R}$, let $T(t)$ denote a unitary translation group on~$\mathcal{E}$, implementing translations of the distinguished \emph{time} coordinate.   
Note that $T(t)^{+}=T(t)$, so 
$\widehat {T(t)}$ is self-adjoint.  Thus, $\widehat {T(t)}$ gives a self-adjoint quantization of the positive-time 
semigroup ${T(t)}$ on~$\mathcal{H}_{+}$ and of the negative time semigroup on $\mathcal{H}_{-}$.  In particular, 
the generators of these semi-groups are the (positive, self-adjoint) Osterwalder-Schrader Hamiltonians 
$0\leqslant H_{\pm}=H_{\pm}^{*}\, $, and 
	\[
		\widehat{T(t)}
		=  \begin{cases}
		{\rm e}^{- t H_{+}}  & \text{on $\mathcal{H}_{+}$  if $0 \leqslant t $} \; ,  \\
		{\rm e}^{t H_{- }}  & \text{on $\mathcal{H}_{-} $ if $t \leqslant 0$} \;.
		\end{cases}
	\]
We return to the case $X_{1}= S^{1}$ in \S \ref{Sect:Positive Temperature-SF}.

\setcounter{equation}{0}
\section{Classical Gaussian Fields on $\mathbb{R}^d$\label{Sect:ExampleI}}
We begin by discussing the Euclidean version of the one-particle space in quantum theory on $\mathbb{R}^{d-1}$ that corresponds to the free field theory with Hamiltonian $H_{\vec v}$ given in \eqref{Boosted Hamiltonian}.  For the free field the map $H \mapsto H_{\pm\vec {\it v}} = H \pm \vec P \cdot \vec {\it v}$ replaces the one-particle Hamiltonian  $\mu$ acting on the one-particle subspace 
$\mathcal{H}_{1}$ by 
\begin{equation}
	\label{mu-beta-Defn}
		{\mu}_{\pm}
		=\mu\pm\vec p\cdot \vec {\it v}\;.
\end{equation}
Here $\vec p=-i\nabla_{\vec x}$ and $\vec {\it v} \in\mathbb{R} ^{d-1}$ is a given constant vector of length 
less than one. Write 
	\begin{equation}
	\label{delta-v}
		 \vec {\it v}=\vec n\,\tanh\beta\; 
		 \quad  \text{and} \quad
		 \delta = \vec p \cdot \vec {\it v}\;,
	\end{equation} 
where  $\vec n\in\mathbb{R} ^{d-1}$ is a unit vector and $\beta\in\mathbb{R}$.  Throughout this work, we assume 
that $m > 0$, so $| \delta | <\mu\tanh |\beta|$.  As $1-\tanh^{2}\beta = \cosh^{-2}\beta=(1-\vec {\it v}^{\,2})^{-1}$, 
we infer that $\mu^{2}/ \cosh^{2}\beta\leqslant \mu^{2}-\delta^{2}$.  Thus,
\begin{equation}
	\label{Elem-Bound-1}
	0
	<  m\sqrt{1-\vec {\it v}^{\,2}} 
	 \leqslant \mu\sqrt{1-\vec {\it v}^{\,2}} 
	\leqslant \mu_{\pm}\;.		
\end{equation}
For $\vec {\it v}\neq 0$,  the two-point function $D_{\vec {\it v}}$ is complex, rather than real and positive.  Nevertheless, the 
hermitian part of the associated heat kernel is strictly positive.  Furthermore, $D_{\vec {\it v}}$ has two reflection 
planes (reflection in the time-axis and in the $\vec {\it v}$-axis) that are reflection-positive. The corresponding configuration space is given by $\boldsymbol{X} = \mathbb{R}^d$. For these reasons, this example fits into the framework introduced in~\cite{JJM Framework}.

For a non-interaction system, we obtain all information from the Gaussian expectation of classical fields.  
The expectation of the product of two fields defines the classical two-point function $D_{\vec {\it v}}$:
	\[
		D_{\vec {\it v}} ((s_1, {\vec x}_1), (s_2, {\vec x}_2))= \langle \varphi (0, \vec x_1)\Omega_0 , 
		{\rm e}^{-(s_1-s_2) (H + \vec P\cdot \vec {\it v})}\varphi (0, \vec x_2) \Omega_0 \rangle 
		\; . 
	\]
\subsection{The Two-Point Function $D_{\vec {\it v}}\, $\label{Sect:Real Propagator}}
Let $x=(t,\vec x)\in\mathbb{R} ^{d}$ and, in Fourier space, let $k=(E,\vec k)\in\mathbb{R} ^{d}$. The introduction of a covariance operator~$D_{\vec {\it v}}$ on~$\mathcal{E} _{1}$ corresponds to the substitution $\mu \mapsto {\mu}_{+}=\mu+\delta$ on the one-particle space. To simplify notation, denote the multiplication operators in Fourier space by 
\[
\mathcal{F} \mu\mathcal{F}^{*}=(\vec k^{2}+m^{2})^{1/2} \quad \text{and} \quad  
\mathcal{F} \,\vec p \,\mathcal{F}^{*}=\vec k \; , 
\]
where $\mathcal{F}$ denotes Fourier transformation. Therefore, in Fourier space,  $\mathcal{F} \delta \mathcal{F}^{*} =\vec k\cdot\vec {\it v}$. Consider the substitution
\begin{equation}
		(2\pi)^{d/2} \widetilde C(k)
		= \frac{1}{E^{2} +\mu^{2}}
		  \;  \mapsto \;   
  \frac{1}{(E + i\delta)^2+\mu^2} = 		(2\pi)^{d/2} \widetilde D_{\vec {\it v}}(k)\;.
	\label{Example 1 Propagator}
\end{equation}
The corresponding operator $D_{\vec {\it v}}$  on $L_{2}(\mathbb{R} ^{d} ; dx)$ has  the integral kernel
\begin{equation}
	\label{D Fourier}
		D_{\vec {\it v}}(x-x')
		= 
		\frac{1}{(2\pi)^{d}}\,\int_{\mathbb{R} ^{d}}
			\frac{1}{(E + i\delta)^{2}+\mu^{2}}\,
			{\rm e}^{ik\cdot (x-x')}\,dk\;.
\end{equation}

\goodbreak
\begin{proposition}[\bf Elementary Properties of $\mathbf{D}_{\vec {\it v}}$]
\label{Prop:ElementaryPropoerites DLorentz} 
With the above conventions:
\begin{enumerate}
\item[$i.)$] {\label{D-in-Fourier}}
The integral kernel of $D_{\vec {\it v}}$ has the representation  
\begin{equation}
        \label{Modified_Kernel}
		\qquad \quad 
		D_{\vec {\it v}}(x-x') = \frac{1}{(2\pi)^{(d-1)}}\int_{\mathbb{R}^{d-1}}
            			{\rm e}^{- | t-t' | \mu+(t-t')\delta + i \vec k(\vec x-\vec x')}\,
            			\frac{d\vec k}{2\mu}   
\end{equation}
with $\delta$ given in \eqref{delta-v};

\item[$ii.)$]{ \label{D-symmetric}} The decomposition of $D_{\vec {\it v}}=K_{\vec {\it v}}+i L_{\vec {\it v}}$   into hermitian and 
skew-hermitian parts on $L_{2}(\mathbb{R}^{d} ; dx)$ yields two hermitian operators $K_{\vec {\it v}}$ 
and $L_{\vec {\it v}} \, $ with real-valued and symmetric kernels. If $0 < m$, then  $0 < K_{\vec {\it v}}$. Thus,
\begin{align*}
		D_{\vec {\it v}} & = D_{\vec {\it v}}^{\rm T} \\
		0 < K_{\vec {\it v}} & = K_{\vec {\it v}}^{*}= K_{\vec {\it v}}^{\rm T}= \overline{ K_{\vec {\it v}}}\; \\	
		 L_{\vec {\it v}} & = L_{\vec {\it v}}^{*}= L_{\vec {\it v}}^{\rm T}= \overline{ L_{\vec {\it v}}}\; ; and,
\end{align*}

\item[$iii.)$] {\label{Hermitian-Reflection}}   Let $\vartheta$ denote time inversion acting as a unitary on 
$L_{2}(\mathbb{R} ^{d} ; dx)$, and let $\pi_{\vec n}$ denote the unitary on $L_{2}(\mathbb{R} ^{d} ; dx)$ 
implementing reflection in the spatial plane normal to $\vec n$.   The operators $\vartheta D_{\vec {\it v}}$,  
$ D_{\vec {\it v}}\,\vartheta$,  $\pi_{\vec n}D_{\vec {\it v}}$, and  $ D_{\vec {\it v}}\,\pi_{\vec n}$
are self-adjoint on $L_2(\mathbb{R} ^d  ; dx)$.
\end{enumerate}
\end{proposition}

\begin{proof}
The first statement is a consequence of the Cauchy Integral Theorem:  
	\[
		\frac{1}{2\pi} \int_{-\infty}^{\infty} 
		\frac{{\rm e}^{iEt} }{(E+i\delta)^{2}+\mu^{2}}\,
		\,dE
		=
		\begin{cases}
		\frac{{\rm e}^{-t(\mu-\delta) } }{2\mu}	&\text{if $t>0$} \\
		\frac{{\rm e}^{t(\mu+\delta)}}{2\mu} 	&\text{if $t<0$} 
		\end{cases}
		\ = \frac{{\rm e}^{-  | t | \mu+t\delta}}{2\mu}
		 \;.
	\]

The second statement is a consequence of the properties of $\widetilde D_{\vec {\it v}}(k)$ in Fourier space. 
If  $k\mapsto-k$, then  $E\mapsto -E$, $\delta \mapsto -\delta$, and $\mu \mapsto \mu$.  Therefore, $\widetilde D_{\vec {\it v}}(-k)=\widetilde D_{\vec {\it v}}(k)$, and consequently $D_{\vec {\it v}}=D_{\vec {\it v}}^{\rm T}$, $K_{\vec {\it v}}=K_{\vec {\it v}}^{\rm T}$ and $L_{\vec {\it v}}=L_{\vec {\it v}}^{\rm T}$  are symmetric, as claimed.  Both $K_{\vec {\it v}}$ and $L_{\vec {\it v}}$ act as multiplication operators by real-valued functions in Fourier space, 
so they are hermitian and real. Their explicit form is 
\[
		(2\pi)^{d/2}\,\widetilde K_{\vec {\it v}}(k)
		=\frac{E^2+\mu^2-\delta^2}
			{(E^2+(\mu-\delta)^2)\,(E^2+(\mu+\delta)^2)}			\;,
\]
and
\begin{equation}
	\label{K-L Form}
		(2\pi)^{d/2}\,\widetilde L_{\vec {\it v}}(k)
		=\frac{-2E\delta}
			{(E^2+(\mu-\delta)^2)\,(E^2+(\mu+\delta)^2)}			\;.	
\end{equation}
The bound \eqref{Elem-Bound-1} shows $\widetilde K_{\vec {\it v}}(k)$ is non-vanishing  as long as $m\neq0$. 
Thus, $0 < K_{\vec {\it v}}$, and $K_{\vec {\it v}}$ is invertible. Also,
\begin{equation}
	\label{L-Fourier}
        		\frac{\widetilde L_{\vec {\it v}}(k)}{\widetilde K_{\vec {\it v}}( k)}
        			= \frac{-2E\delta}{E^{2}+\mu^{2}-\delta^{2}}\;.
\end{equation}

Finally, consider the third statement. Time reflection in Fourier space leaves $\mu,\delta$, and 
$\vec k$ invariant and sends $E\mapsto -E$. Thus, under time inversion, 
	\[ 
		\widetilde D_{\vec {\it v}}(k)\mapsto \overline{\widetilde D_{\vec {\it v}}(k)} \; . 
	\]
In configuration space this implies that $\vartheta D_{\vec {\it v}}\vartheta = D_{\vec {\it v}}^*\; $.  Hence $\vartheta D_{\vec {\it v}}$ 
and  $D_{\vec {\it v}}\vartheta$ are self-adjoint.

In Fourier space, spatial-reflection $\pi_{\vec n}$ acts as follows: it leaves $E$ invariant, and it sends $\vec
k \mapsto \pi_{\vec n} \vec k= \vec k -2(\vec n\cdot \vec k) \vec n$.  This is a consequence of
\begin{align*}
       \pi_{\vec n}\,\vec x\cdot \vec k
       &= (\vec x -2 (\vec x\cdot \vec n)\vec n)\cdot\vec k
       = \vec k\cdot\vec x - 2(\vec k\cdot \vec n)(\vec x\cdot\vec n)
       \\
       &=\vec x\cdot (\vec k -2(\vec k\cdot \vec n)\vec n)\;.
\end{align*}
Under this transformation, $\mu \mapsto \mu$ and $\delta  \mapsto  -\delta$.  Thus, $\widetilde D_{\vec {\it v}}(k) \mapsto  \overline{\widetilde D_{\vec {\it v}}(k)}$, and in configuration space 
$\pi_{\vec n}\,D_{\vec {\it v}}\,\pi_{\vec
n}=D_{\vec {\it v}}^*$.  Thus, $\pi_{\vec n}D_{\vec {\it v}}$ and  $D_{\vec {\it v}}\,\pi_{\vec n}$ are self-adjoint.
\end{proof}

We now provide  bounds on the self-adjoint real and imaginary parts of $D_{\vec {\it v}} =K_{\vec {\it v}}+iL_{\vec {\it v}} $.  
Denote the absolute value by $| D_{\vec {\it v}} | = (D_{\vec {\it v}}^{*}\,D_{\vec {\it v}})^{1/2} $.

\begin{proposition}\label{Prop:ComplexCovarianceBounds} 
The operators $K_{\vec {\it v}}$, $L_{\vec {\it v}} $, $D_{\vec {\it v}}$, and $C$ on $L_{2}(\mathbb{R}^{d} ; dx)$ mutually 
commute and  satisfy
\[
        K_{\vec {\it v}}
        \leqslant | D_{\vec {\it v}} |
        \leqslant (\cosh\beta) K_{\vec {\it v}} \;.
\]
Moreover,
\begin{equation}
    \label{L-K_RelativeBound}
          \left( \tfrac{1}{2\cosh^{2}\beta} \right) C
          < K_{\vec {\it v}}
          < (\cosh^{4}\beta)C 
        \;, \text{and} \; 
        \sup_{k}  \left| \tfrac   {\widetilde L_{\vec {\it v}}(k)} {\widetilde K_{\vec {\it v}}( k)} \right|  
        =
        \sinh | \beta | \;,  
\end{equation}
as well as,
\begin{equation}
		(2\cosh^{2}\beta)^{-1} C
		< | D_{\vec {\it v}}| 
       		< (\cosh^{5}\beta)C\;.
	\label{AbsD Bound} 
\end{equation}
\end{proposition}

\begin{proof}
The operators $K_{\vec {\it v}}$, $L_{\vec {\it v}} $, $D_{\vec {\it v}}$, and $C$ are all translation-invariant, so they commute.  
Furthermore, $\|  K_{\vec {\it v}}  \|  = (2\pi)^{d/2} \sup_{k} | \widetilde  K_{\vec {\it v}}(k) | $. Note that 
\begin{align*}
		(E^{2}+(\mu-\delta)^{2}) (E^{2}+(\mu+\delta)^{2})
		& = E^{4} + 2E^{2}(\mu^{2}+\delta^{2}) + (\mu^{2}-\delta^{2})^{2}
		\\
		&< E^{4} + 4 E^{2}\mu^{2} + \mu^{4}
		\\
		&< 2(E^{2}+\mu^{2})^{2}\;.
\end{align*}
From  \eqref{Elem-Bound-1} and \eqref{K-L Form} one then infers  the lower bound  for $K_{\vec {\it v}}$ 
in \eqref{L-K_RelativeBound}.  To establish the upper bound  on  $K_{\vec {\it v}}$, use  
\begin{align}
		E^{4} + 2E^{2}(\mu^{2}+\delta^{2}) + (\mu^{2}-\delta^{2})^{2}
		& >  E^{4} + 2 E^{2}\mu^{2} + \frac{\mu^{4}}{\cosh^{4}\beta}
		\nonumber \\ & \geqslant \left( \frac{ E^{2}+\mu^{2} }{\cosh^{2}\beta} \right)^{2} , \nonumber
\end{align}
which entails
\begin{align}
		(2\pi)^{d/2}\,\widetilde K_{\vec {\it v}}(k)
		& < \cosh^{4}\beta\,\frac{( E^{2}+\mu^{2}-\delta^{2})}{(E^{2}+\mu^{2})^{2}} \nonumber
		\\ & \leqslant (2\pi)^{d/2}\,(\cosh^{4}\beta)  \widetilde C(k)\;. \nonumber
\end{align}
The upper bound on  $K_{\vec {\it v}}$ then follows.   We also use the explicit forms in \eqref{K-L Form} to 
bound the ratio  \eqref{L-Fourier}.  From \eqref{Elem-Bound-1} we conclude that 
\begin{eqnarray}
	\label{L-Fourier-Bound}
        \left|   \frac{\widetilde L_{\vec {\it v}}(k)}{\widetilde K_{\vec {\it v}}( k)} \right|
        & =&  \frac{2  |  E\, \vec k\cdot \vec n| 
        		\frac{1}{\cosh\beta}}
		{E^{2}+\mu^{2}-\delta^{2}}\, \sinh | \beta |
        	\nonumber
	\\
		& \leqslant &
        \frac{E^{2}+{\frac{\vec k^{2}}{\cosh^{2}\beta}}}
        		{E^{2}+\mu^{2}-\delta^{2}}\,\sinh | \beta |
	\nonumber \\
	& \leqslant & \frac{E^{2}+{\frac{\vec k^{2}}{\cosh^{2}\beta}}}
        		{E^{2}+\frac{\mu^{2}}{\cosh^{2}\beta}}\,\sinh |\beta|
	<  \sinh |\beta | \;.
\end{eqnarray}
In fact, one can approach the bound \eqref{L-Fourier-Bound} by choosing 
	\[
		\vec k\cdot \vec n=| \vec k |=-E\cosh\beta>0 \; .  
	\]
Then
	\[
        		\frac{{\widetilde L_{\vec {\it v}}(k)}}
            	{\widetilde K_{\vec {\it v}}(k)}
        		= \frac{2E^2\sinh\beta}{2E^2+m^2}
        			\to
        		\sinh\beta  \quad \text{as} \quad E\to\infty\;.
    	\]
This shows that the upper bound in  \eqref{L-Fourier-Bound}  is the best possible, so the equality 
in~\eqref{L-K_RelativeBound} holds. 

Finally, we bound $| D_{\vec {\it v}} | = (K_{\vec {\it v}}^{2} + L_{\vec {\it v}}^{2})^{1/2}$.  Since $K_{\vec {\it v}}$ and 
$L_{\vec {\it v}}$ are self-adjoint and commute, the bound \eqref{L-K_RelativeBound} yields
\begin{eqnarray*}
		 K_{\vec {\it v}} 
		&\leqslant& | D_{\vec {\it v}} | 
		\\ & = & ({\mathbb {1}} +( L_{\vec {\it v}} K_{\vec {\it v}}^{-1})^{2})^{1/2}K_{\vec {\it v}}
		\\
		&\leqslant& (1+\sinh^{2}\beta)^{1/2} \, K_{\vec {\it v}} \\ & = & (\cosh\beta) K_{\vec {\it v}}\;,
\end{eqnarray*}
where $\mathbb 1$ denotes the identity operator. 
\end{proof}	

\subsection{Time-Reflection Positivity\label{Sect:TimeRP}}

Consider the positive-time half-space $\boldsymbol{X}_{+}=  \mathbb{R}_{+}  \times\mathbb{R}^{d-1}$; the negative-time half-space  $\boldsymbol{X}_{-}$ is defined similarly. Let $L_{2,+} \, = \, L_2(\boldsymbol{X}_\pm ; dx)$ 
denote the subspace of~$L_2(\mathbb{R}^d ; dx)$ consisting of functions supported in $\boldsymbol{X}_\pm$.

\begin{proposition}
\label{Proposition:Neutral Time_RP} 
The operators $\vartheta D_{\vec {\it v}}$ and $D_{\vec {\it v}}\,\vartheta$ have  positive expectations on 
$L_{2,+}$.  The corresponding Oster\-walder-Schrader Hamiltonians ${\mu}_{+}$ 
for~$\vartheta D_{\vec {\it v}}$ and ${\mu}_{-}$ for $ D_{\vec {\it v}}\,\vartheta$ are the Hamiltonians ${\mu}_{\pm}$ 
defined in~\eqref{mu-beta-Defn} acting on $\mathfrak H_{-1/2}(\mathbb R^{d-1})$.
\end{proposition}

\begin{proof} 
We establish positivity of $\vartheta D_{\vec {\it v}}$ on $L_{2,+}$ directly from the form of its integral kernel \eqref{Modified_Kernel}.  
The operator $\vartheta D_{\vec {\it v}}$ is hermitian by Proposition~\ref{Prop:Sigma-Properties-Lorentz-D}, and its integral kernel 
on $L_{2,+}\times L_{2,+}$ is 
	\begin{equation}
		(\vartheta D_{\vec {\it v}})(x,x')
			= \frac{{\rm e}^{-(t+t')(\mu+\delta)}}{2\mu}(\vec x-\vec x')\;,
	\label{theta D Kernel}
	\end{equation}
which exhibits its positivity and shows that the Osterwalder-Schrader Hamiltonian is $\mu_+ = \mu+\delta \, $, acting on the Sobolev space  $\mathfrak{H}_{-\frac{1}{2}}(\mathbb{R}^{d-1})$ with inner product 
	\begin{equation}
	\label{Sobolev-Space3}
		\bigl\langle \cdot , \cdot \bigr\rangle_{\mathfrak{H}_{-\frac{1}{2}}(\mathbb{R} ^{d-1})}
			= \left \langle \tfrac{1}{\sqrt{2\mu}}\ \cdot\ , \ 
				\tfrac{1}{\sqrt{2\mu}}\ \cdot 
				\right \rangle_{L_{2}(\mathbb{R} ^{d-1} ; d \vec x \, )}\;.
	\end{equation}
Let $f\in L_{2,+}$ be smooth, and consider  $f_t(\vec x)=f(t,\vec x)$ to be a family of functions of $\vec x \in \mathbb{R}^{d-1}$.  
It follows that 
	\[
		\langle f, \vartheta D_{\vec {\it v}} g \rangle_{L_{2}(\mathbb{R} ^{d})}
		= \Bigl\langle \int_{0}^{\infty} {\rm e}^{-t {{\mu}_{+}}} f_{t}\,dt  \; , \int_{0}^{\infty} {\rm e}^{-t {{\mu}_{+}}} g_{t}\,dt \Bigr\rangle_{\mathfrak{H}_{-\frac{1}{2}}(\mathbb{R} ^{d-1})}\;.
	\]
The expression \eqref{Modified_Kernel} shows that the integral kernel of $D_{\vec {\it v}}\,\vartheta$ is 
	\[
		(D_{\vec {\it v}}\,\vartheta )(x,x') = \frac{{\rm e}^{-(t+t')(\mu-\delta)}}{2\mu}(\vec x-\vec x')\;,
	\]
which exhibits reflection positivity and shows that the Osterwalder-Schrader Hamiltonian for the operator $D_{\vec {\it v}}\,\vartheta$ is 
$\mu_- =\mu-\delta$. In this case,
\begin{align}
		\langle f,  D_{\vec {\it v}}\,\vartheta\, g \rangle_{L_{2}(\mathbb{R} ^{d})}
		& = \left \langle \int_{0}^{\infty} {\rm e}^{-t {{\mu}_{-}}} f_{t}\,dt \; , 
		\int_{0}^{\infty} {\rm e}^{-t {{\mu}_{-}}} g_{t}\,dt \right \rangle_{\mathfrak{H}_{-\frac{1}{2}}(\mathbb{R} ^{d-1})}\;. \nonumber
	\end{align}
\end{proof}

\begin{proposition} Let $\mathcal K_{+,0}\subset \mathcal K_{+} $ be the dense subset defined as the linear span of $C^{\infty}_{0}(S^{1}_{+}) \times C^{\infty}_{0}({\mathbb R}^{d-1})$. Define the {\em Osterwalder-Schrader quantization maps} $\wedge_\pm : \mathcal K_{+,0} \to  \mathfrak{H}_{-\frac{1}{2}}(\mathbb{R} ^{d-1})$,
	\begin{equation}
	\label{O-S quantization maps-flat space}
	\widehat{f}^{\; \pm} = \int_{0}^{\infty} {\rm e}^{-t {{\mu}_{\pm}}} f_{t}\,dt  \; .  
	\end{equation}
Then, it follows that
\begin{align}	\langle f, \vartheta D_{\vec {\it v}} g \rangle_{L_{2}(\mathbb{R} ^{d})}
		& = \bigl\langle \widehat{f}^{\; +} , \widehat{g}^{\; +} \bigr\rangle_{\mathfrak{H}_{-\frac{1}{2}}} \\
	\langle f,  D_{\vec {\it v}}\,\vartheta\, g \rangle_{L_{2}(\mathbb{R} ^{d})}
		& = \bigl\langle \widehat{f}^{\; -} , \widehat{g}^{\; -} \bigr\rangle_{\mathfrak{H}_{-\frac{1}{2}}}\;.
	\end{align}
\end{proposition}

\subsection{The Classical Gaussian Field\label{Sect:GaussianNeutralField}}
The neutral field  $\Phi(x)$  acts on~$\mathcal{E} $ as a sesquilinear form defined as  a  linear function 
of commuting coordinates 
\[
	\widetilde Q (k)=\widetilde Q(-k)^{*} \; .  
\]
The latter operators are  linear functions of the creation and annihilation operators on $\mathcal{E} $;
see \S II of \cite{JJM Framework}.  We set
\begin{equation}
	\label{Field_Form}
        		\Phi(x) = (2\pi)^{-d/2} \int \widetilde Q (k)\, \widetilde{\sigma}(k) \,{\rm e}^{ik\cdot x}\,dk \;,
\end{equation}
and note that    
\[
        \Phi(x)^{*}   = (2\pi)^{-d/2} \int \widetilde Q (k)\,\overline{ \widetilde{\sigma}(-k)} \,{\rm e}^{ik\cdot  x}\,dk\;.
\]
The expectations of products of such fields in $\Omega_{0}^{E}$ obey a Gaussian recursion relation,
\[
		S_{n}(f) = (n-1) S_{2}(f)\,S_{n-2}(f),
\]
where $S_{n} (f)= \langle \Omega_{0}^{E} , \Phi(f)^{n}\Omega_{0}^{E} \rangle$. Permutation symmetry ensures that $\langle \Omega_{0}^{E}, \Phi(f_{1})\cdots\Phi(f_{n})\Omega_{0}^{E} \rangle$ 
is fixed uniquely through polarization.  The expectation of the product of two fields equals the propagator 
$D_{\vec {\it v}} \, $, with integral kernel	
\begin{align*}
		\langle \Omega_{0}^{E}, \Phi(x)\,\Phi(x') \Omega_{0}^{E} \rangle
		&=  D_{\vec {\it v}}(x-x')
		\\
		&= (2\pi)^{-d} \,
			\int \widetilde\sigma(k)\,\widetilde \sigma(-k)
				\,{\rm e}^{ik(x-x')}\,dk
		\\
		&=  (\sigma \sigma^{\rm T})(x,x')\;.
\end{align*}
Furthermore, the estimate \eqref{AbsD Bound} shows that  $\Phi(f)\Omega_{0}^{E}$ satisfies the bound 
\begin{align*}
		\|   \Phi(f)\Omega_{0}^{E} \|^{2}
		&=\langle \Omega_{0}^{E}, \Phi(f)^{*} \Phi(f) \Omega_{0}^{E} \rangle
		\\ & = \langle f,\overline{\sigma} \sigma^{\rm T} f \rangle
		\\ & = \|  \sigma^{\rm T} f  \|^{2}_{L_{2}}
		\nonumber \\
		&= \|   | D_{\vec {\it v}}| ^{1/2} f  \|^{2}_{L_{2}}
		\\ & \leqslant  (\cosh^{5}\beta) \, \|  C^{1/2}f  \|^{2}_{L_{2}}\;.
\end{align*}

While $D_{\vec {\it v}}$ does not determine $\sigma_{\vec {\it v}}$ uniquely, an elementary solution is to define 
$\sigma_{\vec {\it v}}$ as a square root of  $D_{\vec {\it v}}$.  Proposition~\ref{Prop:ElementaryPropoerites DLorentz} shows that~$D_{\vec {\it v}}$ has a positive real part, so we can define its square root as also having a positive real part.  
In Fourier space, this square root depends continuously on~$k$.  Write 
\begin{align}
        \sigma_{\vec {\it v}}
        = D_{\vec {\it v}}^{1/2}
\end{align}        
or, in Fourier space,
\begin{align} 
        \widetilde \sigma_{\vec {\it v}}(k)
        = (2\pi)^{d/4}\,\widetilde D_{\vec {\it v}}(k)^{1/2}\;.
    \label{D-as-Root}
\end{align}
In configuration space, one has for this example,
\begin{eqnarray*}
		\sigma_{\vec {\it v}}
		&=& \left( (-i \tfrac{\partial}{\partial t} 
		+ \nabla_{\vec x} \cdot \vec {\it v})^{2} -\nabla_{\vec x}^{2} + m^{2} \right)^{-1/2}\nonumber \\
		&=& \left( -\Delta + m^{2} 
		+ (\nabla_{\vec x}\cdot \vec {\it v} )^{2} 
		-2i\tfrac{\partial}{\partial t} (\nabla_{\vec x}\cdot \vec {\it v}) \right)^{-1/2}\;.
\end{eqnarray*}
Correspondingly, the formula for $D_{\vec {\it v}}$ in configuration space is 
	\begin{equation}
	\label{Dv Configuration Space}
		D_{\vec {\it v}}
		=  \left( -\Delta + m^{2} + (\nabla_{\vec x}\cdot \vec {\it v} )^{2} -2i\tfrac{\partial}{\partial t} 
			(\nabla_{\vec x}\cdot \vec {\it v}) \right)^{-1}\;.
	\end{equation}

\begin{proposition}[\bf Properties of
the  Classical Field]\label{Prop:Sigma-Properties-Lorentz-D} Let $D_{\vec {\it v}}$ have the form presented in~\eqref{D Fourier}, and let the operator $\sigma_{\vec {\it v}}$ be given by \eqref{D-as-Root}.  Then  
\begin{equation}
\label{Sigma-D-transformation Properties}
    	\sigma_{\vec {\it v}}
	=\sigma_{\vec {\it v}}^{\rm T}\;,\quad
        \vartheta\,\sigma_{\vec {\it v}}^{\rm T}\,\vartheta
        = \sigma_{\vec {\it v}}^*
        \quad \text{and}\quad
        \pi_{\vec {\it v}}\,\sigma_{\vec {\it v}}^{\rm T}\,\pi_{\vec {\it v}}
        = \sigma_{\vec {\it v}}^*\;.
\end{equation}
Also,
	\begin{equation}
	\label{Dv Reflection-Adjoint Relation}
		\vartheta D_{\vec {\it v}} \vartheta
		= D_{\vec {\it v}}^{*}\;.
	\end{equation}
The field  $\Phi$ transforms under time and spatial reflections as
\begin{equation}
        \Theta\,\Phi(x)\,\Theta
        		= \Phi(\vartheta x)^*\;
        		 \quad \text{and} \quad
       		\Pi_{\vec n}\,\Phi(x)\,\Pi_{\vec n}
        		= \Phi(\pi_{\vec n} x)^*\;.
	\label{Field Transformation}
\end{equation}
The field  $\Phi(x)$ is hermitian if and only if  $\vec {\it v}=0$.
\end{proposition}

\begin{proof} The operator $D_{\vec {\it v}}=D_{\vec {\it v}}^{\rm T}$ is symmetric; thus, in Fourier space 
$\widetilde D_{\vec {\it v}}(-k)=\widetilde D_{\vec {\it v}}(k)$. The square root $\widetilde \sigma_{\vec {\it v}}(k)$ 
has a positive real part and is continuous in $k$, so it satisfies $\widetilde \sigma_{\vec {\it v}}(-k)= \widetilde \sigma_{\vec {\it v}}(k)$.  
Hence, the operator $\sigma_{\vec {\it v}}$ is also symmetric.  The field $\Phi(x)$ is hermitian if 
$\widetilde \sigma(k)=\overline{\widetilde \sigma(-k)}$, namely if the operator~$\sigma$ is real.  Therefore, the field $\Phi(x)$ 
is hermitian on $\mathcal{E} $ only in the case that $\vec {\it v}=0$. 
 
In Proposition \ref{Prop:ElementaryPropoerites DLorentz}, we showed that
$\vartheta D_{\vec {\it v}}$ is self-adjoint, so 
	\[
		\widetilde \sigma_{\vec {\it v}}(\vartheta k)^2=\overline{\widetilde\sigma_{\vec {\it v}}(k)}^2 \; . 
	\] 
As  $(2\pi)^{d/2}\widetilde D_{\vec {\it v}}(k)=\widetilde\sigma_{\vec {\it v}}(k)^2$ has a positive real part, its square root with 
positive real part also satisfies $\widetilde \sigma_{\vec {\it v}}(\vartheta k)=\overline{\widetilde\sigma_{\vec {\it v}}(k)}$.  The 
Fourier transform of this relation is equivalent to the second identity in \eqref{Sigma-D-transformation Properties}. 
The proof of the third identity is similar. The relation \eqref{Dv Reflection-Adjoint Relation} follows 
from \eqref{Sigma-D-transformation Properties} and $D_{\vec {\it v}}=\sigma_{\vec {\it v}}^{2}$.  We have shown the equivalence 
of the transformation properties \eqref{Field Transformation} and \eqref{Sigma-D-transformation Properties} for fields of 
the form~\eqref{Field_Form} in Propositions II.1 and II.5  of  \cite{JJM Framework}.  
\end{proof}	

\subsection{The Gaussian Quantum Field\label{Sect:GaussianQuantumField}}
The time-zero quantum field results from the quantization 
	\[
		\varphi(\vec x) = \widehat{\Phi(0,\vec x)}
	\]
of the time-zero classical field and acts on the Fock space
	\[
	\label{Fock Space}
        \mathcal{H}
        		= \mathbb{C}  \oplus
        			\bigoplus_{n=1}^\infty \,\mathcal{H}_n\;,
         		\quad \text{where} \quad 
        		\mathcal{H}_n
        			= \underbrace{\mathcal{H}_1 \otimes_s
                			\cdots\otimes_s \mathcal{H}_1}_{n\ \rm factors}\;,
	\]
with $ \mathcal{H}_{1}= \mathfrak H_{-1/2}(\mathbb R^{d-1})$. Recall that $\mathfrak{H}_{-\frac{1}{2}}(\mathbb R^{d-1})$ 
is the Sobolev space defined in \eqref{Sobolev-Space3}, which is just the usual one-particle space for the real, free 
scalar field. The Gaussian nature of the Fock space~$  \mathcal{E}  $, together with the fact 
that~$\Phi(f)$ maps $  \mathcal{E}  _{+}$ into $  \mathcal{E}  _{+} \, $, implies that 
	\[
	\varphi(h)=\widehat{\Phi(0,h)}=\widehat{\Phi(f)}\; , \qquad f= \delta \otimes h\; .  
	\]
For a real test-function $h$, the reflection property determined by Proposition~\ref{Prop:Sigma-Properties-Lorentz-D} yields
\begin{align*}
		\| \varphi(h)^{n}\Omega_{0} \|^{2}_{\mathcal{H}}
		&=\langle \Phi(f)^{n}\Omega_{0}^{E} , \Theta \Omega_{0}^{E} \rangle_{  \mathcal{E}  }  \\
		&=\langle\Omega_{0}^{E}, (\Theta\Phi(f)^{*}\Theta)^{n}\Phi(f)^{n} \Omega_{0}^{E} \rangle_{  \mathcal{E}  } \\
		&= \langle \Omega_{0}^{E}, \Phi(f)^{2n} \Omega_{0}^{E} \rangle_{  \mathcal{E}  }
		\\ & = (2n-1)!!\langle h, h \rangle^{n}_{\mathfrak{H}_{-\frac{1}{2}}(\mathbb{R}^{d-1})}\;.\ 
\end{align*}
Expressed in  terms of the creation operators for the free field $a^{*}(\vec x)$ 
one has that, for the Fock-space zero-particle vector~$\Omega_{0}$, 
\begin{align*}
		\| \varphi(h)\Omega_{0} \|^{2}_{\mathcal{H}}
		& = \langle h,h \rangle_{\mathfrak{H}_{-\frac{1}{2}}(\mathbb{R}^{d-1})} 
		\\ & = \|  a^{*}(h)\Omega_{0}  \|^{2}_{\mathcal{H}}	
		\\ & = \|  a^{*}((2\mu)^{-1/2}h)\Omega_{0}  \|^{2}_{L^2 (\mathbb{R}^{d-1})}\;.	
\end{align*}
This shows that the time-zero field has the same expectations as the time-zero \emph{free} field.  Furthermore, the time-zero quantum fields generate an abelian algebra. 

The expectations of products of classical fields on $\mathcal{E}$ satisfy Gaussian recursion relations, so their quantizations  also satisfy  Gaussian recursion relations. The expectations $\langle \Omega_{0}, \varphi(h_{1})\cdots\varphi(h_{n}) \Omega_{0} \rangle$ can be obtained from the expectations of $\langle \Omega_{0},  \varphi(h)^{n} \Omega_{0} \rangle$ by polarization.  In this case,
	\[
		\| \varphi(h)\Omega_{0} \|^{2}
		= \langle h_{1},h_{1} \rangle_{\mathfrak{H}_{-\frac{1}{2}}}
		+  \langle h_{2},h_{2} \rangle_{\mathfrak{H}_{-\frac{1}{2}}}
		=  \langle h,h \rangle_{\mathfrak{H}_{-\frac{1}{2}}}\;,
	\] 
as the scalar product in $\mathfrak{H}_{-\frac{1}{2}}$ is hermitian. Thus, the time-zero field on~${\mathcal H}$ is hermitian 
and has the form
	\[
		\varphi(\vec x)
		=  (2\pi)^{-(d-1)/2}
		\int \left( a(\vec k)^{*}+a(-\vec k) \right) {\rm e}^{-i\vec k\cdot\vec x}
%
%
%
		\,d\vec k\;.
	\]

The Hamiltonian $H_{\pm} = H_{\rm free} \pm \vec P\cdot \vec {\it v}$ acts on the $n$-particle subspace as a direct sum of the $1$-particle Hamiltonians  $\mu_{\pm}=\mu\pm\delta$. Recall that $H_{\rm free}$ and $\vec P$ are the free-field Hamiltonian and momentum operator on $\mathcal{H}$. Setting
\begin{align*}
\varphi_\pm (t, \vec x) &= {\rm e}^{it H_\pm}\varphi (\vec x) {\rm e}^{-it H_\pm}\;,  
\end{align*}
we obtain Wightman functions 
\begin{align*}
&{\mathcal W}^{(n)} \bigl(\Lambda^{-1}_\pm (t_1, \vec 0) +  (0, \vec x_1) ,\ldots, \Lambda^{-1}_\pm  (t_n, \vec 0) + (0, \vec x_n) \bigr) \\
& \qquad = \langle \Omega_0 , \varphi_\pm (t_1, \vec x_1) \cdots \varphi_\pm (t_n, \vec x_n) \Omega_0 \rangle \;,
\end{align*}
where the Lorentz transformation $\Lambda_{\pm}$ gives the boost by velocity $\pm\vec v$.   

\subsection{Spatial Reflection Positivity}
In order to establish spatial reflection positivity, we proceed as in Section \ref{Sect:TimeRP} but evaluate the Fourier transform 
in a spatial direction.  Let $k=(E,\vec k)\in\mathbb{R} ^{d}$ and let  $\vec k^{\perp}\in\mathbb{R} ^{d-2}$ denote the 
component of $\vec k\in\mathbb{R} ^{d-1}$ in the subspace of dimension $d-2$ orthogonal to the vector $\vec n$.  
Also, let $\nu=\nu(\vec n,  E,  \vec k^{\perp})$ be the positive square root
	\begin{equation*}
		\nu = \left(E^{2} + \tfrac{\vec k ^{\perp\,2}   + m^{2}}{\cosh^{2}\beta } \right)^{1/2}
		\geqslant (E^{2}+ \tfrac{m^{2}}{\cosh^{2}\beta } )^{1/2}
		>   |  E  | \;.
	\end{equation*}
In particular, $\nu\pm E\tanh\beta>0$. Define the one-particle Sobolev 
space~$\widetilde { \mathfrak{H} }_{-\frac{1}{2}} (\mathbb{R} ^{d-1})$ as the Hilbert space of functions  with coordinates $(E,\vec k^{\perp})\in\mathbb{R} ^{d-1}$ and with inner product 
\begin{equation}
	\label{Spatial-Sobolev Space}
		\bigl \langle \cdot , \cdot \bigr\rangle_{\widetilde{\mathfrak{H}}_{-\frac{1}{2}} (\mathbb{R} ^{d-1})}
= \left \langle \tfrac{1}{\sqrt{2\nu}} \, \cdot ,  \tfrac{1}{\sqrt{2\nu}} \, \cdot \right \rangle_{L_{2}(\mathbb{R} ^{d-1} ; d E \, d \vec k^{\perp} )}\;.
\end{equation}
Let  $\nu$ also denote the corresponding pseudo-differential operator acting in configuration space,
	\begin{equation*}
		\nu
		= \left( - \tfrac{\partial^{2}}{\partial t^{2} } +
			\tfrac{-\nabla_{\vec x}^{2} + ( \vec n\cdot\nabla_{\vec x})^{2}  +m^{2} }{\cosh^{2}\beta } \right)^{1/2}\;.
	\end{equation*}
Finally, let $U(s)$ denote translation in the coordinate direction. 

\begin{proposition}
\label{Theorem:Spatial-RP} The
operators $\pi_{\vec n} D_{\vec {\it v}}$ and $D_{\vec {\it v}}\,\pi_{\vec n}$ have both positive 
expectations on $L_2(\boldsymbol{X}_{\vec n+}  ; d E \, d \vec  k^{\perp} )$.  
The corresponding Osterwalder-Schrader Hamiltonians $ {\nu}_+$ and $ {\nu}_-$   
(for $\pi_{\vec n}  D_{\vec {\it v}}$ and for $ D_{\vec {\it v}}\,\pi_{\vec n} $) both act on the 
Sobolev space of functions \textcolor{black}{$\widetilde{\mathfrak{H}}_{-\frac{1}{2}}(\mathbb{R} ^{d-1})$} 
defined in \eqref{Spatial-Sobolev Space}.   In momentum space, the explicit forms of the Hamiltonians are    
\begin{equation}
	\label{Spatial-Hamiltonian}
	 {\nu}_{\pm}	= ( \cosh^{2}\beta )  ( \nu \pm E\tanh\beta)\;.
\end{equation}
\end{proposition}

\begin{remark}\rm
In case $d=2$, one has  $\vec k^{\perp}=0$, so \textcolor{black}{$\nu$ is only a function of the energy, $m$ and $\beta$. In this case,}
\[
		\nu
		= \left( E^{2} + \tfrac{m^{2}}{\cosh^{2}\beta } \right)^{1/2}\;,
\]
and
\begin{equation}
	\label{nu(E)}
		{\nu}_{\pm} 
		= \cosh^{2}\beta \left( \left( E^{2} 
			+ \tfrac{m^{2}}{\cosh^{2}\beta} \right)^{1/2} \pm E \tanh\beta \right) \;.
\end{equation}
This has a similar form to ${\mu}_{\pm}$ of \eqref{mu-beta-Defn}, 
but with an overall multiple of $\cosh^{2}\beta$ and with a mass modified  by the factor $(\cosh\beta)^{-1}$. Hence, in configuration space, 
\[
		\nu
		= \left( -\tfrac{\partial^{2}} {\partial t^{2}}  + \tfrac{m^{2}}{\cosh^{2}\beta} \right)^{1/2}\;,
\]
and
\[
		\mathcal{F}^{*} \,  {\nu}_{\pm} \, \mathcal{F}
		= \cosh^{2}\beta \left( ( -\tfrac{\partial^{2}}{\partial t^{2}}  
		+ \tfrac{m^{2}}{\cosh^{2}\beta})^{1/2}\pm i \tfrac{\partial}{\partial t}  \tanh\beta \right) \;.
\]
For $d>2$, one must also scale the remaining spatial variables 
by $\cosh\beta$.
\end{remark}

\begin{proof}
Under rotations, the operator $D_{\vec {\it v}}$ transforms by the rotation 
of~$\vec n$, so it is no loss of generality to assume that $\vec n$ points in the 
direction of the coordinate~$x_{1}$.  In  Fourier space $\delta =k_{1}\tanh\beta$.   Write the inverse of $(2\pi)^{d/2} \widetilde D_{\vec {\it v}}(k)$ as 
\begin{align*}
		(E+i\delta)^{2}+\mu^{2}
		&= \tfrac{1}{\cosh^{2}\beta} 
			\Bigl( k_{1}^{2}  
			+ ( iE  \sinh 2 \beta) k_{1}
			\nonumber \\
			& \qquad  
			+ \bigl( E^{2} + \vec k^{\,\perp\,2} + m^{2} \bigr)\cosh^{2}\beta  \Bigr)  \nonumber \\ 
		&=\tfrac{1}{\cosh^{2}\beta} ( k_{1}-ik_{+} ) ( k_{1}-ik_{-}) \;,	
\end{align*}
where the roots $0< k_{+}, -k_{-}$ are given by 
	\[
		k_{\pm}
		= ( \pm \nu  - E\tanh\beta ) \cosh^{2}\beta
	\; .
	\]
With this information, one can evaluate the integral 
\begin{align*}
		(2\pi)^{-1} \int_{-\infty}^{\infty} 
		\frac{{\rm e}^{ik_{1}\xi}}{(E+i\delta)^{2}+\mu^{2}}\,
		 \,d k_{1} & =\frac{\cosh^{2}\beta}{k_{+}-k_{-}}\,
		\begin{cases}
			{\rm e}^{-k_{+}\xi} \\
		{\rm e}^{-k_{-}\xi} 
		\end{cases}
		\\ & = \frac{\cosh\beta}{2\nu}\,
		\begin{cases}
			{\rm e}^{-k_{+}\xi}\;, 	&\text{if $\xi>0$} \\
			{\rm e}^{-k_{-}\xi}\;, 	&\text{if $\xi<0$} \;. 
		\end{cases}
		 \;
\end{align*}
Replace $\xi$ by $x_{1}-x'_{1}$ and apply the reflection $\pi_{\vec n}$.  Infer that the 
integral kernel $( \pi_{\vec n}D_{\vec {\it v}} ) (x,x')$ of $\pi_{\vec n}D_{\vec {\it v}} \, $, acting on functions 
in $L_{2}(\mathbb{R} ^{d}_{\vec n+};  dx )$, is 
	\[
		( \pi_{\vec n}D_{\vec {\it v}} ) (x,x')
		= \left(  \frac{{\rm e}^{k_{-}
		(x^{\phantom{\prime}}_{1}+x'_{1})}}{2\nu}\,\right)(t-t', x^{\phantom{\prime}}_{2}-x'_{2}, 
		\ldots, x^{\phantom{\prime}}_{d}-x'_{d})\;.
	\]
Thus, conclude that spatial reflection positivity holds for $\pi_{\vec n} \mathcal{D}_{\vec {\it v}}$, 
and that its one-particle Osterwalder-Schrader Hamiltonian is 
	\[
		{\nu}_{+}
		= -k_{-} = (\nu + E\tanh\beta)\cosh^{2}\beta\;,
	\]
which acts naturally on the one-particle Hilbert space $\widetilde {\mathfrak{H}}_{-\frac{1}{2}}$ defined 
in \eqref{Spatial-Sobolev Space}.
\goodbreak

Repeating the same argument for $D_{\vec {\it v}}\,\pi_{\vec n} \, $,  find that 
	\[
		(D_{\vec {\it v}}\,\pi_{\vec n})(x,x')
		= \left( \frac{{\rm e}^{-k_{+}
		(x^{\phantom{\prime}}_{1}+x'_{1})}}{2 \nu} \right)(t-t', x^{\phantom{\prime}}_{2}-x'_{2}, 
		\ldots, x^{\phantom{\prime}}_{d}-x'_{d})\;.
	\]
This shows that reflection positivity holds for $D_{\vec {\it v}}\,\pi_{\vec n} \, $, 
and that its Osterwalder-Schrader Hamiltonian is
	\[
		{\nu}_{-} = k_{+} = (\nu - E\tanh\beta)\cosh^{2}\beta\;,
	\]
which acts on the same one-particle space $\widetilde {\mathfrak{H}}_{-\frac{1}{2}}$ 
as~${\nu}_{+}$. 
\end{proof}	

\subsection{Quantization of Spatial Reflection Positivity}

We use coordinates such that $x = (t, x_{\vec n}^{\,\perp}, x_{\vec n})\in \mathbb{R}^d$ to introduce the $x_{\vec n}=0$  quantum field~$\widetilde\varphi(t)$ as the quantization of $\Phi(t,x^{\,\perp}, 0)$ with respect to  the inner product determined by the matrix elements of  $\Pi_{\vec n}$ 
on $\widetilde   {\mathcal{E}}_{+} \equiv  \mathcal{E}_{\vec n+}$, namely,
	\[
		\widetilde \varphi(t, x^{\,\perp}) = \widetilde {\Phi(t,x^{\,\perp}, 0)}.
	\]
 The quantum field $\widetilde\varphi(t)$ acts on the Fock space
	\[
        \widetilde{\mathcal{H}}
        		= \mathbb{C}  \oplus
        			\bigoplus_{n=1}^\infty \,\widetilde{\mathcal{H}}_n\;,
         		\quad \text{where} \quad 
        		\widetilde{\mathcal{H}}_n
        			= \underbrace{\widetilde{\mathcal{H}}_1 \otimes_s
                			\cdots\otimes_s \widetilde{\mathcal{H}}_1}_{n\ \rm factors}\;,
	\]
where $\widetilde{\mathcal{H}}_1=\widetilde{\mathfrak{H}}_{-\frac{1}{2}} $ is the spatial one-particle Sobolev space defined 
in~\eqref{Spatial-Sobolev Space}.  The scalar $1 \in  \mathbb{C}$ is the standard zero-particle state $\widetilde{\Omega}_{0}$ in $\widetilde{\mathcal{H}}$. Note that  $\Pi$ maps 
$\widetilde {\mathcal{E}}_{+}$ to $\widetilde {\mathcal{E}}_{-}$, 
and also $\pi_{\vec n} (g\otimes \delta)=g\otimes\delta$.  Thus, for $f=g\otimes \delta$ real-valued,  
	\[
	\widetilde \varphi (g)=\int \widetilde \varphi(t, x^{\,\perp}) g(t, x^{\,\perp}) \, dt \, d x^{\,\perp} =\widetilde{\Phi(f)}
	\]
and
\begin{align*}
		\| \widetilde\varphi(g)^{n} \widetilde{\Omega}_{0} \|^{2}_{\mathcal{H}}
		&=\langle \Phi(f)^{n}\Omega_{0}^{E} , \Pi \Phi(f)^{n}\Omega_{0}^{E} \rangle_{  \mathcal{E}  }  \\
		&=\langle \Omega_{0}^{E}, (\Pi\Phi(f)^{*}\Pi)^{n}\Phi(f)^{n}\Omega_{0}^{E} \rangle_{  \mathcal{E}  } \\
		&=\langle \Omega_{0}^{E}, \Phi(f)^{2n} \Omega_{0}^{E} \rangle_{  \mathcal{E}  }
		\\ & = (2n-1)!!  \langle g,g \rangle_{\widetilde{ \mathfrak{H}}_{-\frac{1}{2}}}^{n}\;.\ 
\end{align*}
As in  \S III.1 of \cite{JJM Framework}, we infer that the field~$\widetilde \varphi$ can be expressed in terms of  creation operators~$\tilde a^{*}$ and (their adjoint)
annihilation operators $\tilde a$, using  the function~$\nu(E)$ defined in~\eqref{nu(E)}. 
Specializing to the two-dimensional case, we have
	\[
		\widetilde\varphi(t)
		=  (2\pi)^{-1/2}  
		\int  \frac{dE}{\sqrt{2\nu(E)}} \bigl( \tilde a(E)^{*}+ \tilde a(-E) \bigr) \,{\rm e}^{-iEt}  \;.
	\]

\setcounter{equation}{0}
\section{Classical Fields on the Cylinder $\boldsymbol{X}= \mathbb{R} \times  \mathbb{T}^{d-1}$\label{The Spatially Compactified Case}}

In this section we study periodization of the spatial directions. We are especially interested in the spacetime $\boldsymbol{X} =   \mathbb{R} \times  \mathbb{T}^{d-1}$ with  $\mathbb{T}^{d-1} = S^{1} \times \cdots \times S^{1}$ denoting the $d-1$ dimensional torus. Let $\Lambda=\prod_{j=1}^{d-1}\ell_{j}$ denote the spatial volume, where $\ell_{j}$ is the circumference of the $j^{\text{th}}$ constituent circle.

Define the quantum-mechanical Fock space
\begin{equation}
	\label{HLambda}
	        {\mathcal H}_{\Lambda}
        = \mathbb{C}  \oplus
        \bigoplus_{n=1}^\infty \,\mathcal{H}_n\;,
         \quad \text{where} \quad 
        \mathcal{H}_n
        = \underbrace{\mathcal{H}_1 \otimes_s
                \cdots\otimes_s \mathcal{H}_1}_{n\ \rm factors}\;,
\end{equation}
with $ \mathcal{H}_{1}= L_{2}(\Lambda ; d \vec x \, )$. As in previous sections, the scalar $1 \in \mathbb{C}$ denotes the zero-particle vector $\Omega_0$.

The  positive Hamiltonians~$H_{\pm}(\Lambda)$ arise as the quantization of the one-particle Hamiltonian ${\mu}_{\pm}$ given in \eqref{mu-beta-Defn}. The form of the one-particle Hamiltonian ${\mu}_{\pm}(\vec k)$ on the compactified spatial torus $\Lambda$ is the same as that on Fourier space $\mathbb R^{d-1}$. For the model on the torus $\mathbb{T}^{d-1}$, however, the momenta $\vec k$ lie in the lattice $\mathcal K_{\Lambda}=\bigoplus_{j=1}^{d-1} \frac{2\pi}{ \ell_{j}} \mathbb Z$ dual to~$\mathbb{T}^{d-1}$.

The time-zero field 
\[ 
\varphi(\alpha)=\int\varphi(\vec x)\alpha(\vec x) \, d\vec x \; , \qquad \alpha \in \mathfrak{H}_{-\frac{1}{2}} (\mathbb{T}^{d-1}) \; .  
\] 
acting on ${\mathcal H}_{\Lambda}$ has domain $\mathcal D(H_{\pm}(\Lambda))^{1/2}\subset \mathcal 
D(H_{\text{free}}(\Lambda))^{1/2}$. Thus\textcolor{red}{,} the imaginary-time field 
	\[
		\varphi_I^{\pm}(t, \alpha)
		= {\rm e}^{-tH_{\pm}} \,\varphi(\alpha)\,{\rm e}^{tH_{\pm}}\;,
	\]
has the dense domain ${\rm e}^{-(t+\epsilon)H_{\pm}}{\mathcal H}_{\Lambda}$ for any $\epsilon>0$. As a form on $C^{\infty}(H_{\pm})\times C^{\infty}(H_{\pm})$,
	\begin{align*}
		 \varphi_I^{\pm}(t,\vec x) 
		&  = \frac{1}{\sqrt{\Lambda}}\sum_{k\in\mathcal K_{\Lambda}}
		\frac{1}{\sqrt{2\mu(\vec k)}}  \left( a(\vec k)^{*}{\rm e}^{-t{\mu}_{\mp} (\vec k)}   +  
		a(-\vec k){\rm e}^{t{\mu}_{\pm} (\vec k)} \right) {\rm e}^{-i\vec k \cdot \vec x}\;.
	\nonumber
	\end{align*}
Introducing the anti-time-ordering operator $\mathbb{A}$ for products of imaginary-time fields 
where, for example,
	\[
		\mathbb{A} \varphi_I^{\pm}(x)\varphi_I^{\pm}(x')
		= \theta(t'-t) \varphi_I^{\pm}(x)\varphi_I^{\pm}(x')
		+  \theta(t-t') \varphi_I^{\pm}(x')\varphi_I^{\pm}(x)\;, 
	\]
the two-point function $D_{\vec {\it v}, \Lambda} $ can be expressed as
	\[
	D_{\vec {\it v}, \Lambda}  (x-x')=  \langle \Omega_0 \, \mathbb{A} \varphi_I^{\pm}(x)\varphi_I^{\pm}(x')  \Omega_0 \rangle_{{\mathcal H}_{\Lambda}} \; . 
	\]
We present similar formulas for the completely compactified case in a later section.

\begin{remark}
\label{Gibbsremark}
The free Hamiltonian  $H_{\pm}(\Lambda)$ on the spacetime $\mathbb R\times \Lambda$ is positive 
and has a trace-class 
heat kernel, which allows us to define the partition function
\begin{align*}
		\mathfrak {Z}_{\pm,\beta,\Lambda} & = {\rm Tr}\; {\rm e}^{-\beta H_{\pm}(\Lambda)}
		\\ & = \prod_{\vec k\in\mathcal K_{\Lambda}} \frac{1}{1-{\rm e}^{-\beta {\mu}_{\pm}(\vec k)}}
		\\ & = \prod_{\vec k\in\mathcal K_{\Lambda}} \left( 1+\rho_{\pm}(\vec k) \right)\;.
\end{align*}
The corresponding {\em Gibbs states}  are  
	\[
		\langle \ \cdot\ \rangle_{\pm,\beta,\Lambda}
		= \frac{ {\rm Tr} (\ \cdot \ {\rm e}^{-\beta H_{\pm}(\Lambda)} ) }
			{\mathfrak{ Z}_{\pm,\beta,\Lambda}}\;.
	\]
Cyclicity of the trace shows that $\langle \ \cdot\ \rangle_{\beta,\pm,\Lambda}$ is invariant under the adjoint action of the 
unitary group ${\rm e}^{itH_{\pm}(\Lambda)}$.  Furthermore,  $\Vert \langle \ \cdot\  \rangle_{\beta,\pm,\Lambda} \Vert=1$.  

Positivity of the Hamiltonian $H_{\pm}$ implies that the functions 
	\[
		F^{\pm}_{A,B}(t)=
		\langle A {\rm e}^{it H_\pm} B \rangle_{\pm,\beta,\Lambda} \; , \qquad A, B \in {\mathcal B}({\mathcal H}_\Lambda) \; , 	
	\]
extend to analytic functions in the strip $z=t+is$ with  $0<s<\beta$. Using the H\"older inequality for Schatten 
norms, one concludes that the function $F_{\pm}(z)$ is bounded uniformly inside the strip by 
$\vert F_{\pm}(z)\vert \leqslant \Vert A\Vert \,\Vert B\Vert$. Moreover, $F^{\pm}_{A,B}(t)$ satisfies the {\em KMS  property} 
	\[
		F^{\pm}_{A,B}(t+i\beta)
		= \langle B {\rm e}^{-it H_\pm} A  \rangle_{\pm,\beta,\Lambda}	\;  . 
	\]
 This property characterizes thermal states \cite{Kubo}\cite{MS}\cite{HHW}. 
\end{remark}

\setcounter{equation}{0}
\section{A Feynman-Kac Formula for ${\mathscr P}(\varphi)_2$ Models\label{Sect:FK}}

In this section we establish a  formula relating expectations of functions of complex fields to certain matrix elements of the heat kernel ${\rm e}^{-t H_{\it{v}}}$.  We call this a generalized Feynman-Kac formula, as the classical side of the identity is a sesquilinear form that is not densely defined.  However, this classical form suffices to obtain a  quantization which is the densely-defined heat kernel of the quantum-mechanical Hamiltonian.   We use this generalized Feynman-Kac formula to establish useful bounds on the quantum-mechanical  matrix elements.  

In most usual cases one defines the classical side of the  Feynman-Kac identity as matrix elements of a densely-defined form that determines an operator on a Hilbert space over path space.  On our examples with interaction, the action functional $\mathfrak A(\Phi)$ is a normal operator on the space $\mathcal E$ of classical fields.  However, we do not have complete information about the domain of $\Theta e^{-\mathfrak A}$, where $ \Theta$ denotes the unitary time reflection on classical fields.
We anticipate that this framework could be helpful in other contexts.

\subsection{Background}
Consider a  spacetime cylinder with the time coordinate $t\in\mathbb R$ and spatial coordinate $ x\in S^{1}$ 
parameterized by  $x\in  [-\tfrac{\ell}{2}  ,\tfrac{\ell}{2} ] $ for $\ell > 0$.  The ${\mathscr P}(\varphi)_2$-Hamil\-tonian is 
\begin{equation}
	\label{Total Hamiltonian}
       H
       = H_{\text{free}}  + H_{\rm int }- E\;,
       \quad\text{where}\quad
       H_{\rm int} 
       = \int_{-\ell/2}^{\ell/2} d x \; \;  
       {:} {\mathscr P}(\varphi(x))  {:} \;,
\end{equation}
where the polynomial ${\mathscr P}$ is bounded from below, and~$:\cdot:$ denotes normal ordering with respect 
to the Fock space vacuum.  The  constant~$E$ equals the infimum of the spectrum of 
$H_{\rm free}+H_{\rm int}$, and $E\leqslant 0$.   
Define 
\begin{align}
	\label{Hv}
    		H_{{\it v}} & = H + P {\it v}   \;,
\end{align}
where ${\it v}=\tanh \beta$. Note that $H_{0}$ refers to the case ${\it v}=0$, and not to the free Hamiltonian that we denote $H_{\rm free}$. 

Glimm and Jaffe showed in  \cite{Glimm-Jaffe-Uniqueness} that $H$ with a spatial cutoff, rather than on a spatial circle, \emph{i.e.}, with periodic boundary conditions, has a  ground state $\Omega$ that is unique up to a phase.  A similar proof applies to $H$ with periodic boundary conditions. Since the Hamiltonian $H$ and the
momentum $P$ commute, they can be simultaneously diagonalized. Both $H_{\rm free}$ and $P$ have purely 
discrete spectra and compact resolvents.  Following Heifets and Osipov, we have:

\begin{proposition}
The Hamiltonian $H_{\it v}$ has pure discrete spectrum, and the heat kernel ${\rm e}^{-t H_{\it v}}$ is trace class on $\mathcal{H}$ for $t > 0$.
\end{proposition}

\begin{proof}
Glimm and Jaffe \cite{Glimm-Jaffe-Uniqueness} proved that given any $\epsilon>0$, there exists $M_{\epsilon}<\infty$ such that 
$\epsilon H_{\rm free} + H_{\rm int}+M_{\epsilon}\geqslant0$.  Moreover, 
	\[
		| {\it v}|\,H_{\rm free}\geqslant \pm P {\it v} \; .  
	\]
Choose $\epsilon>0$,  so   $1-2\epsilon>|{\it v}|$. Hence,  $(1-2\epsilon) H_{\rm free} +P{\it v}\geqslant0$.  It follows that    
	\begin{align*}
		 H_{{\it v}}
		&= \left((1-2\epsilon) H_{\rm free} +P{\it v}\right)+ \left( \epsilon H_{\rm free} + H_{\rm int} +M_{\epsilon}\right) 
			+ \epsilon H_{\rm free} -E-M_{\epsilon}\nonumber\\
		&\geqslant  \epsilon H_{\rm free} -E-M_{\epsilon}\;.
	\end{align*}
In other words, $\epsilon H_{\rm free}\leqslant H_{{\it v}} + M_{\epsilon} +E$ and $H_{{\it v}}$ is relatively compact with respect 
to~$H_{\rm free}$.  Since  $H_{0}$  has pure discrete spectrum, so does $H_{{\it v}}$.  As the heat kernel ${\rm e}^{-t H_{\rm free}}$ 
on the Hilbert space $\mathcal H$ is trace class for all $t>0$, it follows that the heat kernel ${\rm e}^{-tH_{{\it v}}}$ is trace 
class on the Hilbert space~$\mathcal H$ for all $t>0$.  
\end{proof}

\subsection{Operators, Forms, and the Feynman-Kac Formula\label{Sect:ExtendedFK}}
In this section we investigate a Feynman-Kac formula for the matrix elements 
	\begin{equation}
	\label{FKForm}
		\langle  \widehat \Omega , {\rm e}^{-TH_{{\it v}}}\, \widehat {\Omega'} \rangle_{\mathcal H}
		= \langle \Omega , \Theta  {\rm e}^{-\mathfrak A} \,{\mathcal T}(T)\Omega' \rangle_{\mathcal E}\;, 
		\qquad T \in \mathbb{R}^+ \; , 
	\end{equation}
of the heat kernel ${\rm e}^{-TH_{{\it v}}}$, representing them as classical expectations of the time-re\-flection $\Theta$ times the 
exponential ${\rm e}^{-\mathfrak A}$ of an action~$\mathfrak A$  for a time interval of length $T$ 
(corresponding to $t\in [-\frac{T}{2},\frac{T}{2}]$), combined with the free unitary time-translation 
$t \mapsto \mathcal T(t)$ on ${\mathcal E}$. Up to an  additive constant $M$\textcolor{red}{,}  the action~$\mathfrak A$ for the non-linear perturbation of the 
Hamiltonian is the integral\footnote{In the case of a two-dimensional, cylindrical spacetime $\boldsymbol{X}=\mathbb{R} \times S^{1}$, 
we  use the notation $(t,x)$ for a point in time and space, dropping the vector on the spatial component $\vec x$.} of a density   ${:}{\mathscr P}(\Phi(t, \vec x)){:}$, namely,
	\[
		{\mathfrak A} = \int_{-T/2}^{T/2} dt \int_{-\ell/2}^{\ell/2} dx \, {:}{\mathscr P}(\Phi(t, x)){:}  +\ell M  T\;.
	\]
The value of the  constant $M$ is unimportant for our argument, except for the fact that it is finite, and it provides a 
normalizing factor that appears in all expectations.  The function $\mathfrak A$, as well as its  adjoint, are operators 
on $\mathcal E$ with the dense 
domain ${\mathcal D}\subset {\mathcal E}$ consisting of polynomials in $\Phi(f)$ applied to~$\Omega_{0}^{E}$, 
where $f\in C^{\infty}_{0}$. Write $\mathfrak A=\mathfrak C +i \mathfrak D$, where $\mathfrak C$ and 
$\mathfrak D$ are symmetric.  Likewise\textcolor{red}{,} $\mathfrak A^{*}=\mathfrak C-i\mathfrak D$. Similarly, the operators 
\begin{align*}
		{\mathfrak A}_{+} & = \int_{0}^{T/2} dt \int_{-\ell/2}^{\ell/2} dx \, {:}{\mathscr P}(\Phi(t, x)){:}  + \tfrac{1}{2} \ell M  T\; \\
		{\mathfrak A}_{-} & = \int_{-T/2}^{0} dt \int_{-\ell/2}^{\ell/2} dx \, {:}{\mathscr P}(\Phi(t, x)){:}  +\tfrac{1}{2} \ell M  T\; \\
		\widetilde{\mathfrak A}_{+} & = \int_{-T/2}^{T/2} dt \int_{0}^{\ell/2} dx \, {:}{\mathscr P}(\Phi(t, x)){:}  +\tfrac{1}{2} \ell  M  T \; \\		\widetilde{\mathfrak A}_{-}  & = \int_{-T/2}^{T/2} dt \int_{-\ell/2}^{0} dx \, {:}{\mathscr P}(\Phi(t, x)){:}    +\tfrac{1}{2} \ell M  T\;,
\end{align*}
as well as their real and imaginary parts ${\mathfrak C}_{\pm}, {\mathfrak D}_{\pm}, \widetilde {\mathfrak C}_{\pm}$, and $\widetilde {\mathfrak D}_{\pm}$, respectively, are all densely-defined, commuting operators on the domain $\mathcal D$.   
	
\begin{proposition}
The operators $\mathfrak A$,  ${\mathfrak A}_{\pm}$, and $\widetilde {\mathfrak A}_{\pm}$, as well as their adjoints,  
have normal, mutually commuting closures.  We also denote them by $\mathfrak A$, \emph{etc}.  The polar decomposition of 
the exponential of $-\mathfrak A$ is ${\rm e}^{-\mathfrak A}={\rm e}^{-{\mathfrak C}} {\rm e}^{-i{\mathfrak D}}$.  Similarly,  
${\rm e}^{-\mathfrak A_{\pm}}={\rm e}^{-{\mathfrak C}_{\pm}} {\rm e}^{-i{\mathfrak D}_{\pm}}$ and  
${\rm e}^{-\widetilde{\mathfrak A}_{\pm}}={\rm e}^{-\widetilde{\mathfrak C}_{\pm}} {\rm e}^{-i\widetilde{\mathfrak D}_{\pm}}$.  
Furthermore, $\Theta \mathfrak A_{\pm}
=\mathfrak A_{\mp}^{*} \Theta$ and $\Theta {\rm e}^{-\mathfrak A_{\pm}}={\rm e}^{-\mathfrak A_{\mp}^{*}} \Theta$, 
so 
\[ 
\Theta {\rm e}^{-\mathfrak A}={\rm e}^{-\mathfrak A_{+}^{*}} \Theta {\rm e}^{-\mathfrak A_{+}} \; .
\]
\end{proposition}

\begin{proof}   The fact that all these operators are defined on the domain~$\mathcal D$ is  a standard estimate relying on 
the fact that the covariance $D_{{\it v}}$ is bounded from above by the covariance $C$; see \S\ref{Sect:Real Propagator}. The proof 
of the normality and commutativity of the closures is similar to the proof of Proposition 2.1.2 in \cite{Les Houches}.  The 
Fock zero-particle state $\Omega_{0}^{E}$ is cyclic for the maximally-abelian von Neumann algebra generated by  bounded 
functions of the fields $\Phi(f)$.  This algebra commutes with the symmetric operators $\mathfrak A, \mathfrak C, \mathfrak D$, 
{\emph{etc}}. It follows that ${\mathfrak C}$ and ${\mathfrak D}$ are essentially self-adjoint, their closures commute, and $\mathfrak A$ is normal. As a consequence of the definition of the fields, the relation $\Theta \mathfrak A_{\pm} =\mathfrak A_{\mp}^{*} \Theta$ is true on the domain $\mathcal D$.  It extends to the closures by continuity, and hence to the exponentials defined by the functional 
calculus.   Similar arguments apply to the other operators.
\end{proof}

\bigskip
While  the vectors $A_{T/2}\Omega_{0}^{E}$, with $A_{T/2}(\Phi)=A(\Phi_{T/2})$ and $\Phi_{T/2}(t,x)=\Phi(t+\frac{T}{2},x)$, may not be in the operator domain of  ${\rm e}^{-\mathfrak A}$, we  require that they be in the domain of $\Theta {\rm e}^{-\mathfrak A}$ or of 
$\Pi {\rm e}^{-\mathfrak A}$  as sesquilinear forms.  The Feynman-Kac formula and the Schwarz inequalities we use 
involve such  matrix elements.  It is for this reason that we use the adjective {\em extended} for the Feynman-Kac formula.  

\begin{theorem} [Feynman-Kac Density] \label{Theorem:Form Domain E} The form domain of the operators 
$\Theta {\rm e}^{-\mathfrak A}$ and $\Pi {\rm e}^{-\mathfrak A}$ includes $\mathcal D_{0}\times \mathcal D_{0}
\subset \mathcal E\times \mathcal E$, where $\mathcal D_{0}=\{\Phi(f_{T/2})^{n}\Omega_{0}^{E}\}$ with  $f\in C^{\infty}_{0}$  
supported in the time-interval $[0,1]\times [0,\frac{\ell}{2}]$.  For $\Omega, \Omega' \in \mathcal D_{0}$, the Feynman-Kac formula \eqref{FKForm} holds.  Moreover, one has the Schwarz inequalities
	\begin{equation}
		\langle \Omega, \Pi {\rm e}^{-\mathfrak A} \Omega' \rangle_{\mathcal E}
		\leqslant \langle \Omega, \Pi {\rm e}^{-\mathfrak A} \Omega \rangle_{\mathcal E}^{1/2}
		\langle \Omega', \Pi {\rm e}^{-\mathfrak A} \Omega' \rangle_{\mathcal E}^{1/2}\;,
	\label{Schwarz1}
	\end{equation}
and
	\begin{equation}
		\langle \Omega, \Theta  {\rm e}^{-\mathfrak A} \Omega' \rangle_{\mathcal E}
		\leqslant \langle \Omega, \Theta {\rm e}^{-\mathfrak A} \Omega \rangle_{\mathcal E}^{1/2}
		\langle \Omega', \Theta {\rm e}^{-\mathfrak A} \Omega' \rangle_{\mathcal E}^{1/2}\;.
	\end{equation}
\end{theorem}

\begin{lemma}\label{Lemma:Bound2}
Let $A,B, A+B$ be essentially self-adjoint operators on a Hilbert space $\mathcal H$, each bounded from above. Let  $C$ be self-adjoint and  commuting with the closures of $A, B$,  and $A+B$.  Then 
	\begin{equation}
	\Vert C{\rm e}^{A+B}\Vert
	\leqslant \Vert {\rm e}^{A}\Vert\ \Vert C{\rm e}^{B}\Vert\;.
	\label{CAB Bound}
	\end{equation} 
\end{lemma}

\begin{proof}
First, consider the case that $C=\mathbb{1}$.  As $A$ and $B$ are bounded from above, the operators 
${\rm e}^{A}$, ${\rm e}^{B}$, and ${\rm e}^{A+B}$ are each bounded, too.  Moreover, essential self-adjointness assures that Chernoff's version of the Trotter product formula applies; see the Corollary to the 
Theorem in \cite{Chernoff}. Therefore, 
	\[
		{\rm e}^{A+B}
	= {\rm st.}\hskip-.3em\lim_{n\to\infty} \bigl({\rm e}^{A/n}\,  {\rm e}^{B/n} \bigr)^{n}\;.
	\]
Consequently,
\begin{align*}
	\Vert {\rm e}^{A+B}\Vert
	& = \bigl\|  {\rm st.}\hskip-.3em\lim_{n\to\infty} \bigl({\rm e}^{A/n}\,  {\rm e}^{B/n}\bigr)^{n} \bigr\|
	\\ & \leqslant \lim_{n\to\infty} \Vert {\rm e}^{A/n}\Vert^{n}\ \Vert {\rm e}^{B/n}\Vert^{n}\;.
\end{align*}
If $M_{A}$ denotes the upper bound of the spectrum of $A$, then the spectral theorem implies 
$\Vert  {\rm e}^{A/n}\Vert = {\rm e}^{M_{A}/n}$, and similarly for $B$.  The bound \eqref{CAB Bound} for 
$C=\mathbb{1}$ now follows.

Consider the case that $C\neq \mathbb{1}$.  Unless  $C{\rm e}^{B}$ is bounded, there is nothing to prove.  
As $\Vert T\Vert = \Vert  TT^{*}\Vert^{1/2}$, one infers 
	\[
		\Vert C {\rm e}^{A+B}\Vert
		= \Vert C {\rm e}^{2A+2B}C\Vert^{1/2}
		= \Vert C^{2} {\rm e}^{2A+2B}\Vert^{1/2}\;.
	\]
As in the proof with $C=\mathbb{1}$,  one has on the domain $\mathcal D\times \mathcal D$ 
with $\mathcal D$ equal to the domain of $C^{2}$, and
	\[
		C^{2}\, {\rm e}^{2A+2B}
		 = C^{2} \, {\rm st.}\hskip-.3em\lim_{n\to\infty} \bigl({\rm e}^{2A/n}\, {\rm e}^{2B/n}\bigr)^{n}
		 =  {\rm w.}\hskip-.3em\lim_{n\to\infty} \bigl( {\rm e}^{2A/n}\,  C^{2/n}{\rm e}^{2B/n} \bigr)^{n}\;.
	\]
Thus\textcolor{red}{,} 
	\[
		\Vert C {\rm e}^{A+B}\Vert
			\leqslant \lim_{n\to\infty} \Vert {\rm e}^{2A/n}\Vert^{n/2}\ \Vert C^{2/n}{\rm e}^{2B/n}\Vert^{n/2}\;.
	\]
As above,  $\Vert  {\rm e}^{2A/n}\Vert = {\rm e}^{2M_{A}/n}=\Vert {\rm e}^{A}\Vert^{2/n}$.  
As $B$ and $C^{2}$ commute, they can be simultaneously diagonalized, so $\Vert C^{2/n}{\rm e}^{2B/n}\Vert
=\Vert C {\rm e}^{B}\Vert^{2/n}$. Hence, the general case of \eqref{CAB Bound} also holds.
\end{proof}

Let $H_{{\it v}}$ be given by \eqref{Hv}.

\begin{lemma}\label{Lemma:Heat-Kernel Analyticity}
The heat kernel ${\rm e}^{-tH_{{\it v}}}$ is 
the boundary value of an analytic function of ${\it v}$ in the disk $\vert {\it v}\vert <1$. 
\end{lemma}

\begin{proof}
Choose a maximum value of $|{\it v}|=\Gamma<1$, and choose $0<\epsilon<1-\Gamma$.   Note that 
	\[
		\left. \frac{d^{n}}{d{\it v}^{n}} \,{\rm e}^{-tH_{{\it v}}} \right\vert_{{\it v}=0}
		= (-tP)^{n} {\rm e}^{-tH_{0}}.
	\]	
Let $A=-t(\epsilon H_{\rm free}+H_{I}-E)$, $B= -t(1-\epsilon) H_{\rm free}$, and $C=(-t{\it v}P)^{n}$.  The 
operators $A$, $B$, and $A+B=-tH$ are essentially-self adjoint on $C^{\infty}(H_{\rm free})$; see \cite{Rosen-HOE}.  For given $t,\Gamma$ they are bounded  from above,  and also $P$ commutes with $A, B$ and their closures. Moreover,
\begin{align*}
		 \frac{|{\it v}|^{n}}{n!} \,
		\left\Vert \left. \frac{d^{n}}{d{\it v}^{\prime\,n}} \,{\rm e}^{-tH_{{\it v}'}} \right\vert_{{\it v}'=0}	\right\Vert
		& = \frac{1}{n!} \,\left\Vert  (t{\it v} P)^{n} {\rm e}^{-tH_{0}}   \right\Vert
		\\ & =	 \frac{1}{n!}  
		\left \Vert C {\rm e}^{(A+B)} \right\Vert\;.
\end{align*}
Thus\textcolor{red}{,} we can apply Lemma \ref{Lemma:Bound2}  to obtain 
	\[
		 \frac{|{\it v}|^{n}}{n!} \,
		\left\Vert \left. \frac{d^{n}}{d{\it v}^{\prime\,n}} \,{\rm e}^{-tH_{{\it v}'}} \right\vert_{{\it v}'=0}	\right\Vert
		\leqslant
		 \frac{1}{n!}\,\left\Vert {\rm e}^{A}\right \Vert\,  
		\left \Vert C {\rm e}^{B} \right\Vert  \;.
	\]
Since $|{\it v}|\leqslant \Gamma$ and $|P|^{n} \leqslant H_{\rm free}^{n}$, we infer 
\begin{align*}
		\Vert C\,{\rm e}^{B}\Vert
		& = \Vert (t{\it v} P)^{n}\,{\rm e}^{-t(1-\epsilon)H_{\rm free}}\Vert
		\\ & \leqslant \left( \frac{|{\it v}|}{1-\epsilon} \right)^{n}\, n!
		\\ & \leqslant \left( \frac{\Gamma}{1-\epsilon} \right)^{n}\, n!\;.
\end{align*}
Combining these bounds, 
\begin{align*}
		 \frac{|{\it v}|^{n}}{n!} \,
		\left\Vert \left. \frac{d^{n}}{d{\it v}^{\prime\,n}} \,{\rm e}^{-tH_{{\it v}'}} \right\vert_{{\it v}'=0}	\right\Vert
		& \leqslant \frac{1}{n!}\,\Vert {\rm e}^{A}\Vert  \,\Vert C\,{\rm e}^{B}\Vert
		\\ & \leqslant 
		\left( \frac{\Gamma}{1-\epsilon} \right)^{n} \,\Vert {\rm e}^{A} \Vert \;.
\end{align*}
As $\Gamma<1-\epsilon$, and  ${\rm e}^{A}$ is bounded, the Taylor series 
\begin{align}
		\left. {\rm e}^{-tH_{{\it v}}}
		= \sum_{n=0}^{\infty}  \frac{{\it v}^{n}} {n!} \left( \frac{d^{n} }{d{\it v}^{\prime \,n} }\right)  
		{{\rm e}^{-tH_{{\it v}'}}} \right|_{{\it v}'=0}
\end{align}
is norm-convergent in the disk $\vert {\it v}\vert \leqslant \Gamma$. But $\Gamma<1$ is arbitrary, so the heat kernel is analytic in ${\it v}$ throughout the open unit disk.  
\end{proof}

\bigskip
\noindent {\bf Proof of Theorem \ref{Theorem:Form Domain E}.}
We only consider $\Theta {\rm e}^{-\mathfrak A}$, as the argument for $\Pi {\rm e}^{-\mathfrak A}$ is similar.  
In the case that ${\it v}=0$, the proposition holds as a consequence of the standard Feynman-Kac formula.  The bound is just 
the Schwarz inequality in the reflection-positive inner product defined by one or the other Osterwalder-Schrader quantization.  
From  Lemma \ref{Lemma:Heat-Kernel Analyticity}\textcolor{red}{,} we infer that 
	\[ 
		{\it v} \mapsto \langle \Omega, \Theta {\rm e}^{-\mathfrak A} \Omega' \rangle_{\mathcal E}
			=\bigl\langle \widehat\Omega, {\rm e}^{-T H_{{\it v}} } \widehat{\Omega'} \bigr\rangle_{\mathcal H}
	\] 
is the boundary value of an analytic function 
of ${\it v}$ in the unit disk.  We can identify this function for ${\it v}$ purely imaginary and small in magnitude as a matrix 
element expectation of the heat kernel ${\rm e}^{-TH_{0}}$ in the vector~$\widehat\Omega$ and a translate of 
the vector~$\widehat{\Omega'}$.  Once the expectation is defined for real ${\it v}$, the Schwarz inequality follows from the 
positivity of the reflection-positive inner product.
\qed

\setcounter{equation}{0}
\section{The Spectrum Condition on the Cylinder\label{Section:Spectral_Condition}} 
In this section we consider quantum fields on the two-dimensional, cylindrical spacetime  $\mathbb{R} \times  S^{1}$.  We take $t\in \mathbb{R}$ as the time coordinate, parameterized by the real line, and
$x\in [-\tfrac{\ell}{2}  ,\tfrac{\ell}{2} ]$ as the spatial coordinate, parametrized by the circle. We consider the ${\mathscr P}(\varphi)_2$-Hamil\-tonian $H$ given in \eqref{Total Hamiltonian} and the Hamiltonian $H_{{\it v}}=H+P{\it v}$ defined in \eqref{Hv}.

Fourier analysis shows that the spectrum condition $0 \leqslant H_{v}$ holds for free fields.  Glimm and Jaffe  conjectured that the spectrum condition also holds for Hamiltonians  defined on the spatial circle~\cite{Glimm-Jaffe-Spectrum}.   Heifets and Osipov later proved this both for ${\mathscr P}(\varphi)_2$ models~ \cite{Osipov} and for the Yukawa$_2$ model~\cite{Osipov2}.   
While Glimm and Jaffe believed to have a proof of the spectrum condition in~\cite{Glimm-Jaffe-Spectrum}, they had not established uniqueness of the ground state $\Omega_{ v}$ for $H_{v}$; see the footnote on  p.~1584 of \cite{Glimm-Jaffe-Spectrum-c}.   Our construction with complex classical fields~\cite{JJM Framework}  provides a context for the Heifets-Osipov argument showing that $\Omega_{ v}$ is unique. 

A consequence of this spectrum condition for theories on a spatial circle is analyticity of the Schwinger  functions in the spatial directions for $|x_{i}-x_{i+1}|< |t_{i}-t_{i+1}|$.

\begin{proposition}[\bf Spectrum Condition on a Cylinder]
\label{Theorem:PeriodicSpectrum} 
For real ${\it v}$, with $|{\it v}|<1$, one has  $0\leqslant H_{{\it v}}$. Any  ground state $\Omega_{{\it v}}$ of $H_{{\it v}}$ is a multiple of $\Omega$, 
and $H_{{\it v}}\Omega=P\Omega=0$. 
\end{proposition}

\begin{proof}
The proof of the spectrum condition $0\leqslant H_{{\it v}}$ has two distinct parts.   The first part is to reduce the proof to the bound 
\begin{eqnarray}
		\langle \Psi , {\rm e}^{-TH_{{\it v}}} \Psi \rangle_{\mathcal{H}}
		&\leqslant&  \langle \Omega_{0}, {\rm e}^{-TH_{{\it v}}}\Omega_{0} \rangle_{\mathcal{H}}^{1/2} \ 
		\langle \Psi', {\rm e}^{-TH_{{\it v}}} \Psi' \rangle_{\mathcal{H}}^{1/2}\;
	\label{AssumeDecay-0}
\end{eqnarray}
for a dense set of vectors $\{ \Psi \}$, with $\Psi'=\Psi'(\Psi)$, and with   $\Omega_{0}$ denoting the Fock zero-particle vector.    The second part is to prove \eqref{AssumeDecay-0}.
\\ \\
\noindent {\em First Part:} 
We give a variation of the argument in Heifets and Osipov~\cite{Osipov}.  As recalled in
the beginning of \S\ref{Sect:FK}, 
the Hamiltonian $H$ has a unique ground state~$\Omega$, up to a phase, and $H_{{\it v}}$ has a negative 
ground-state energy $E_{{\it v}}$.   
Assume  \eqref{AssumeDecay-0}. Then 
\begin{align}
	\label{AssumeDecay-1} 
	\langle \Psi, {\rm e}^{-T(H_{ {\it v} }-E_{ {\it v} })} \Psi \rangle_{\mathcal{H}}
		&\leqslant   \langle \Omega_{0}, {\rm e}^{-T(H-E_{ {\it v} })}\Omega_{0} \rangle_{\mathcal{H}}^{1/2} \ 
			\langle \Psi', {\rm e}^{-T(H_{ {\it v} }-E_{ {\it v} })} \Psi' \rangle_{\mathcal{H}}^{1/2} \nonumber\\
		&\leqslant  {\rm e}^{TE_{ {\it v} }/2} \, \Vert  \Psi' \Vert\;,
\end{align} 
where we have used $P\Omega_{0}=0$ to bound the first expectation and $0 \leqslant H_{ {\it v} }-E_{ {\it v} }$ to bound the second one.  
Thus, for a dense set of vectors $\{ \Psi \}$, the expectation of ${\rm e}^{-T(H_{ {\it v} }-E_{ {\it v} })}$  decays with the 
exponential rate ${\rm e}^{-T|E_{ {\it v} }|/2}$ as $T\to\infty$.  Hence\textcolor{red}{,} $\frac12|E_{ {\it v} }| \leqslant H_{ {\it v} }-E_{ {\it v} }$. However, $H_{ {\it v} }-E_{ {\it v} }$  has a zero energy ground state, so it follows that $E_{ {\it v} }=0$.  

Now consider whether the ground state vector is unique.  Denote one  ground state vector of $H_{ {\it v} }$ by $\Omega$, 
and let $\Omega'$  be a second ground state.  For each value of $ {\it v} $ and for $0<\epsilon< 1-\tanh | {\it v} |$, one has ${\it v} _{\pm}$ for which $\tanh {\it v} _{\pm}=\tanh {\it v} \pm\epsilon$.  Hence, $0\leqslant H_{ {\it v} _{+}}H_{ {\it v} _{-}}$, that is,
 	\begin{equation}
		\epsilon^{2} P^{2}\leqslant H_{ {\it v} }^{2}\;.
	\label{PHBound}
	\end{equation}
Taking the expectation of \eqref{PHBound} in the vector $\Omega'$ shows $P\Omega'=0$.  It follows that
$\Omega'$ is also a ground state of $H$, whose  ground state is unique.  Thus, $\Omega'=\alpha\Omega$, where $\alpha$ is a phase.

\bigskip
\noindent{\em Second Part:} 
Let $A(\varphi)$ denote a polynomial function of the imaginary-time field, with each field averaged with a 
$C^{\infty}_{0}$-function supported in the  rectangle $(t,x)\in [0,\epsilon]\times [0, \frac{\ell}{2}]$. Then $\Psi=A(\varphi)\Omega_{0}
= ( A(\Phi)\Omega_{0}^{E} )^{\wedge}$ is the quantization from temporal-reflection positivity. 
Note that Theorem~\ref{Theorem:Quantization Domain}  
ensures that the states $\Psi$ are dense in $\mathcal H$.  The vector ${\rm e}^{-TH_{ {\it v} }/2}\Psi$ is the quantization of 
${\rm e}^{-{\mathfrak A}_{+}} A_{T/2}\Omega_{0}^{E}$. 
The Feynman-Kac formula based on temporal-reflection positivity, elaborated in \S\ref{Sect:ExtendedFK}, is
 	\begin{equation}
		\langle \Psi, {\rm e}^{-T H_{ {\it v} }} \Psi \rangle_{\mathcal H}
		= \langle A_{T/2} \Omega_{0}^{E} , \Theta {\rm e}^{-{\mathfrak A}} \,A_{T/2} \Omega_{0}^{E} \rangle_{{\mathcal E}}\;.
	\label{ExtendedFK}
	\end{equation}  

The action  $\mathfrak A= {\mathfrak A}_{+} + {\mathfrak A}_{-}$ is the sum of positive-time and negative-time 
parts~${\mathfrak A}_{\pm}$, with ${\mathfrak A}_{+}^{*\Theta}= \Theta {\mathfrak A}_{+}^{*}\Theta={\mathfrak A}_{-}$ and ${\mathfrak A}_{+}^{*\Pi}= \Pi {\mathfrak A}_{+}^{*}\Pi={\mathfrak A}_{-}$.  It is also a 
sum  of parts  $\widetilde{\mathfrak A}_{\pm}$   localized at positive or negative spatial coordinate $x$, namely,
${\mathfrak A} =  \widetilde{\mathfrak A}_{+} + \widetilde{\mathfrak A}_{-}$, with  $\Pi \widetilde{\mathfrak A}_{+}^{*}\Pi
=\widetilde{\mathfrak A}_{-}$. Let $A''= A_{T/2}^{* \Theta} \,A_{T/2}$, $A'=A^{*\Pi}A$, and $\Psi'= (A' \Omega_{0}^{E} )^{\wedge}$, where $A^{*\Theta}_{T/2} = \Theta A^{*}_{T/2} \Theta$ and $A^{* \Pi} = \Pi A^{*} \Pi$. Spatial-reflection positivity, along with invariance of $\Omega_{0}^{E}$ under $\Theta$ and $\Pi$ and time-translation, implies
 	\begin{align}
		\langle \Psi, {\rm e}^{-T H_{ {\it v} }} \Psi \rangle_{\mathcal H}
		&= \langle \Omega_{0}^{E}, \Theta {\rm e}^{-{\mathfrak A}} \,A''\Omega_{0}^{E} \rangle_{{\mathcal E}}
		= \langle \Omega_{0}^{E}, \Pi  {\rm e}^{-{\mathfrak A}} \,A''\Omega_{0}^{E} \rangle_{{\mathcal E}}\nonumber\\
		&\leqslant  \langle \Omega_{0}^{E}, \Pi  {\rm e}^{-{\mathfrak A}} \,\Omega_{0}^{E} \rangle_{{\mathcal E}}^{1/2}
		\langle A''\Omega_{0}^{E}, \Pi  {\rm e}^{-{\mathfrak A}} \,A''\Omega_{0}^{E} \rangle_{{\mathcal E}}^{1/2}\nonumber\\
		&=\langle \Omega_{0}^{E}, \Theta  {\rm e}^{-{\mathfrak A}} \,\Omega_{0}^{E} \rangle_{{\mathcal E}}^{1/2}
		\langle A'_{T/2}\Omega_{0}^{E}, \Theta  {\rm e}^{-{\mathfrak A}} \,A'_{T/2}\Omega_{0}^{E} \rangle_{{\mathcal E}}^{1/2}\nonumber\\
		&= \langle \Omega_{0}, {\rm e}^{-T H_{ {\it v} }} \Omega_{0} \rangle_{\mathcal H}^{1/2}
		\langle \Psi', {\rm e}^{-T H_{ {\it v} }} \Psi' \rangle_{\mathcal H}^{1/2}\;.
	\label{DoubleRPSchwarz}
	\end{align}
Note that we use the identity 
\begin{align*}
A^{\prime\prime * \Pi}A'' = A_{T/2}^{\prime *\Theta}A'_{T/2}
=A_{T/2}^{\Theta\Pi}A_{T/2}^{*\Theta}A_{T/2}^{*\Pi}A_{T/2}^{\phantom{\Theta}}.
\end{align*}
In the Feynman-Kac formula and the inequality  \eqref{DoubleRPSchwarz}, we have used 
Proposition \ref{Theorem:Form Domain E}.  In  \eqref{Schwarz1}\textcolor{red}{,} we take  
$\Omega_{1}=\Omega_{0}^{E}$ and $\Omega_{2}=A''\Omega_{0}^{E}$.
\end{proof}

\setcounter{equation}{0}
\section{Classical Fields on the Cylinder $\boldsymbol{X}= S^1 \times \mathbb{R}^{d-1}$\label{Sect:Positive Temperature-SF}}
In this section we study periodization of time for the classical field. We are especially interested 
in the spacetime $\boldsymbol{X}  =  S^{1} \times  \mathbb{R}^{d-1}$ with  $S^{1}$ parameterized\footnote{Here and in other sections describing positive-temperature fields, we denote 
$\beta=\frac{1}{k_{\text{B}}T}$ as the inverse temperature. This is in contrast with the notation in \S \ref{Section:Spectral_Condition},
 where  $\tanh|\beta|=|\vec {\it v}|$ denotes the velocity of a Lorentz boost.} by $t \in [-\frac{\beta}{2}, \frac{\beta}{2}]$. As investigated  by H{\o}egh-Krohn \cite{HK}, these classical fields yield a positive temperature state of the quantum field. Define the positive-time and nega\-tive-time half-circles $S^{1}_{+}=[0, \tfrac{\beta}{2}]$ and $S^{1}_{-}=[-\tfrac{\beta}{2}, 0]$, and the positive and negative time half-spaces 
	\begin{equation}
		\boldsymbol{X}_{\pm}=S^{1}_{\pm}\times \mathbb R^{d-1} \; . 
		\label{PostiveTimeCylinder}
	\end{equation}
Their intersection 
	\begin{equation}
		\boldsymbol{X}_{+} \cap \boldsymbol{X}_{-} =
		( \{0\} \times \mathbb R^{d-1} ) \cup ( \{\tfrac{\beta}{2}\} \times 
		\mathbb R^{d-1} ) 
		\label{IntersectionCylinder}
	\end{equation}
consists of two disjoint copies of $\mathbb R^{d-1}$ forming the boundary $\partial  \boldsymbol{X}_{+}$ of $\boldsymbol{X}_{+}$.

\subsection{Reflection Positivity\label{Sect:Reflection Positivity-SF}}
As proved in Pro\-position VI.3 of  \cite{JJM Framework}, reflection positivity of 
$D_{\vec {\it v}}$ carries over to reflection positivity of 
	\begin{equation}
    		{D}^{c}(x,x') = \sum_{n=-\infty}^\infty  
		{D_{\vec {\it v}}}(t-t'+n\beta , \vec x - \vec x\, ' )\;.
	\label{Dc Definition}
	\end{equation}
Thus, temporal-reflection positivity, $0\leqslant  \vartheta D_{\vec {\it v}}$ on $L_{2,+}$, established in 
Proposition \ref{Proposition:Neutral Time_RP}, ensures  				
	\[
		0 \leqslant\vartheta D^{c}
			\quad\text{on}\quad
		\mathcal K_{\pm}= L_{2}(\boldsymbol{X}_{\pm}) \; . 
	\]
Therefore, one can quantize functions supported in the positive and negative-time half-spaces $\boldsymbol{X}_{\pm}$, 
using the reflection positive kernels $\vartheta D^{c}$ or $D^{c}\vartheta$ on $\mathcal K_{\pm}$
for the neutral scalar field. 

For the two-point function  $D_{\vec {\it v}}$ on ${L}_{2}(\mathbb R^{d})$ defined in \eqref{D Fourier} periodization  
 yields, using \eqref{Modified_Kernel},   
\begin{align}
		D^{c}(x,x') \kern -1mm
		&= \kern -1mm
		\sum_{n \in  \mathbb{Z}}   \frac{{\rm e}^{- {\mu} \, | t-t' + n \beta |  +\delta(t-t' + n\beta) }  }{2\mu}  
		  (\vec x-\vec x \, ')
		  \label{DcKernel}\\ 
		  &= \kern -1mm
		  \begin{cases}
		  	\frac{1}{2\mu} \left( \frac{{\rm e}^{-(t-t') {\mu}_{-}}}{1-{\rm e}^{-\beta  {\mu}_{-}} }
			+ \frac{{\rm e}^{-(\beta-(t-t')) {\mu}_{+}}}{1-{\rm e}^{-\beta {\mu}_{+}} }
			 \right) (\vec x-\vec x^{\,\prime})\;,
			& \text{if } t>t' , \\
			&\\
		  	 \frac{1}{2\mu} \left( \frac{ {\rm e}^{(t-t') {\mu}_{+} } } {1-{\rm e}^{-\beta {\mu}_{+}} }
			+ \frac{{\rm e}^{-(\beta+(t-t'))  {\mu}_{-}}}{1-{\rm e}^{-\beta  {\mu}_{-}} }
			\right)  (\vec x-\vec x^{\,\prime})\;,
			& \text{if } t<t'   ,\\
		  \end{cases}\nonumber
\end{align}
where ${\mu}_{\pm}= \mu\pm\delta$ is defined in \eqref{mu-beta-Defn}.  
\goodbreak

\smallskip
\paragraph{\bf Interpretation}
In two different ways, one can connect a point with negative time, $-s\in [-\frac{\beta}{2},0]$, to a point with positive time, $s'\in[0, \frac{\beta}{2}]$.  A trajectory of minimal length can pass through time $t=0$, or it can pass through time $t=\frac{\beta}{2}$.  The minimal length of a trajectory through $t=0$ is $s+s'$, while 
that through $t=\frac{\beta}{2}$ is $\beta-s-s'$.  A trajectory may also wind $n \geqslant 1$ times around the circle, adding a length of $n \beta$ to the minimum. Each of these possibilities contributes a term to the Green's function with an exponential  decay rate equal to the minimal length times a corresponding Hamiltonian.  In fact, the full Green's function is a sum of such terms. In particular, the minimal trajectory through $t=0$ contributes the heat kernel ${\rm e}^{-(s+s') \mu_{+}}$ to the Green's function.   As the length of the trajectory $s+s'$ increases with the increase of either $s$ or $s'$, the corresponding decay rate increases as well; therefore the Hamiltonian ${\mu}_{+}$ is positive. The factor 
	\[
	\sum_{n=0}^{\infty} {\rm e}^{-n\beta{\mu}_{+}}
	=(1-{\rm e}^{-\beta{\mu}_{+}})^{-1}
	\] 
in the inner product reflects a correction from including terms labelled by trajectories that make  $n \geqslant 0$ multiple circuits around the circle.  This factor affects the normalization of the state rather than the Hamiltonian. 

The second term  in the Green's function arises from terms labelled by the trajectories through $t=\frac{\beta}{2}$. The corresponding minimal trajectory gives the heat kernel ${\rm e}^{-(\beta-s-s') \mu_{-}}$.  The separation $\beta-s-s'$  decreases 
with increasing $s,s'\in [0,\frac{\beta}{2}]$, so the corresponding Hamiltonian  $-{\mu}_{-} $ is negative.   The correction 
factor $1+\rho_{-}= ( 1-{\rm e}^{-\beta{ \mu}_{-}})^{-1}$ arises from the sum over  circuits from $s'$ to $-s$ that 
travel the minimal distance, but with the constraint that they pass  through time $t= \frac{\beta}{2}$, followed by  multiple complete circuits around the circle.   

\bigskip 
The operators ${\mu}_{\pm}= \mu\pm\delta$ acting on  the Hilbert space $L_{2}(\mathbb R^{d-1})$ are self-adjoint and positive. They satisfy the lower bound $m\sqrt{1-\vec {\it v}^{\,2}}\leqslant  {\mu}_{\pm}$ of \eqref{Elem-Bound-1}.  Hence,
\begin{align*}
		\Vert {\rm e}^{-\beta {\mu}_{\pm}}\Vert
		\leqslant {\rm e}^{-m\beta\sqrt{1-\vec {\it v}^{\,2}}}
\end{align*}
and
\begin{align*}		
		\Vert\left(\mathbb{1}-{\rm e}^{-\beta {\mu}_{\pm}}\right)^{-1}\Vert
		\leqslant   \left( 1-  {\rm e}^{-m\beta\sqrt{1-\vec {\it v}^{\,2}}} \right)^{-1}\;.
\end{align*}
Consequently, for fixed $t$ and $t'$, the sum over $n\in\mathbb Z$ in \eqref{DcKernel} is norm-convergent in the 
sense of operators on $L_{2}(\mathbb R^{d-1})$.  It is common to define  
	\[
		\rho_{\pm}
		=\frac{{\rm e}^{-\beta  \mu_{\pm}}}
			{\mathbb{1}-{\rm e}^{-\beta  \mu_{\pm}}} \;, 
		\qquad\text{so}\quad \mathbb{1}+\rho_{\pm}
		=  \frac{1}{ \mathbb{1}-{\rm e}^{-\beta \mu_{\pm}}}\; .
	\]
Note that 
	\begin{equation}
		\label{rho Bound} 
		\Vert \rho_{\pm}\Vert
		\leqslant 
		( {\rm e}^{m\beta\sqrt{1-\vec {\it v}^{\,2}}} -1)^{-1}  \; . 
	\end{equation} 		
This allows us to rewrite  \eqref{DcKernel}, for $x,x'\in \boldsymbol{X}_{+}$, as 
	\begin{equation}
		( \vartheta {D}^{c} )( x, x')
		=  \frac{1}{2\mu} \left( (\mathbb{1}+\rho_{+}) 
		{\rm e}^{-(t+t') \mu_{+}} 
		+ \rho_{-} {\rm e}^{(t+t')  \mu_{-}} \right)
				 (\vec x-\vec x')	  \;. 
	\label{vartheta DcKernel}
	\end{equation}
The operators  $\rho_{\pm}$ commute with $\mu$, so  their  norms on ${\mathfrak H}_{-1/2}(\mathbb R^{d-1})$ agree with 
their norms on $L_{2}(\mathbb R^{d-1})$. In Fourier space the function $\mu(\vec k)$  is real and even, so the operator 
$\mu$ in configuration space is real, acting on either $L_{2}(\mathbb R^{d-1})$ or on~${\mathfrak H}_{-1/2}(\mathbb R^{d-1})$.  
Similarly, $\delta(k)$ is real and odd, so the operator $\delta$ in configuration space is purely imaginary, \emph{viz.}, $\overline\delta=-\delta$. Consequently,  $\overline {\mu_{\pm}}= \mu_{\mp}$ and, 
setting $f_{t}(\vec x)=f(t,\vec x)$, one has
	\begin{align}
	\label{Norm 1}
	\bigl\langle  f, \vartheta D^{c} g \bigr\rangle_{\mathcal K}
	&= \Big\langle \int_{0}^{\beta/2}  {\rm e}^{-t{\mu}_{+} } f_{t}\,dt \; , 
	( \mathbb{1}+\rho_{+} )    
	\int_{0}^{\beta/2}  {\rm e}^{-t'{\mu}_{+} } g_{t'}\,dt' \Bigr\rangle_{\mathfrak{H}_{-\frac12}}
	\nonumber \\
	&\; \! + \! \Bigl\langle \int_{0}^{\beta/2}  \overline{{\rm e}^{t {\mu}_{+} } \rho_{+}^{1/2} } f_{t}\,dt \; , \! \int_{0}^{\beta/2}
	\overline{{\rm e}^{t {\mu}_{+} } \rho_{+}^{1/2} } \, g_{t'}\,dt'  \Bigr\rangle_{\mathfrak{H}_{-\frac12}}  \;.
	\end{align} 

\begin{remark} Let $s\in \mathbb{R}$, $0 \leqslant t \leqslant \frac{\beta}{2}$ and $\alpha \in \mathfrak{H}_{-\frac12}$. Eq.~\eqref{vartheta DcKernel}
suggests to consider the analytic map 
\[
	s+it \mapsto \bigl( {\rm e}^{i (s+it) {\mu}_+ } ( \mathbb{1}+\rho_{+})^{1/2} \alpha \; , \; 
		{\rm e}^{ - i (s+it) {\mu}_-  } \rho_{-}^{1/2}  \alpha \bigr) \in \mathfrak{H}_{-\frac12} \oplus \mathfrak{H}_{-\frac12}
	\; .
\]
Note that the action of the time-evolution $s \mapsto {\rm e}^{i s {\mu}_+ }$ in the second component is oriented toward 
the past\textcolor{red}{,} and $\mu_+$ is replaced by its complex conjugate ${\mu}_- = \overline {\mu_{+}}$. Both aspects can be avoided 
by using a complex conjugate Hilbert space in the second component, as suggested by the equivalent  formula for 
$\bigl\langle  f, \vartheta D^{c} g \bigr\rangle_{\mathcal K}$ given in Eq.~\eqref{Norm 1a}\textcolor{red}{;} see 
Section~\ref{Sect:Thermal Quantization Maps-SF}.
\end{remark}

\subsection{Estimates on the Kernels\label{Sect:Time Compactification Estimates-SF}}
The  operator $D^{c}$ can be decomposed into its hermitian and skew-hermitian parts on $\mathcal K$,  
	\[
		D^{c} = K^{c} + i L^{c}\;.
	\]
The operators $D^{c}, K^{c}, L^{c}$ and their   kernels $D^{c}(x,x')$, $K^{c}(x,x')$, \emph{etc}., have  elementary Fourier representations, 
almost identical to the expression for the continuum Fourier transform, such as  \eqref{Dc Fourier Representation} 
in the case of $D_{\vec {\it v}}(x,x')$.  As a consequence, this entails operator bounds similar to those established 
in \S\ref{Sect:Real Propagator} prior to compactification. We formulate these properties in the following proposition. Let us denote the lattice of energy values dual to $[0,\beta]$ by ${\mathcal L}_{\beta}= \frac{2\pi}{\beta} \mathbb{Z}$,  known as \emph{Matsubara frequencies} in the physics literature.

\begin{theorem} \label{Prop:Compactified ComplexCovarianceBounds} The kernel $D^{c}(x,x')$ has the representation 
	\begin{equation}
		D^{c}(x,x')
		= \frac{1}{\beta}\sum_{E\in \mathcal L_{\beta}} \frac{1}{(2\pi)^{d-1}}\int_{\mathbb R^{d-1}} d\vec k\,  
		\frac{{\rm e}^{iE(t-t')+i\vec k(\vec x-\vec x')}}{(E+i\delta)^{2}+\mu(\vec k)^{2}}\,\;.
	\label{Dc Fourier Representation}
	\end{equation}
The operators $K^{c}$, $L^{c} $, $D^{c}$, $\vert D^{c}\vert$, and $C^{c}$ mutually commute and  satisfy
\[
        K^{c}
        \leqslant | D^{c}|
        \leqslant ( 1-\vec {\it v}^{\,2})^{-1/2}\,K^{c} \;.
\]
In particular, $D^{c}$ is invertible. Moreover,
\begin{equation}
    \label{L-K_RelativeBound Compact}
         \tfrac12  \left(1-\vec {\it v}^{\,2} \right) C^{c}
          < K^{c}
          < (1-\vec {\it v}^{\,2})^{-2}\,C^{c}  
\end{equation}
and $\sup_{k}  \left| \tfrac   {\widetilde L^{c}(k)} {\widetilde K^{c}( k)} \right|  
        = \tfrac{\vert \vec {\it v}\vert}{\sqrt{1-\vec {\it v}^{\,2}}}$. Also,   
\begin{equation}
		\tfrac12  \left(1-\vec {\it v}^{\,2} \right)C^{c}
		< | D^{c}| 
       		<  (1-\vec {\it v}^{\,2})^{-5/2}\,C^{c}\;.
	\label{AbsDc Bound} 
\end{equation}
\end{theorem}

\begin{proof}
The operator $D^{c}(x,x')$ is periodic in $t-t'$ with period $\beta$. To establish its Fourier representation, we  consider the Fourier series 
in the variable $\xi=t-t'$.  Similar to the proof of Proposition II.4 of \cite{Twist Positivity}, but for operators, 
	\begin{align*}
		(\mathbb{1}+ \rho_-) \int_{0}^{\beta}
		 {\rm e}^{-({\mu}_{-} +iE )\xi} \,d\xi 
		 &= \frac{1}{{\mu}_{-}+iE}\;,\\
		 \int_{0}^{\beta}
		 {\rm e}^{({\mu}_{+} -iE )\xi} \rho_+ \,d\xi 
		 &= \frac{1}{{\mu}_{+}-iE}\;,
	\end{align*}
and ${\mu}_{+}+{\mu}_{-}=2\mu$.  Use the representation \eqref{DcKernel} for $\xi \geqslant 0$.   
One has the $E$-dependent operator $D_E$ on $\mathfrak H_{-\frac12}$ defined by  
	\[
		 D_E (x,x') = \int_{0}^{\beta} D^{c}(x,x') {\rm e}^{-iE\xi} \, d\xi \, .
	\]
It follows that 
	\begin{align*}
		 D_E &= \frac{1}{2\mu}\, \left( 
			\frac{1}{{\mu}_{-}+iE}
			+ \frac{1}{{\mu}_{+}-iE}			
		 \right)\\
		 &= \frac{1}{({\mu}_{-}+iE)({\mu}_{+}-iE)}
		\\ & = \frac{1}{(E+i\delta)^{2}+\mu^{2}}\;.
	\end{align*}
In terms of the integral kernel $D_E (\vec x-\vec x^{\,\prime})$,  one has 
	\[
		D^{c}(x,x')
		= \frac{1}{\beta} 
		\sum_{E\in \mathcal L_{\beta}} \,{\rm e}^{iE(t-t')}
				\,D_E(\vec x-\vec x^{\,\prime})  \;,
	\] 
which is equivalent to \eqref{Dc Fourier Representation}. 

The operators $K^{c}$, $L^{c} $, $D^{c}$, and $C^{c}$  are all translation invariant and hence mutually commute, 
as well as with their adjoints and functions defined by the spectral theorem.  The remaining statements are estimates on the operators that follow by comparing their Fourier representations.  Since the Fourier representation of $D^{c}(x,x')$ 
agrees with that of $D_{\vec {\it v}}(x,x')$ for specific values of $E$, the inequalities established in \S\ref{Sect:Real Propagator} 
hold here as well, after translation of the constants to the present notation by substituting $\tanh|\beta|\mapsto |\vec {\it v}|$.
\end{proof}

\begin{corollary} The operator $D^{c}$ extends to $\mathfrak H_{-1}(\boldsymbol{X})$.  
\end{corollary}

\begin{proof}
Proposition \ref{Prop:Compactified ComplexCovarianceBounds} ensures that $D^{c}$ has a square root $D^{c\,\frac12}$.  
Denote the constant $( 1-\vec {\it v}^{\,2})^{-5/2}$ in \eqref{AbsDc Bound} of Proposition \ref{Prop:Compactified ComplexCovarianceBounds}
by~$M^{2}$. It follows that 
	\begin{align}
	\label{extension estimate}
		 \Vert  D^{c\,\frac12} f\Vert_{\mathcal K}
		& = \langle f, \vert D^{c}\vert f\rangle_{\mathcal K}^{1/2}
		\\ & \leqslant M \langle f, C^{c} f\rangle_{\mathcal K}^{1/2}
		\nonumber \\ & =  M\Vert f \Vert_{\mathfrak H_{-1} (\boldsymbol{X}^c)}\;. \nonumber
	\end{align}
Hence\textcolor{red}{,} $D^{c}$ extends to $\mathfrak H_{-1}(\boldsymbol{X})$.  
\end{proof}

\bigskip

Note that  the function $\delta_{s}\otimes\alpha$ has Fourier representation proportional to ${\rm e}^{-is E} \,\widetilde \alpha (\vec k)$.  Thus\textcolor{red}{,}  	\[
		\Vert \delta_{s}\otimes \alpha \Vert_{\mathfrak H_{-1}(\boldsymbol{X}^c)}^{2}
		=
		\frac{1}{\beta}\sum_{E\in \mathcal L_{\beta}}\int_{\mathbb R^{d-1}}\frac{\vert \widetilde \alpha(\vec k)
		\vert^{2}}{E^{2}+\mu(\vec k)^{2}} \,d\vec k\;.
	\]
This  sum is real and positive, so there is  a constant $\widetilde M=1+O(\frac{1}{\beta})$ such that it can be estimated 
by a multiple of  the Riemann integral it approximates.  Then  
\begin{align*}
		\Vert \delta_{s}\otimes \alpha \Vert_{\mathfrak H_{-1}(\boldsymbol{X}^c)}^{2}
		& \leqslant
		\frac{ \widetilde M}{2\pi}\int_{E\in \mathbb R} dE  \int_{\mathbb R^{d-1}} 
		\frac{\vert \widetilde \alpha(\vec k)\vert^{2}}{ E^{2}+\mu(\vec k)^{2}} \,d\vec k 
		\\ & =\widetilde M
		 \Vert \alpha \Vert^{2}_{\mathfrak H_{-\frac12}(\mathbb R^{d-1})}\;.
\end{align*}
Thus\textcolor{red}{,}  for $\alpha\in\mathfrak H_{-\frac12}(\mathbb R^{d-1})$, the generalized function $\delta_{s}\otimes \alpha$ 
is an element of the Sobolev space $ \mathfrak H_{-1}(\boldsymbol{X})$. 

\subsection{Thermal Quantization Maps\label{Sect:Thermal Quantization Maps-SF}}

We can write \eqref{Norm 1} as 
	\begin{align}
	\label{Norm 1a}
	\bigl\langle  f, \vartheta D^{c} g \bigr\rangle_{\mathcal K}
	&=  \Big\langle \int_{0}^{\beta/2}  
		{\rm e}^{-t{\mu}_{+} } f_{t}\,dt \; , 
	( \mathbb{1}+\rho_{+})    
	\int_{0}^{\beta/2}  {\rm e}^{-t'{\mu}_{+} } g_{t'}\,dt' \Bigr\rangle_{\mathfrak{H}_{-\frac12}}
	\nonumber \\
	& \;  +  \Bigl\langle \int_{0}^{\beta/2}  
	{\rm e}^{t {\mu}_{+} } \rho_{+}^{\frac{1}{2}} \,  \overline{g_{t}} \,dt \; ,  \int_{0}^{\beta/2}
	{\rm e}^{t {\mu}_{+} } \rho_{+}^{\frac{1}{2}} \, \overline{f_{t'}}\,dt'  \Bigr\rangle_{\mathfrak{H}_{-\frac12}} \;,  
	\end{align}
where $\overline{f}$ denotes the complex conjugate of the function $f$. The two terms appearing on the right hand side 
in \eqref{Norm 1a} are related
to the two disjoint components of the boundary $\partial \boldsymbol{X}_+$ of $\boldsymbol{X}_+$ as discussed in the interpretation provided in 
Section~\ref{Sect:Reflection Positivity-SF}. 
The special form of these two terms 
can be accommodated for by the following points:
\begin{itemize}
\item[$i.)$]
Identify the one-particle  Hilbert space $\mathcal H_{1}$  with
	\begin{equation} 
	\label{direct sum}
	 \mathfrak H_{-\frac12} (\mathbb R^{d-1})\oplus \overline{\mathfrak H_{-\frac12}(\mathbb R^{d-1})}
	\; , 
	\end{equation}
where the second factor in the direct sum denotes the Hilbert space that is complex-conjugate\footnote{Let ${\mathfrak h}$ 
be a complex Hilbert space of functions.  Then the {\em conjugate Hilbert space} $\overline{\mathfrak h}$ is defined as the 
Hilbert space ${\mathfrak h}$ with the complex structure $-i$ and the inner product 
$\langle h_{1}, h_{2}\rangle_{\overline{\mathfrak h}} = \langle h_{2}, h_{1}\rangle_{\mathfrak h}$. 
There is a natural $\mathbb{C}$-{\em linear} map  ${\mathfrak h}\mapsto \overline{{\mathfrak h}}$ given by $f\mapsto \overline f$, the complex-conjugate 
of the function $f$.},\footnote{In Dirac's notation, if $| f \rangle \in {\mathfrak h}$, then $\langle g | \in \overline{\mathfrak h}$. Clearly, 
$| \lambda f \rangle =  \lambda | f \rangle$ and $\langle \mu g | = \overline{\mu} \langle g | $ for $\lambda, \mu \in \mathbb{C}$.
Thus\textcolor{red}{,} the map $| f \rangle \mapsto \langle  \overline{f} |$ is {\em linear}.}  
to $\mathfrak H_{-\frac12}(\mathbb R^{d-1})$; and,
\item[$ii.)$]
Define two bounded linear maps $\kappa_\pm \colon \mathfrak H_{-\frac12} (\mathbb R^{d-1}) \to \mathcal H_{1}$ (see \cite{AW}) by 
\begin{equation}
\kappa_\pm \colon \alpha    \mapsto
\left( (\mathbb{1}+\rho_{\pm})^{1/2}\alpha, 
\overline{\rho_{\pm}^{1/2} \alpha} \, \right)\;.
\label{Commutative 2}
\end{equation}
\end{itemize}
This will allow us to define the quantization maps $\wedge_\pm$. Before we do so, we mention the following property of $\kappa_\pm$. 

\begin{proposition}
Define $\ell_\pm = {\mu}_{\pm} \oplus (-  {\mu}_{\mp})$. The maps 
\[ 
	\mathbb{R} \ni s  \mapsto \kappa_\pm ({\rm e}^{i s {\mu}_{\pm}} {\alpha} )  
	= {\rm e}^{i s  \ell_\pm }  \kappa_\pm ( {\alpha} ) \; ,
\]
extend analytically to the strip $\{ s+it  \in  \mathbb{C} \mid 0 <  t < \frac{\beta}{2} \}$. Moreover, they satisfy {\em one-particle $\beta$-KMS conditions}: For $\alpha, \alpha' \in \mathfrak H_{-\frac12} (\mathbb R^{d-1})$ and $s\in \mathbb{R}$, we have
\begin{equation}
		\label{o-p-kms-condition}
		\langle \kappa_\pm ( \alpha)  ,  \kappa_\pm ( {\rm e}^{(is - \beta)
		{\ell}_\pm } \alpha' )  \rangle_{\mathcal H_1} 
		= \langle {\kappa}
		( {\rm e}^{is {\ell}_\pm }  	\alpha' ) , 
		 {\kappa}( \alpha)  \rangle_{\mathcal H_1} \; .
\end{equation}
\end{proposition}

\bigskip
We can now define two one-particle quantization maps 
$\wedge_\pm \colon \mathcal K_{+,0}\mapsto {\mathcal H}_{1}$  by setting  	
	\begin{equation}
	\label{thermal quantization map 1}
		\widehat{f}^{\; \pm}(\vec x) =  
		\int_{0}^{\beta/2}  
		{\rm e}^{-t  \ell_{\pm} } \kappa_\pm (f_t(\vec x))\, dt \; . 
	\end{equation}
$\mathcal K_{+,0}\subset \mathcal K_{+} $ is the dense subset defined 
as the linear span of $C^{\infty}_{0}(S^{1}_{+})\times C^{\infty}_{0}({\mathbb R}^{d-1})$. 

\begin{proposition}\label{Prop:Quantization Map}
The maps $\wedge_\pm$ agree with the Osterwalder-Schrader quantizations 
defined by  $\vartheta D^{c}$ and $D^{c}\vartheta$.  Namely,~for~$f,g\in{\mathcal K}_{+,0}$, 
\begin{align*}
\langle f,\vartheta D^{c} g \rangle_{\mathcal K}
		& = \bigl\langle {\widehat{f}}^{\; +}, {\widehat{g}}^{\; +} \bigr\rangle_{\mathcal{ H}_{1}} \\ 
		\langle f, D^{c}\vartheta g \rangle_{\mathcal K}
		& = \bigl\langle \widehat{f}^{\; -}, \widehat {g}^{\; -}   \bigr\rangle_{\mathcal{ H}_{1}}. 
\end{align*}		
They extend by continuity to sharp-time  test functions $f=\delta_{s}\otimes\alpha$, with $s\in[0,\frac{\beta}{2}]$ and 
$\alpha\in{\mathfrak H}_{-\frac12}(\mathbb{R}^{(d-1)})$. Explicitly, 
	\begin{eqnarray}
		 \bigl\langle \widehat{ \delta_{s}\otimes \alpha}^\pm,
		\widehat{ \delta_{s'}\otimes \alpha'}^\pm \bigr\rangle_{\mathcal H_{1}} 
		= \begin{cases} \langle   
			 {\alpha}, ( \vartheta {D}^{c} ) ( s, s') \,
			  {\alpha'}
		\rangle_{{\mathfrak H}_{-\frac12}}\;, & \text{in case $+$} \; ,\nonumber\\
		\langle   
			 {\alpha}, \overline{ ( \vartheta {D}^{c} )( s, s')}\,
			  {\alpha'}
		\rangle_{{\mathfrak H}_{-\frac12}} \;,  & \text{in case $-$} \;. 
		\end{cases}
		\qquad 
	\label{Sharp-Time Scalar Products}		
	\end{eqnarray}
\end{proposition}

\begin{proof} We have 
	\begin{align*}
	\bigl\langle  f, \vartheta D^{c} g \bigr\rangle_{\mathcal K}
	&=  \Big\langle \int_{0}^{\beta/2}  {\rm e}^{-t\boldsymbol{\ell}_+ } \kappa_+ (f_{t}) \,dt \; , 	  
	\int_{0}^{\beta/2}  {\rm e}^{-t'\boldsymbol{\ell}_+ } \kappa_+ (g_{t'})\,dt' \Bigr\rangle_{\mathcal{H}_{1}} \\ 
	\bigl\langle  f, D^{c} \vartheta g \bigr\rangle_{\mathcal K}
	\kern -1mm
	&=  \Big\langle \int_{0}^{\beta/2}  {\rm e}^{-t\boldsymbol{\ell}_- } \kappa_- (f_{t}) \,dt \; , 	  
	\int_{0}^{\beta/2}  {\rm e}^{-t'\boldsymbol{\ell}_- } \kappa_- (g_{t'})\,dt' \Bigr\rangle_{\mathcal{H}_{1}} \: . 
	\nonumber 
	\end{align*}
This verifies the first two statements. Next, note that $\vartheta K^{c}=K^{c}\vartheta$ and $\vartheta L^{c}=-L^{c}\vartheta$, so  $\vartheta D^{c\,\frac12}=(D^{c\,\frac12})^{*}\vartheta$.  Hence,
	\[
		 \langle f, \vartheta D^{c}f \rangle_{\mathcal K}
		= \langle  f, \vartheta D^{c\,\frac12} D^{c\,\frac12} f \rangle_{\mathcal K} 
		= \langle D^{c\,\frac12} f, \vartheta D^{c\,\frac12} f\rangle_{\mathcal K}\;.
	\]
As $\vartheta$ is unitary, one can use the Schwarz inequality in $\mathcal K$, as well as inequality~\eqref{AbsDc Bound} of Proposition \ref{Prop:Compactified ComplexCovarianceBounds}.  Moreover,
we have seen in \eqref{extension estimate}
that there is a constant $M<\infty$ such that for all $f\in\mathcal K_{+,0}$,
	\begin{equation}
	\label{H_1 bound}
		\Vert {\widehat f}^{\; \pm} \Vert_{\mathcal H_{1}} 
		\leqslant M \Vert f\Vert_{{\mathfrak H}_{-1}(\boldsymbol{X})}\;.
	\end{equation}
As $\mathcal K_{+,0}$ is dense in $\mathfrak H_{-1}(\boldsymbol{X})$, \eqref{H_1 bound} ensures that the 
maps $\wedge_\pm$ extend by continuity to maps 
from ${\mathfrak H}_{-1}(\boldsymbol{X}_{+})$ to ${\mathcal H}_{1}$. For $s,s' \in S^{1}_{+}$ fixed, one can interpret $\vartheta D^{c}(x,x')$ 
as defining a transformation on the Sobolev space ${\mathfrak H}_{-\frac12}(\mathbb R^{d-1})$, namely, 	
	\begin{equation}
		(\vartheta {D}^{c})( s, s')
		=(\mathbb{1}+\rho_{+}) {{\rm e}^{-(s+s') {\mu}_{+}}}
			+\overline{ \rho_{+}{{\rm e}^{(s+s') {\mu}_{+}}}
			} \;
		\label{vartheta Dc 2}
	\end{equation}
for $ s,s'\in S^{1}_{+}$.  Similarly, for $ s,s'\in S^{1}_{+}$, 
	\begin{equation}
		({D}^{c}\vartheta )( s, s')
		=(\mathbb{1}+\rho_{-}) {{\rm e}^{-(s+s'){\mu}_{-}}}
			+\overline{ {{\rm e}^{(s+s'){\mu}_{-}}  \rho_{-}}
			} \;.
	\label{Dc vartheta  2}
	\end{equation}
Also, 
$		\overline{\vartheta D^{c}(s,s')}
		= D^{c}\vartheta(s,s')
$.
Thus\textcolor{red}{,} \eqref{Sharp-Time Scalar Products} follows from \eqref{vartheta Dc 2} and \eqref{Dc vartheta  2}.
\end{proof}

We summarize our results in a commutative diagram that relates quantization, compactification, and the Araki-Woods maps~$\kappa_\pm$:
\smallskip
\begin{equation*}
\begin{CD}
\mathfrak H_{-1}(\mathbb R^{d}_{+})     
@>{\rlap{$\scriptstyle{\ \ \ \ \text{compactification}}$}\phantom{\text{very very very very long}}}>>
\mathfrak H_{-1}(S^{1}_+ \times  \mathbb{R}^{d-1})  \\
@VV{\text{$\wedge_\pm$}}V        @VV{\text{$\wedge_\pm$}}V\\
\mathfrak H_{-\frac12}(\mathbb R^{d-1})    
@>{\rlap{$\scriptstyle{\ \ \ \ \ \ \ \ \text{$\kappa_\pm$}}$}\phantom{\text{very very very very long}}}>>
\mathfrak H_{-\frac12} (\mathbb R^{d-1})\oplus \overline{\mathfrak H_{-\frac12}(\mathbb R^{d-1})} \; \; . 
\end{CD}
\qquad
\end{equation*}

\begin{remark}	\label{thermal quantization maps 3,4}
The maps $\kappa_\pm$ (and thus also the maps $\wedge_\pm$) are linear. We might as well define two anti-linear maps 
$\kappa'_\pm \colon \mathcal K_{+,0}\mapsto {\mathcal H}_{1}$ by setting 
	\begin{equation}
		\label{kappa prime}
		\kappa'_\pm (\alpha) 
		= \left(  {\rho}_\pm^{1/2} 
			\alpha, 
			\overline{(\mathbb{1}+ {\rho}_\pm)^{1/2} 
		\alpha} \right)   \; .	
	\end{equation}
For $\alpha \in \mathfrak H_{-\frac12}(\mathbb R^{d-1})$, the map ${\mathbb R} \ni s \mapsto  \kappa'_\pm ({\rm e}^{i s {\mu}_\pm} \alpha )  = {\rm e}^{i s {\ell}_\pm} \kappa'_\pm ( \alpha )$ extends analytically to the strip $\{ s+it \in  \mathbb{C} \mid - \frac{\beta}{2} < t  < 0 \}$. 
This leads to two anti-linear quantization maps $\#_\pm$, 
namely,
	\[
		f^{\#_\pm}  (\vec x) =  
		\int_{0}^{-\beta/2}  
		{\rm e}^{-t {\ell_{\pm}}} \kappa'_\pm (f_t(\vec x))\, dt \; . 
	\]
\end{remark}

\subsection{Time Translation and its Unbounded Quantization\label{Sect: Time Translation and its Unbounded Quantization}}
Let $T(s)$ denote the unitary time translation group  on $\mathcal{K}=L_{2}(\boldsymbol{X})$, by
	\[
		T(s)f_{t}=f_{t-s}
		\;,
		\quad\text{or}\quad
		(T(s)f)(t,\vec x)=f(t-s,\vec x)\;.
	\]
The periodicity of time causes a problem for the quantization of~$T(s)$.  The function $T(s)f_{t}$, for   $0\leqslant s$, is supported at positive-time (\emph{i.e.}, in the time-interval $[0,\frac{\beta}{2}]$), only if $f_{t}$ is  supported in the time-interval $[0,\frac{\beta}{2}-s]$. The domain of~$\widehat{T(s)}^\pm$ does not include all of $\widehat{\mathcal{K}}_{+}^\pm$, and consequently  $\widehat{T(s)}^\pm$ must be unbounded. Recall ${\ell}_\pm =   {\mu}_{\pm} \oplus ( -{\mu}_{\mp})$.

\begin{proposition}\label{Prop VII.7}
Let $s\in(0, \frac{\beta}{2})$ and let $\mathcal D_{s}$ be the linear span of   $\delta_{t}\otimes\alpha$ for $t\in[0, \frac{\beta}{2} - s]$.  The quantizations $\widehat{T(s)}^\pm$ of $T(s)$ with domains $\widehat{\mathcal{D}}_{s}^\pm$ have self-adjoint closures on $\mathcal H_{1}$. Explicitly, these are given by
	\begin{equation}
		\widehat {T(s)}^\pm 
		= {\rm e}^{-s \boldsymbol{\ell}_\pm}.
	\label{Liouville 1}
	\end{equation}
The spectrum of ${\ell}_+$ and ${\ell}_-$ is $\mathbb R \setminus (-M,M) $, where $M=m\sqrt{1-\vec {\it v}^{\,2}}$.   
\end{proposition}

\begin{remark}  Returning from the rescaled time to the proper time amounts to replacing 
${\ell}_\pm$ by  $ (1-\vec {\it v}^{\,2})^{-1/2} {\ell}_\pm$. The latter has spectrum  
$\mathbb R \setminus (-m,m) $.
\end{remark}

\begin{proof}
The fact that the quantizations of $\mathcal D_{s}$ are dense in $\mathcal H_{1}$ follows from Proposition \ref{Lm:Sharp-Time Quantization}. The matrix elements of  $\widehat {T(s)}^\pm$  in sharp-time vectors follow from Proposition \ref{Prop:Quantization Map}. 
If one restricts $\alpha$ to have its Fourier transform supported on a fixed compact domain, then both~${\mu}_{\pm}$, as well as 
$\widehat {T(s)}^\pm$, are bounded operators on such a subspace. Such subspaces of functions 
are dense in 
$\mathfrak H_{-\frac12}$, so both  $\widehat {T(s)}^+$ and $\widehat {T(s)}^-$  are essentially self-adjoint.  The spectral properties  
follow from those of ${\mu}_{\pm}$ established in \S\ref{Sect:ExampleI}.
\end{proof}

\begin{proposition}\label{Lm:Sharp-Time Quantization}
Let ${\mathcal D}_{s_{1},s_{2}}$ denote the linear span  of generalized functions of the form $\delta_{s}\otimes\alpha$, with  
$\alpha\in\mathfrak H_{-\frac12}$ and  $s=s_{1}$ or $s=s_{2}$ for $s_{1}\neq s_{2}$.   Then  
$\widehat {\mathcal D}_{s_{1},s_{2}}^\pm$ are   dense in ${\mathcal H}_{1}$.  
\end{proposition}

\begin{proof}
We show that the range under quantization of two distinct sharp  
times $s_{1}, s_{2}$ gives a core of  ${\mathcal H}_{1}$. On the contrary, suppose there exists a 
unit vector $\chi\in\mathcal H_{1}$ with components $\chi_{1}, \chi_{2}$, 
which is  orthogonal to all vectors of the form
$\widehat{\delta_{s_{j}}\otimes\alpha}^+$ for $j=1$ and $j=2$. According to this assumption,    
	\begin{align*}
		\bigl\langle \chi, \widehat{\delta_{s}\otimes\alpha}^+ \bigr\rangle_{\mathcal H_{1}}
		&= \langle \chi_{1}, (\mathbb{1}+\rho_{+})^{1/2} 
		{\rm e}^{-s  \mu_{+}}\alpha \rangle_{\mathfrak{ H}_{-\frac12}}\nonumber \\
		& \quad + 
		\bigl\langle 
		\chi_{2}, (\mathbb{1}+\rho_{-})^{1/2}  
		{\rm e}^{-(\frac{\beta}{2} -s)  \mu_{-}}\alpha \bigr\rangle_{\mathfrak{ H}_{-\frac12}} \\ & =0
	\end{align*}
for all $\alpha\in\mathfrak H_{-\frac12} (\mathbb{R}^{d-1})$ and for $s=s_{1}, s_{2}$. Taking adjoints in the inner products, as ${\mu}_{\pm}^{*}={\mu}_{\mp}$ and ${\rho}_{\pm}^{*}={\rho}_{\mp}$, we infer
\begin{align}
		 (\mathbb{1}+\rho_{-})^{1/2} {\rm e}^{-s_{j} \mu_{-}}\chi_{1}
		 =-  (\mathbb{1}+\rho_{+})^{1/2}  {\rm e}^{-(\frac{\beta}{2} -s_{j})  \mu_{+}}
		 \chi_{2} \label{chi1chi2}
\end{align}
for $j=1,2$.  There is no loss of generality to assume $s_{1}< s_{2}$, so 
$ {\rm e}^{-(s_{2}-s_{1}) {\mu}_{-}}$ is bounded.  Thus, we arrive at the following system of equations
	\begin{align*} 
		 (\mathbb{1}+\rho_{-})^{1/2} {\rm e}^{-s_{2} {\mu}_{-}}\chi_{1}
		 &=  - {\rm e}^{-(s_{2}-s_{1})   \mu_{-}}
		   (\mathbb{1}+\rho_{+})^{1/2}  {\rm e}^{-(\frac{\beta}{2} -s_{1}){\mu}_{+}}
		 \chi_{2} \; , \\
		 (\mathbb{1}+\rho_{-})^{1/2} {\rm e}^{-s_{2} {\mu}_{-}}\chi_{1}
		 &=  -  (\mathbb{1}+\rho_{+})^{1/2}  {\rm e}^{-(\frac{\beta}{2} -s_{2}) {\mu}_{+}}
		 \chi_{2}\;. 
	\end{align*}
Eliminating $\chi_{1}$ yields
	\[
		{\rm e}^{-(s_{2}-s_{1})   \mu_{-}}
		   (\mathbb{1}+\rho_{+})^{1/2}  {\rm e}^{-(\frac{\beta}{2} -s_{1}){\mu}_{+}}
		 \chi_{2} =   (\mathbb{1}+\rho_{+})^{1/2}  {\rm e}^{-(\frac{\beta}{2} -s_{2}) {\mu}_{+}}
		 \chi_{2}
		 \;.
	\]
Note that $\mu_{+}$, $\mu_{-}$, $\rho_{+}$ and $\rho_{-}$ all commute. Thus, multiplying both sides with 
$(\mathbb{1}+\rho_{+})^{-1/2}  {\rm e}^{(\frac{\beta}{2} -s_{2}) {\mu}_{+}} $, we arrive at
	\begin{equation}
	\label{Can Not Be}
		  \chi_{2} = {\rm e}^{-(s_{2}-s_{1})
		 ( {\mu}_{+} + {\mu}_{-} ) } \chi_{2}
		 = {\rm e}^{-2(s_{2}-s_{1})\mu}\chi_{2}
		\;,
	\end{equation}
where we use ${\mu}_{+} + {\mu}_{-} =2\mu$.   But $0<s_{2}-s_{1}$ and $0<m\leqslant \mu$, so  
$\Vert {\rm e}^{-2(s_{2}-s_{1})\mu}\Vert<1$.   Thus\textcolor{red}{,}  \eqref{Can Not Be} can only hold in  case  $\chi_{2}=0$, and \eqref{chi1chi2} implies $\chi_1 =0$, since $ (\mathbb{1}+\rho_{-})^{1/2} {\rm e}^{-s_{j} \mu_{-}}$ is not singular.
Hence, $\chi\equiv0$, which contradicts 
the assumption $\Vert \chi\Vert =1$.
\end{proof}

\subsection{The Tomita-Takesaki Operators\label{The Tomita-Takesaki Operators-SF}} 

We introduce a time-re\-flection operator $\theta$ that leaves 
$\boldsymbol{X}_{\pm}$ invariant by 
	\begin{equation*}
		\theta \colon (t,\vec x) \mapsto (\tfrac{\beta}{2} -t,\vec x)\; .
	\end{equation*}
Thus,  $\theta$  reflects the time   about $t=\pm\frac{\beta}{4}$, depending on whether $t$ is positive or negative. It follows that
	\begin{equation*}
		\theta L_{2}(\boldsymbol{X}_{\pm})  = L_{2}(\boldsymbol{X}_{\pm}) \;.
	\end{equation*}
Acting on $L_{2}(\boldsymbol{X})$, or on $L_{2}(S^{1})\times 
{\mathfrak H}_{-1/2}(\mathbb R^{d-1})$, the operator $\theta$ is  a self-adjoint, real, symmetric, idempotent, and commutes with $\vartheta$, 
	\begin{equation*}
		\theta^{*}  = \theta = \overline \theta\;,\quad
		\theta^{2} = \mathbb{1}\;,\quad
		\vartheta\theta=\theta\vartheta\;.
	\end{equation*}
Consequently, $\theta $ commutes with $\vartheta D^{c}(t,t')$.
The map $f \mapsto  \theta \overline{f}$ maps $f_t \mapsto \overline{f_{\frac{\beta}{2}-t}}$, and thus induces an anti-linear involution whose quantization is the Tomita-Takesaki modular conjugation. In order to verify this claim, we define the relevant modular objects. The spaces 
	\[ 
	\mathcal L_\pm =
	\kappa_\pm \bigl(\mathfrak H_{-\frac12}(\mathbb R^{d-1})\bigr)
	\]
are real subspaces in $\mathcal H_{1}=\mathfrak H_{-\frac12} (\mathbb R^{d-1})\oplus \overline{\mathfrak H_{-\frac12}(\mathbb R^{d-1})}  $. 
Multiplication by $(i \oplus -i ) $ preserves the subspaces $ {\mathcal L}_\pm$, 
but multiplication by $\boldsymbol{i} = (i \oplus i ) $ does not. Moreover,
\begin{itemize}
\item[$i.)$] $ {\mathcal{L}}_\pm \cap \boldsymbol{i} {\mathcal{L}}_\pm  = \{0\}$; 
\item[$ii.)$] $ {\mathcal{L}}_\pm + \boldsymbol{i}  {\mathcal{L}}_\pm$ is dense in~$\mathcal H_{1}$. 
\end{itemize} 
It is interesting to note that 
	\begin{equation*}
	\boldsymbol{i} \mathcal L_\pm =
	\kappa'_\pm \bigl(\mathfrak H_{-\frac12}(\mathbb R^{d-1})\bigr)\; . 
	\end{equation*}
Eckmann and Osterwalder~\cite{EO} have shown that, whenever a real subspace of a Hilbert space satisfies $i.)$ and $ii.)$, one can define an anti-linear operator $s_\pm$ by setting 
\begin{align*}
			s_\pm:    {\mathcal{L}}_\pm  
				& +   \boldsymbol{i}   {\mathcal{L}}_\pm  
			 \to    {\mathcal{L}}_\pm   
				  +   \boldsymbol{i}  {\mathcal{L}}_\pm  \\
				  {k} & +   \boldsymbol{i}  {k}'
				\, \, \, \mapsto -  {k}  +   \boldsymbol{i} 		
				 {k}'. 
\end{align*}
The operator $s_{\pm}$ are closable. 
The polar decompositions of their closure 
\begin{equation}
\label{tomita}
\overline {s}_\pm = j \delta_\pm^{1/2} \;,
\end{equation}
define the modular conjugation $j$, the modular operators $\delta_{\pm}^{1/2}$, and the one-particle Liouvillian $L_{\pm}$. In our case,  $j$ maps $(f, \overline{g})$ to $(-g, -\overline{f})$. Thus,
\begin{equation}
\label{modular}
j \circ \kappa_\pm =  -  \kappa'_\pm \; , \qquad 
\text{and} \quad j  {\mathcal L}_\pm = \boldsymbol{i} {\mathcal L}_\pm \;. 
\end{equation}
The modular operator $\delta^{1/2}_\pm$ is related to the one-particle Liouvillian,
\begin{equation}
\label{tomita modular operator}
\delta^{1/2}_\pm = {\rm e}^{- \beta \ell_\pm /2} \; , 
\quad\text{with}\quad
\ell_\pm= {\mu}_\pm \oplus (-  {\mu}_{\mp})\;.
\end{equation}
We summarize our results.

\begin{proposition}
Let ${\mathscr C}$ denote the anti-linear operator of complex conjugation 
${\mathscr C} f=\overline{f}$.
Then one has the commutative diagram:
\begin{displaymath}
    \xymatrix{ 
    \mathfrak H_{-1}(\boldsymbol{X}_+) 
     \ar[r]^{ \theta {\mathscr C} } \ar[d]^{\wedge_\pm} \ar[dr] ^{\#_\pm}
     & \mathfrak H_{-1}(\boldsymbol{X}_+) 
      \ar[d]^{\wedge_\pm} \\
               \mathcal H_{1}
                 \ar[r]^{j}& 
                \mathcal H_{1}     \; .  
                 } 
\end{displaymath}
\end{proposition}
\goodbreak

\subsection{The Araki-Woods Fock Space\label{Sect: The Araki-Woods Fock Space}} 
The one-particle space $\mathcal H_{1}= \mathfrak H_{-\frac12} (\mathbb R^{d-1})\oplus \overline{\mathfrak H_{-\frac12}(\mathbb R^{d-1})}$ gives rise to a Fock space of the form \eqref{Fock Space}. The elements of $ (h, h') \in \mathcal H_{1}$ have two components, 
namely $h \in \mathfrak H_{-\frac12} (\mathbb R^{d-1})$ and 
$h' \in \overline{\mathfrak H_{-\frac12}(\mathbb R^{d-1})}$. 
The  Fock {\em annihilation operator} 
	\[
	a (h \oplus h') = \Bigl( \int_{\mathbb R^{d-1}} h(\vec x) a (\vec x) d\vec x \Bigr) 
	\oplus  \Bigl( \int_{\mathbb R^{d-1}} h'(\vec x') \, a(\vec x') \, d \vec x' \Bigr)
	\]
(which is actually a densely-defined bilinear form on $\mathcal{H}\times\mathcal{H}$) 
has non-vanishing matrix elements from $\mathcal{H}_n$ to $\mathcal{H}_{n-1}$.  In the Fourier representation it acts as 
	\begin{align*}
	& (a(\vec k \oplus \vec k')f)_{n-1}(\vec k_1 \oplus \vec k'_1, \ldots, \vec k_{n-1} \oplus \vec k'_{n-1})
	\\
	& \qquad =\sqrt{n}\,f_n(\vec k \oplus \vec k',\vec k_1\oplus \vec k'_1,\ldots, \vec k_{n-1} \oplus \vec k'_{n-1}) \;.  
	\end{align*}
and satisfies $[a(\vec k_1\oplus \vec k_1'), a(\vec k_2\oplus \vec k_2')]=0$. The adjoint creation form $a(\vec k \oplus \vec k')^*$ satisfies the usual canonical
relations, namely,
	\[ [a( \vec k_1\oplus \vec k_1'), a^*(\vec k_2 \oplus  \vec k_2')]=\delta(\vec k_1- \vec k_2)\oplus \delta(\vec k'_1- \vec k'_2) \; .  \] 
The hermitian time-zero field $\varphi(\vec x \oplus \vec x' )$ on~$ \mathbb{R}^{d-1} \times  \mathbb{R}^{d-1}$ is defined as
	\begin{align*}
		\varphi(\vec x \oplus \vec x' )
		& =  (2\pi)^{-\frac{d-1}{2}}
		\int \frac{d\vec k}
		{\sqrt{2 \mu (\vec k ) }}   \left( a^{*}(\vec k) +a(-\vec k ) \right) {\rm e}^{-i \vec k \cdot\vec x } 
		\\
		& \qquad   \oplus				
			 (2\pi)^{-\frac{d-1}{2}}
		\int \frac{d\vec k'}
		{\sqrt{2 \mu (\vec k' ) }} \left( a^{*}(\vec k') +a(-\vec k' ) \right) {\rm e}^{-i \vec k' \cdot\vec x' }    
\; . 
\end{align*}

The Liouvillean $L_\pm$, the momentum operator $\vec P$, 
the modular conjugation~$J$ and the Tomita operator $S$ 
act on $\mathcal{H}_{1}$ as \textcolor{red}{$\ell_\pm$}, ${ \vec p}$, $j$ and $s$, respectively. Note that 
\[
\Omega_0 \in {\mathcal D} (L_\pm) \quad \text{and} \quad  L_\pm \Omega_{0} = 0 \; .
\]
The spectrum of $L_\pm$ is the real line $ \mathbb{R}$, and its zero eigenvalue is simple. The latter follows from the  gap in the spectrum of $\ell_\pm$ \cite[Theorems 1a \& 1b]{Kay1}.

\subsection{Quantization of Field Operators\label{Sect: Quantization of Field Operators}} 
The quantization map \eqref{thermal quantization map 1}
for vectors fixes the quantization map for the field operators, as we require that 
\begin{equation}
		\widehat{\Phi(f)}^{\pm} \widehat{\, \Omega \,}^{\pm}
		= \widehat {\Phi(f) \Omega}^{\pm} \quad \text{for} \quad
		\Omega \in \mathcal{D}(\Phi(f))\cap   \mathcal{E}_{\pm}\;.
	\label{thermal field operator quantization}
\end{equation}
This implies that 
\begin{align}
\label{field quantization condition}
	\langle (\widehat{\Phi(f)}^{\pm} )^n \,  \Omega_{0} ,  (\widehat{\Phi(f)}^{\pm} )^n \,  \Omega_{0} \rangle_{ \mathcal{H}  }  
		&= \langle \Phi(f)^{n}\Omega_{0}^{E}, \Theta_\pm \Phi(f)^{n}\Omega_{0}^{E} \rangle_{  \mathcal{E}  } \nonumber \\
		&
		= (2n-1)!!\langle\widehat {f}^{\pm}, \widehat {f}^{\pm} \, \rangle^{n}_{ {\mathcal H}_1 } \; ,
\end{align}
where $\Theta_\pm$ are the reflections associated to $\vartheta D^{c}$ and $D^{c}\vartheta$, respectively. As before,  ${\mathcal H}_1 = \mathfrak{H}_{-\frac{1}{2}}  \oplus \overline{\mathfrak{H}_{-\frac{1}{2}}}$ and 
	\begin{align*}
	\langle\widehat {f}^{\pm}, \widehat {f}^{\pm} \rangle_{ {\mathcal H}_1}
	&=  \Bigl\| \; \bigl|  \int_{0}^{\beta/2}  
		{\rm e}^{-t{\mu}_{\pm} } ( \mathbb{1}+\rho_{\pm} )^{1/2} f_{t}\,dt \; , 
		\int_{0}^{\beta/2}  {\rm e}^{t {\mu}_{\pm} } \rho_{\pm}^{\frac{1}{2}} \,  \overline{f_{t}} \,dt \bigr\rangle \; \Bigr\|^2_{{\mathcal H}_1} \;.  
	\end{align*}
The Gaussian nature of the Fock space~$  \mathcal{E}  $ 
together with Proposition~\ref{Prop:Quantization Map} imply that~\eqref{field quantization condition}
is satisfied if we set
\[
\widehat{\Phi(f)}^{\pm} = \varphi(\widehat {f}^{\pm} ) \; ,  \qquad   f \in \mathcal{E}_1 \cap   \mathcal{E}_{\pm} \; . 
\]
Note that $\widehat {f}^{\pm} $ has two components, and therefore $\widehat {f}^{\pm} $
can be viewed as a function on $ \mathbb{R}^{d-1} \times  \mathbb{R}^{d-1}$. For certain sharp-time functions, $f$ and $g$, the quantized field operators take a special form.

 \begin{proposition}
 \label{thermal quantized sharp-time fields}
 For $f= \delta \otimes \alpha$ with $\alpha \in \mathfrak{H}_{-\frac{1}{2}} (\mathbb R^{d-1})$\textcolor{red}{,} we have
 \begin{align}
	\label{thermal time zero field}
	\widehat{\Phi(f)}^{\pm}  & = \widehat{\Phi(0,\alpha)}^{\pm}
	=\varphi \bigl( \kappa_\pm (\alpha)\bigr) \;  . 
 \end{align}
Moreover, for $g= \delta_{\frac{\beta}{2}} \otimes \alpha$, where $ \delta_{\frac{\beta}{2}} (t) =  \delta(t- \frac{\beta}{2})$ denotes the shifted Dirac delta function, we have
\begin{align}
	\label{right-regular-rep}
	 \widehat{\Phi(g)}^{\pm} &= \widehat{\Phi(\tfrac{\beta}{2},\alpha)}^{\pm}
	 = \varphi \bigl(   \kappa'_\pm (\alpha) \bigr) \; .
\end{align} 
\end{proposition}
 
\begin{proof} The identity	
\eqref{thermal time zero field} follows from \eqref{thermal field operator quantization} and \eqref{thermal quantization map 1}. 
The second identity, \eqref{right-regular-rep},  follows from 
	\begin{align*}
		\widehat{g}^{\pm} &=  
		{\rm e}^{- \beta {\ell_{\pm}} /2 } \kappa_\pm (\alpha ) \\
		& = \left( {\rm e}^{- \beta {\mu_{\pm}}/2}(\mathbb{1}+\rho_{\pm})^{1/2}\alpha, \overline{
		{\rm e}^{ \, \beta {\mu_{\pm}}/2} \rho_{\pm}^{1/2} \alpha}\, \right) \\
		& = \left(  {\rho}_\pm^{1/2} \alpha, \overline{(\mathbb{1}+ {\rho}_\pm)^{1/2} \alpha} \right) 
		= \kappa'_\pm (\alpha) \: . 
	\end{align*}
\end{proof}

The bounded functions of the time-zero fields generate abelian 
von Neumann algebras
\[
\mathcal{U}_0^{\pm} = \{ \widehat{\Phi(0,\alpha)}^{\pm}  \mid \alpha \in \mathfrak{H}_{-\frac12} (\mathbb R^{d-1})\}'' \; . 
\]
Similarly, the bounded functions of the time-$\frac{\beta}{2}$ fields generate another two abelian von Neumann algebras,
\[
\mathcal{U}_{\frac{\beta}{2}}^{\pm} = \{ \widehat{\Phi(\frac{\beta}{2},\alpha)}^{\pm}  
\mid \alpha \in \mathfrak{H}_{-\frac12} (\mathbb R^{d-1})\} '' \; .
\] 
Not only do they commute with each other, but also their time translates commute with each other. 
For $\alpha \in  \mathfrak{H}_{-\frac12} (\mathbb R^{d-1})$ and $s \in \mathbb{R}$, 
\begin{align*}
\varphi_\pm (s, \alpha) &= {\rm e}^{is L_\pm}\varphi (\kappa_\pm (\alpha)){\rm e}^{-is L_\pm}\; , \\ 
\varphi'_\pm (s, \alpha) &= {\rm e}^{is L_\pm}\varphi (\kappa'_\pm (\alpha)){\rm e}^{-is L_\pm}\;  . 
\end{align*}
We will also use $\varphi_\pm (\alpha) = \varphi_\pm (0, \alpha)$ and
$\varphi'_\pm (\alpha) = \varphi'_\pm (0, \alpha)$. 

\begin{proposition}
\label{thermal commutator}
For $ \alpha, \alpha' \in \mathfrak{H}_{-\frac12} (\mathbb R^{d-1})$ and $ s, s' \in \mathbb{R}$, \textcolor{red}{,}
\[
	\bigl[\,   \varphi_\pm \bigl(  -s,  \overline{\alpha} \bigr) , 
	  \varphi'_\pm \bigl( s',   \alpha' \bigr)
	  \bigr]= 0 \; . 
\]
In particular, for $\alpha, \alpha' \in \mathfrak{H}_{-\frac12} (\mathbb R^{d-1})$,
\[
	\bigl[\,   \widehat{\Phi(0,\alpha)}^{\pm} , 
	  \widehat{\Phi(\tfrac{\beta}{2},\alpha')}^{\pm}\bigr]= 0.
\]
\end{proposition}

\begin{proof} We compute 
\begin{align*}
	& \langle {\rm e}^{is L_\pm} \varphi \bigl(   \kappa_\pm (\alpha)\bigr) \,  \Omega_{0}
	,  {\rm e}^{it L_\pm}\varphi \bigl(   \kappa'_\pm  (\alpha')\bigr) \,  \Omega_{0} \rangle_{ \mathcal{H}  } 
	\\
	& \qquad = \langle {\rm e}^{is {\ell}_\pm} \kappa_\pm (\alpha)
	,  {\rm e}^{it {\ell}_\pm} \kappa'_\pm  ( \alpha') \rangle_{ \mathcal{H}_1  }
	\\
	& \qquad = \langle {\rm e}^{is {\mu}_\pm} \alpha, {\rho}_\pm^{1/2}(\mathbb{1}+ {\rho}_\pm)^{1/2}
	,  {\rm e}^{it {\mu}_\pm}  \alpha'  \rangle_{ \mathfrak{H}_{-\frac12} (\mathbb R^{d-1}) }  
	\\
		& \qquad \qquad + \langle {\rm e}^{-it {\mu}_\pm} \overline{\alpha}' ,  {\rho}_\pm^{1/2}(\mathbb{1}+ {\rho}_\pm)^{1/2}
	{\rm e}^{- is {\mu}_\pm} \overline{\alpha}  \rangle_{ \mathfrak{H}_{-\frac12} (\mathbb R^{d-1})  } \; .
\end{align*}
Similarly, 
\begin{align*}
	& \langle {\rm e}^{-it L_\pm} \varphi \bigl(   \kappa'_\pm (\overline{\alpha}')\bigr) \,  \Omega_{0}
	,  {\rm e}^{-is L_\pm}\varphi \bigl(   \kappa_\pm  ( \overline{\alpha})\bigr) \,  \Omega_{0} \rangle_{ \mathcal{H}  } 
	\\
	& \qquad = \langle {\rm e}^{- it {\ell}_\pm} \kappa'_\pm (\overline{\alpha}')
	,  {\rm e}^{- is {\ell}_\pm} \kappa_\pm  ( \overline{\alpha}) \rangle_{ \mathcal{H}_1  }
	\\
	& \qquad = \langle {\rm e}^{- it {\mu}_\pm} \overline{\alpha}', {\rho}_\pm^{1/2}(\mathbb{1}+ {\rho}_\pm)^{1/2}
	,  {\rm e}^{- is {\mu}_\pm}  \overline{\alpha}  \rangle_{ \mathfrak{H}_{-\frac12} (\mathbb R^{d-1}) }  
	\\
		& \qquad \qquad + \langle {\rm e}^{is {\mu}_\pm}  \overline{\alpha} ,  {\rho}_\pm^{1/2}(\mathbb{1}+ {\rho}_\pm)^{1/2}
	{\rm e}^{it {\mu}_\pm}  \alpha'  \rangle_{ \mathfrak{H}_{-\frac12} (\mathbb R^{d-1})  }.
\end{align*}
Therefore, the expectation of the  commutator vanishes.  As the commutator is a scalar, it must equal zero.  
The second commutator claimed to vanish is a special case, as the fields on the boundary can obtained from  \eqref{right-regular-rep}. 
\end{proof}

That is, the adjoint action of the unitary group ${\rm e}^{it L_\pm}$, $t \in  \mathbb{R}$,  
on $\mathcal{U}_0^{\pm}$ and $ \mathcal{U}_{\frac{\beta}{2}}^{\pm}$ gives rise to two commuting non-abelian algebras, 
\begin{equation}
\label{vNA-R_0}
\mathcal{R}_0^{\pm} = \{ \varphi_\pm (s,\alpha)  \mid s \in \mathbb{R}, \alpha \in \mathfrak{H}_{-\frac12} (\mathbb R^{d-1})\}''  
\end{equation}
and 
\begin{equation}
\label{vNA-R_b}
\mathcal{R}_{\frac{\beta}{2}}^{\pm} = \{ \varphi'_\pm (s,\alpha)  \mid s \in \mathbb{R}, \alpha \in \mathfrak{H}_{-\frac12} (\mathbb R^{d-1})\}'' \; ,  
\end{equation}
respectively.

\begin{proposition} 
\label{neu}
Let $\alpha_{i}\in \mathfrak{H}_{-\frac12} (\mathbb R^{d-1})$ and $0 \leqslant s_i$, $1\leqslant i\leqslant n$. 
Moreover,  assume that $ \sum_{j= 1}^n s_j \leqslant \frac{\beta}{2}$. Then 
\begin{equation*}
	\label{kl}
		{\rm e}^{- s_{n-1} L _\pm} \varphi_\pm (\alpha_{n-1}) 
		\cdots {\rm e}^{- s_{1}  L_\pm } 
		\varphi_\pm (\alpha_{1})   \Omega_{0} \in  {\mathcal D}  \bigl( \varphi_\pm (\alpha_{n}) \bigr)
\end{equation*}
and 
\begin{equation*}
		\label{kln}
 		 \varphi_\pm (\alpha_{n}) {\rm e}^{- s_{n-1} L_\pm } \varphi_\pm (\alpha_{n-1}) 
		\cdots {\rm e}^{- s_{1}  L_\pm } 
		\varphi_\pm (\alpha_{1})   \Omega_{0} \in {\mathcal D} \bigl({\rm e}^{- s_{n} L_\pm }
		\bigr) \;  . 
\end{equation*}
Furthermore, the linear span of such vectors is dense in ${\mathcal H}$ and
\begin{align}
& {\rm e}^{-s_n L_\pm } \varphi_\pm (\alpha_{n}) {\rm e}^{-  (s_{n-1}+s_n) L_\pm } 
\varphi_\pm (\alpha_{n-1})  
\ldots {\rm e}^{- (s_{1}- s_{2})  L_\pm } 
\varphi_\pm (\alpha_{1}) \Omega_{0}  
\nonumber
\\ 
& \qquad  =\Bigl(\Phi(s_n,\alpha_{n}) \Phi(s_{n-1},\alpha_{n-1})\cdots  \Phi(s_1,\alpha_{1}) \Omega_{0}^{E} \Bigr)^{ \wedge_\pm } \; . 
\label{quantization-of-vectors}
\end{align}
\end{proposition}

\begin{proof} 
Let $\boldsymbol{T}(s)$ be the second quantization of the unitary time translation $T(s)$ introduced 
in Section \ref{Sect: Time Translation and its Unbounded Quantization}. Its quantization is the second quantization of ${\rm e}^{-s \ell_\pm}$, \emph{i.e.}, 
\[
\widehat{\boldsymbol{T}(s)}^\pm = {\rm e}^{-s L_\pm} \; .
\]
It follows that for $0 \leqslant s_i$, $1\leqslant i\leqslant n$, and $ \sum_{j= 1}^n s_j \leqslant \frac{\beta}{2}$,
\[ 
\boldsymbol{T} (s_{n-1}) \Phi(0, \alpha_{n-1} )  \cdots \boldsymbol{T} (s_{1})
\Phi (0, \alpha_{1})\Omega_{0}^{E}  \in 
{\mathcal D} \left( \Phi(0,\alpha_{\alpha_{n}}) \right) \;      
\]
and
\[ 
\Phi(0, \alpha_{n} )\boldsymbol{T} (s_{n-1}) \Phi(0, \alpha_{n-1} )  \cdots \boldsymbol{T} (s_{1})
\Phi (0, \alpha_{1})\Omega_{0}^{E}  \in 
{\mathcal D} \left( \boldsymbol{T} (s_{n}) \right) \; .     
\]
The results now follow from \eqref{quantization-of-vectors}. The fact that the linear span of such vectors is dense in ${\mathcal H}$
is a consequence of Proposition \ref{Lm:Sharp-Time Quantization}.
\end{proof}

\begin{proposition} Let $\alpha_{i}\in \mathfrak{H}_{-\frac12} (\mathbb R^{d-1})$, for $1\leqslant i\leqslant n$. 
If $0\leqslant s_{1}\leqslant \cdots \leqslant s_k \leqslant \frac{\beta}{2} \leqslant s_{k+1}\leqslant \ldots \leqslant s_{n}\leqslant \beta$, then
\begin{align} 
\label{31}
& \Bigl\langle \Omega_{0}^{E} , \prod_{j=1}^{n}  \Phi(s_{j}, \alpha_{j}) \Omega_{0}^{E} \Bigr\rangle
\\
&
= \Bigl\langle {\rm e}^{(s_n-\beta) L_\pm } \varphi_\pm ( \overline{\alpha}_{n})
{\rm e}^{(s_{n-1}-s_{n})  L_\pm} \varphi_\pm ( \overline{\alpha}_{n-1})
\cdots  {\rm e}^{(s_{k+1}-s_{k+2}) L_\pm }
\varphi_\pm (\overline{\alpha}_{k+1}) \Omega_{0} \; ,  \; \quad
\nonumber \\
& \qquad  \qquad {\rm e}^{- s_1 L_\pm } \varphi_\pm (\alpha_{1}) {\rm e}^{(s_1 -  s_2) L_\pm }
\varphi_\pm (\alpha_{2})  \cdots  {\rm e}^{(s_{k-1}-s_{k}) L_\pm }
\varphi_\pm (\alpha_{k}) \Omega_{0} \Bigr\rangle   \, .
\nonumber
\end{align}
Moreover,
\[
\|  {\rm e}^{- (\beta/2) L_\pm}  \varphi_\pm (\alpha_{n})
\cdots
\varphi_\pm (\alpha_{1}) \Omega_{0} \bigr  \|_{\mathcal H}
= \|  \varphi_\pm (\alpha_{n})
\cdots
\varphi_\pm (\alpha_{1})  \Omega_{0} \bigr  \|_{\mathcal H} \; .
\]
\end{proposition}

\begin{proof}
If $0 \leqslant s_{1}\leqslant \ldots\leqslant s_k \leqslant \frac{\beta}{2}$ and $\frac{\beta}{2} \leqslant s_{k+1} \leqslant \ldots \leqslant s_{n} \leqslant \beta$, then according to Proposition \ref{neu}
the right hand side in \eqref{31} is well-defined and equals
\begin{align*}
&  \bigl\langle (  \Phi(\beta-s_n, \overline{\alpha}_{n})
\cdots  \Phi(\beta -s_{k+1}, \overline{\alpha}_{k+1}) \Omega_{0}^{E} \bigr)^{ \wedge_\pm } , \\
& \qquad \qquad \qquad \qquad \qquad \qquad \qquad \qquad
\bigl( \Phi(s_k ,\alpha_{k})   \cdots
\Phi(s_1,\alpha_{1}) \Omega_{0}^{E} )^{ \wedge_\pm }  \bigr \rangle_{\mathcal{H}}
\\
& \qquad = \Bigl\langle  \Bigl(  \prod_{j=k+1}^{n} \Phi(\beta-s_{j}, \overline{\alpha}_{j} )   \Bigr) \Omega_{0}^{E}  ,
\Theta_\pm \Bigl(  \prod_{j=1}^{k} \Phi(s_{j}, \alpha_{j} ) \Bigr) \Omega_{0}^{E}  \Bigr\rangle_{\mathcal E}
\\
& \qquad = \Bigl\langle   \Bigl(  \boldsymbol{T}(\beta) \prod_{j=k+1}^{n} \Phi(-s_{j}, \overline{\alpha}_{j} ) \Bigr)\Omega_{0}^{E}  ,
\Theta_\pm \Bigl(
\prod_{j=1}^{k} \Phi(s_{j}, \alpha_{j} ) \Bigr) \Bigr\rangle_{\mathcal E}
\\
& \qquad = \Bigl\langle  \Bigl(  \prod_{j=k+1}^{n} \Phi(-s_{j}, \overline{\alpha}_{j} ) \Bigr) \Omega_{0}^{E}  ,
\Theta_\pm \Bigl( \prod_{j=1}^{k} \Phi(s_{j}, \alpha_{j} ) \Bigr) \Bigr\rangle_{\mathcal E}
\\
& \qquad = \bigl\langle \Omega_{0}^{E}  , \prod_{j=1}^{n} \Phi(s_{j}, \alpha_{j} ) \Omega_{0}^{E}   \bigr\rangle_{\mathcal E} \; .
\end{align*}
We made use of $\boldsymbol{T}(\beta) =1$, which holds by periodicity. By Proposition \ref{neu} we have
\[
\varphi_\pm (\alpha_{n})  \varphi_\pm ( \alpha_{n-1})
\cdots
\varphi_\pm (\alpha_{1})   \Omega_{0} \in \mathcal{D} \bigl({\rm e}^{- \beta L_\pm / 2}
\bigr)  \; .
\]
Now
\begin{align*}
& \left\| {\rm e}^{- \beta  L_\pm / 2}  \varphi_\pm (\alpha_{n})   \varphi_\pm (\alpha_{n-1})
\cdots
\varphi_\pm (\alpha_{1})   \Omega_{0} \right\|^2_{\mathcal H}
\\
& \quad =
\left\| \Bigl( \boldsymbol{T} (\beta/2) \Phi ( 0, \alpha_{n})
\cdots
\Phi ( 0, \alpha_{1}) \Omega_{0}^{E}
\Bigr)^{\wedge_\pm} \right\|^2_{\mathcal H}
\\
& \quad =
\left\langle  \boldsymbol{T} (\beta/2) \Phi ( 0, \alpha_{n})
\cdots
\Phi ( 0, \alpha_{1}) \Omega_{0}^{E} \, , \, \Theta_\pm \boldsymbol{T} (\beta/2) \; \Phi ( 0, \alpha_{n})
\cdots
\Phi ( 0, \alpha_{1}) \Omega_{0}^{E}  \right\rangle_{\mathcal E}
\\
& \quad =
\left\langle  \Phi ( 0,  \alpha_{n})
\cdots
\Phi ( 0,  \alpha_{1})  \Omega_{0}^{E} \; , \;  \boldsymbol{T}(-\beta/2) \, \Theta_\pm \boldsymbol{T} (\beta/2) \; \Phi ( 0, \alpha_{n})
\cdots
\Phi ( 0, \alpha_{1}) \right\rangle_{\mathcal E}
\\
& \quad =
\left\langle   \Phi ( 0, \alpha_{n})
\cdots
\Phi ( 0, h_{1})  \Omega_{0}^{E} \; , \;  \Theta_\pm \boldsymbol{T} (\beta) \; \Phi ( 0, \alpha_{n})
\cdots
\Phi ( 0, h_{1}) \Omega_{0}^{E} \right\rangle
\\
& \quad =
\left\| \left(  \Phi ( 0, \alpha_{n})
\cdots
\Phi ( 0, \alpha_{1}) \Omega_{0}^{E}
\right)^{\wedge_\pm} \right\|^2_{\mathcal H}
\\
& \quad =
\| \varphi_\pm (\alpha_{n})   \varphi_\pm ( \alpha_{n-1})
\cdots
\varphi_\pm ( \alpha_{1})   \Omega_{0} \|^2_{\mathcal H} \; ,
\end{align*}
again using $\boldsymbol{T}(\beta) =1$.
\end{proof}

\begin{proposition} {\bf (Special case of Theorem 2.5.14. in \cite{BR})}
\label{Tomita theorem}
\quad
\nobreak
\begin{itemize}
\item[$i.)$] The adjoint action of the unitary group $\{ \exp (it L_\pm) \mid t \in  \mathbb{R} \}$ leaves the 
algebras~$\mathcal{R}^\pm_0$ and $\mathcal{R}^\pm_{\beta/2}$ invariant;
\item [$ii.)$]  The identity holds, $	J  \mathcal{R}^\pm_0 J 
=  (\mathcal{R}^\pm_0)' =  \mathcal{R}^\pm_{\beta/2} $; and,
\item [$iii.)$]  The operator $S_\pm$ is closed, its polar decomposition is 
\[
	S_\pm = J {\rm e}^{- \beta L_\pm /2}
\]
and  $ S_\pm  A \Omega_0 = A^* \Omega_0 $ for all $A \in \mathcal{R}^\pm_0$. 
For $B  \in \mathcal{R}^\pm_{\beta/2}$, one has $ S_\pm^*  B \Omega = B^* \Omega_0 $.
\end{itemize}  
\end{proposition}

\begin{proof} Property $i.)$ follows from Proposition \ref{thermal commutator}. 
Property $iii.)$ follows from the fact that $S_\pm$, $J$, and $L_\pm$ are the second quantizations of the 
one particle operators $s_\pm$, $j$, and $\ell_\pm$ which satisfy \eqref{tomita}. 
Finally, Property $ii.)$ follows from \eqref{modular}.
\end{proof}

\begin{corollary} \label{Corollary:KMS} The vector $\Omega_0$ induces a unique  KMS state for the quantum 
dynamical systems $(\mathcal{R}^\pm_{0}, {\rm e}^{it L_\pm} )$
associated to the one-particle Hamiltonians $\mu_{\pm}$ acting on the  Sobolev 
space $ {\mathfrak{H}}_{-\frac12}(\mathbb R^{d-1})$, \emph{i.e.}, 
for bounded operators $A,B \in \mathcal{R}_{0}$ the functions 
	\[
		\mathbb{R} \ni t \mapsto F^\pm_{A, B}(t)= \langle \Omega_{0} , A {\rm e}^{it L_\pm} B \Omega_{0} \rangle
	\]
extend to an analytic functions in the strip $z=t+is$ with  $0<s<\beta$,  with continuous boundary values given by
	\begin{equation}
	\label{KMS condition-sf}
		F^\pm_{A, B}(t+i\beta)
		= \langle \Omega_{0} , B{\rm e}^{-it L_\pm} A \Omega_{0} \rangle \; . 
	\end{equation}
This KMS state is Gaussian and   its two-point function is
	\[
	\left\langle \varphi_\pm ( 0, \alpha )\Omega_{0} , \varphi_\pm  (t',  \alpha') 
	 \Omega_{0} \right\rangle_{ \mathcal{H}  } 	
	= \bigl\langle {\alpha} \, , \, 
	\coth (\beta {\mu}_\pm ) \, 
	{\rm e}^{it' {\mu}_\pm} {\alpha'} \bigr\rangle_{  {\mathfrak{H}}_{-\frac12}   } 
	\;. 
	\]
The GNS representation associated to the pairs $(\mathcal{R}^\pm_{0},\langle \Omega_0 , \, \cdot \, \Omega_0 \rangle )$
are the Araki-Woods representations (see Section \ref{Sect: The Araki-Woods Fock Space}).
\end{corollary}

\begin{proof} The KMS condition  \eqref{KMS condition-sf} follows directly from Theorem \ref{Tomita theorem} (iii).
Uniqueness of the KMS state follows from the commutation relations, the KMS condition and $m>0$. 
The fact that the GNS representation associated to the pair $(\mathcal{R}^\pm_{0},\langle \Omega_0 , \, \cdot \, \Omega_0 \rangle )$
is the Araki-Woods representation follows from the fact that $\Omega_0$ is cyclic for $\mathcal{R}^\pm_{0}$.
\end{proof}

\setcounter{equation}{0}
\section{Classical Fields on the $d$-Torus $\boldsymbol{X}= \mathbb{T}^{d} $\label{Sect:The Fully Compactified Case}}

In this section we study periodization of  both time and spatial directions. Thus\textcolor{red}{,} we are interested 
in the spacetime for the classical field given by $\boldsymbol{X} =   S^1 \times  \mathbb{T}^{d-1} = \mathbb{T}^{d}$ with $S^1$ a circle of circumference $\beta$, and $\mathbb{T}^{d}$ is the $d$-dimensional torus. Let, as before, $\Lambda=\prod_{j=1}^{d-1}\ell_{j}$ denote the spatial volume of the torus~$\mathbb{T}^{d-1}$.

\subsection{The two-point function}
The fully compactified covariances $D^{c}_{\pm, \beta, \Lambda}=D^{c\,*}_{\mp, \beta, \Lambda}$ arise 
form the covariance $D_{\vec {\it v}}$ by compactifying both the time and the spatial coordinates. For the   
kernels this yields
	\begin{align}
		\label{Gibbs-Two-Point-Compact}
		& D^{c}_{\pm, \beta,\Lambda}(x-x') \\
		&= 
			\frac{\theta(t'-t)}{\Lambda}\sum_{\vec k\in \mathcal K_{\Lambda}} 
		\frac{1}{2\mu(\vec k)}\left(
		\frac{{\rm e}^{-(t'-t){\mu}_{\pm}(\vec k)}
		}
		{1-{\rm e}^{-\beta {\mu}_{\pm}(\vec k)}} 
		+ \frac{ {\rm e}^{-(\beta - (t'-t)){\mu}_{\mp}(\vec k)} }
		{{1-{\rm e}^{-\beta {\mu}_{\mp}(\vec k)} }}
		\right) {\rm e}^{-i\vec k\cdot (\vec x-\vec x')} \nonumber\\
	&+	
		\frac{	\theta(t-t')}{\Lambda}\sum_{\vec k\in \mathcal K_{\Lambda}} 
		\frac{1}{2\mu(\vec k)}\left(
		\frac{{\rm e}^{-(t-t'){\mu}_{\mp}(\vec k)}
		}
		{1-{\rm e}^{-\beta {\mu}_{\mp}(\vec k)}} 
		+ \frac{ {\rm e}^{-(\beta - (t-t')){\mu}_{\pm}(\vec k)} }
		{{1-{\rm e}^{-\beta {\mu}_{\pm}(\vec k)} }}
		\right) {\rm e}^{-i\vec k\cdot (\vec x-\vec x')} 
		\;,\nonumber 
	\end{align}
where $\theta(t)$ denotes the characteristic function for the half-line $t\geqslant0$. The kernels  
\[
D^{c}_{+, \beta,\Lambda}(x-x') = \overline{D^{c}_{-, \beta,\Lambda}(x-x')}
\]
have smooth limits as each $\ell_{j}\to\infty$ converging in the limit of infinite volume to the  kernels	
	\[
		D^{c}(x-x')= \langle \mathbb{A}\varphi_I^{+}(x)\varphi_I^{+}(x') \rangle_{\beta,\pm}
	\]
introduced to study the time compactification in  \eqref{DcKernel},  in the same sense that 
a Fourier series approximates a Fourier transform. The sum over each coordinate of $\vec k$ converges to a Riemann 
integral, and the limiting kernels coincide with the operators $D^{c}$ introduced in \eqref{DcKernel}. The corresponding 
anti-time-ordered two-point functions also converge. 
 
The operators $D^{c}$  act on $L_{2}(S^{1}\times \mathbb R^{d-1})$. Likewise, $D^{c}_{+,\beta,\Lambda}$  acts on $L_{2}(S^{1}\times \mathbb{T}^{d-1})$.  In all cases, these operators equal 
the inverse of the corresponding differential operators 
	\[
		D_{\vec {\it v}}^{-1} = {-\Delta + m^{2} + ( \nabla_{\vec x}\cdot \vec {\it v} )^{2} 
	 			- 2i\tfrac{\partial}{\partial t} ( \nabla_{\vec x}\cdot \vec {\it v} ) }\;,
	\]
originally introduced in \eqref{Dv Configuration Space} on $\mathbb R^{d}$, but here acting on these (partially or 
fully) compactified spacetimes.  

As such $D^{c}$ and $D^{c}_{+,\beta,\Lambda}$ are {\em doubly temporally reflection-positive}. The operator $D^{c}$ is {\em doubly spatially reflection-positive} in the direction $\vec n$.  The same is true for $D^{c}_{+,\beta,\Lambda}$ in 
case  $\vec n$ lies along a lattice coordinate direction.

\subsection{Quantization}
The Osterwalder-Schrader quantization from the $d$-torus results in only minor changes\footnote{The one-particle space $ \mathfrak{H}_{-\frac{1}{2}}$ and the Laplace operator 
have to be adapted to periodic boundary conditions.} to the results presented in \S \ref{Sect:Positive Temperature-SF}. 
The Araki-Woods one-particle Hilbert space $\mathcal{H}_1(\Lambda) $, which arises from Osterwalder-Schrader quantization, is now
\[  
\mathcal{H}_1(\beta, \Lambda) =\mathfrak H_{-\frac12} (\mathbb{T}^{d-1})\oplus \overline{\mathfrak H_{-\frac12}(\mathbb{T}^{d-1})} \; . 
\]
The associated Fock space is again of the form \eqref{Fock Space}. The elements of $(h, h') \in \mathcal H_{1}$ have two components, namely $h \in \mathfrak H_{-\frac12} (\mathbb{T}^{d-1})$ and $h' \in \overline{\mathfrak H_{-\frac12}(\mathbb{T}^{d-1})}$. 

The Liouvillean $L_\pm $, the momentum operator $\vec P$, the modular conjugation~$J$ and the Tomita operator $S$ 
act on $\mathcal{H}_{1} (\beta, \Lambda)$  as $\ell_\pm$, ${ \vec p}$, $j$ and $s$. Note that 
\[
\Omega_0 \in {\mathcal D} (L_\pm) \quad \text{and} \quad  L_\pm \Omega_{0} = 0  
\]
still holds. However, the spectrum of $L_\pm$ of is discrete (and symmetric), and the discrete eigenvalue zero is infinitely degenerated.

\begin{remark} The spectral properties can be understood by considering Gibbs states on $\mathbb{R} \times {\mathbb T}^{d-1}$; 
see Remark \ref{Gibbsremark} of \S \ref{The Spatially Compactified Case}.
Using energy eigenvectors $\Psi^\pm_i$, 
	\[ 
		H_\pm  \Psi^\pm_i = E^\pm_i \Psi^\pm_i \qquad \hbox{with} \qquad E_i \in \mathbb{R}^+ \cup \{ 0 \} \: . 
	\]
The Gibbs density matrix\footnote{The following arguments hold true in the {\em interacting} case too.} takes the form
	\begin{equation}
	\label{gibbs-ccc}
		\frac{ {\rm e}^{- \beta H_\pm (\Lambda) } }{ {\rm Tr} \, {\rm e}^{- \beta H_\pm (\Lambda)} }
			= { \sum_ i   {\rm e}^{- \beta E^\pm_i}  \, \, | \Psi^\pm_i \rangle \langle \Psi^\pm_i | 
				\over  \sum_k {\rm e}^{- \beta E^\pm_k} } ,
				\qquad \beta > 0 \; .
	\end{equation}
The GNS representation for the state given by the Gibbs density matrix is of the form
	\begin{equation}
	\label{GNS-rep}
		{\mathcal B}({\mathcal H}_\Lambda) \ni A \mapsto A \otimes \mathbb{1} \; ,
	\end{equation}
acting on the tensor product ${\mathcal H}_\Lambda \otimes {\mathcal H}_\Lambda$ of the Hilbert space 
${\mathcal H}_\Lambda $ introduced in~\eqref{HLambda} with itself.    
The cyclic vector, \emph{i.e.}, the GNS vector, 
	\[
		 |  \sqrt {\rho_\pm } \rangle =  { \sum_i    {\rm e}^{- \beta E^\pm_i /2  }  \over  
			\sqrt{ \sum_k  {\rm e}^{- \beta E^\pm_k } } }  \, \,   \Psi^\pm_i \otimes \Psi^\pm_i  
	\]
induces the Gibbs state, \emph{i.e.}, 
	\[
		\frac{ {\rm Tr} \, {\rm e}^{- \beta H_\pm (\Lambda)} A}{ {\rm Tr} \, {\rm e}^{- \beta H_\pm (\Lambda)} }
			=  \langle  \sqrt {\rho_\pm }\, , \, ( A \otimes \mathbb{1} ) \sqrt {\rho_\pm } \rangle\; , 
			\qquad A \in {\mathcal B}({\mathcal H}_\Lambda) \; .
	\]
The generator
of the time evolution in the GNS representation is  
	\[
		L_\pm = H_\pm (\Lambda) \otimes  \mathbb{1}  - \mathbb{1} \otimes H_\pm  (\Lambda) \; .
	\]
In finite volume $\Lambda$, the Araki-Woods representation on the Fock space over the one-particle space
$\mathcal{H}_1(\beta, \Lambda)$ resulting 
form the Osterwalder-Schrader quantization is unitarily equivalent to the GNS representation \eqref{GNS-rep}. 
However,  in the infinite volume case discussed in \S \ref{Sect:Positive Temperature-SF},  the von Neumann algebras
${\mathcal R}_0$ and ${\mathcal R}_0' = {\mathcal R}_{\beta/2}$ introduced in~\eqref{vNA-R_0} and \eqref{vNA-R_b}, respectively, 
are both factors of type III and, consequently,
their von Neumann tensor product is type III as well, in contrast to 
${\mathcal R}_0 \vee {\mathcal R}_0' = {\mathcal B} ({\mathcal H}_\beta)$.

The fact that the spectrum of $L_\pm$  is discrete  and symmetric,  and the  infinite degeneracy of the eigenvalue zero cause a 
number of problems.
However, it is instructive to consider the map $N^\pm_\Lambda \colon {\mathcal B}({\mathcal H}_\Lambda) \to 
{\mathcal H}_\Lambda \otimes {\mathcal H}_\Lambda$, 
	\[
		A \mapsto {\rm e}^{ - \lambda |  L_\pm  | } ( A \otimes \mathbb{1}   ) \, \, |  \sqrt {\rho_\pm  } \rangle \; .
	\]
A straight forward computation yields
\begin{align*}
	N^\pm_\Lambda (A) & = \sum_{i,j} {\rm e}^{- \lambda | E^\pm_i - E^\pm_j | } ( A_{i,j} \otimes \mathbb{1} )
		{   {\rm e}^{- \beta E^\pm_j /2  } \over \sqrt{ \sum_k  {\rm e}^{- \beta E^\pm_k } } } \, \,   \Psi^\pm_j \otimes \Psi^\pm_j
		\\
		& = \frac{1}{ \sqrt{ \sum_k  {\rm e}^{- \beta E^\pm_k } }} \sum_{i,j} {\rm e}^{- \lambda | E^\pm_i - E^\pm_j | - \beta E^\pm_j /2}
		( A^\pm_{i,j} \Psi^\pm_j \otimes \Psi^\pm_j) \; , 
\end{align*}
where $A^\pm_{i,j} := | \Psi^\pm_i \rangle \langle \Psi^\pm_i | A  | \Psi^\pm_j \rangle \langle \Psi^\pm_j |$ is a rank $1$ 
operator. The sum $\sum_{i,j}$ is convergent for $\lambda >0$; thus, $N^\pm_\Lambda$ is a nuclear map. In fact, $N^\pm_\Lambda$ is nuclear for all $\lambda > 0$ and, consequently, it is an element of all Schatten--von Neumann classes.
\end{remark}

\begin{remark}
The imaginary anti-time-ordered two point function 
	\begin{equation}
		D^{c}_{\pm,\beta,\Lambda}(x-x') = \langle \mathbb{A} \varphi_I^{\pm}(x)\varphi_I^{\pm}(x') \rangle_{\pm,\beta,\Lambda}
	\end{equation}
in the Gibbs state $\langle \, \cdot \,  \rangle_{\pm,\beta,\Lambda}$ defined 
by density matrix \eqref{gibbs-ccc} agrees with the covariances $D^{c}_{\pm,\beta,\Lambda}$ given 
in \eqref{Gibbs-Two-Point-Compact}.  The proof of this statement relies on two basic facts. Firstly,
the {\em pull-through} identity holds:
	\[
		a(f){\rm e}^{-tH_{\pm} } = {\rm e}^{-t H_{\pm} } 
		a({\rm e}^{-t{\mu}_\pm } f) \; . 
	\]
Secondly, cyclicity of the trace, translation invariance of $H_{\pm}$, and the fact that  $H_{\pm}$ commutes  with the number operator $N$, ensures a symmetry of the expectation of creation and annihilation operators. Let  $a(k)^{\#}$ 
denote either $a(k)$ or~$a(k)^{*}$.  Then     
	\[
		\langle a_{\pm}^{\#}(k,t)a_{\pm}^{\#'}(k',t') \rangle_{\beta,\pm,\Lambda}=0\;,
	\]
unless $k=k'$, as well as one $a(k)^{\#}$ being a creation operator, while the other $a(k)^{\#'}$ is an annihilation operator.  
Using these two facts, and the expansion for the time-zero field,  we can evaluate the two-point function in closed form.  
We omit further details.  
\end{remark}

\setcounter{equation}{0}
\section{Some Comments on Partial Wick Rotation}
In this paper we use complex classical fields to construct (interacting) quantum fields in finite and infinite volumes both at zero and positive temperatures.  These fields describe neutral particles in both vacuum and thermal equilibrium states.  
We now indicate how the quantum fields we have constructed in this work are related to the quantum fields considered in more traditional approaches.

\subsection{Flat Space and Spatially Compactified Space}
Let us consider scalar Wightman quantum fields $\varphi (t, \vec x)$ defined on $d$-dimensional, Minkowski  spacetime, acting on the Hilbert space ${\mathcal H}$.  Let $\Omega_{0}  \in {\mathcal H}$ denote the vacuum vector, and let  $H$ and $\vec P$ denote the Hamiltonian and momentum operator, respectively.  The fields are Poincar\'e covariant and $\Omega_{0}$ is Poincar\'{e} invariant.  Hence, the  {\em Wightman functions} are also Poincar\'e-invariant functions on spacetime.  With $(\Lambda,a)$ in the Poincar\'e group, the Wightman function satisfies 
\begin{align*}
	{\mathcal W}^{(n)}(x_1, \ldots, x_n) 
		&= \langle \Omega_0 , \varphi (t_1, \vec x_1) \cdots \varphi (t_n, \vec x_n) \Omega_0 \rangle 
		\\
		&={\mathcal W}^{(n)}(\Lambda^{-1}x_1+a,\ldots, \Lambda^{-1} x_n+a) 
		 \; .
\end{align*}
The elementary {\em positive-energy condition} for the Hamiltonian $H$ entails 
	\[
		0\leqslant H\;
		\quad \text{and}\quad
		H\Omega_0 = 0\;,
	\]
which ensures that the Wightman functions  (as functions of anti-time-ordered variables $t_{j+1}> t_{j}$ to purely imaginary time $it_{j}$) continue analytically to the corresponding {\em Schwinger functions}, {\it viz.}, 
\begin{align*}
	{\mathcal S}^{(n)}(t_1, \vec x_1,\ldots, t_n, \vec x_n) &= {\mathcal W}^{(n)}\bigl( (it_1, \vec x_1),\ldots, (it_n, \vec x_n)\bigr)   \;,
\end{align*}
which play a key role in the construction of interacting quantum field theories. The Wightman functions satisfy the following identity\footnote{Assuming that the $\vec {\it v} = (v, 0,0)$ lies in the $x_1$-direction, the Lorentz transformation representing this boost 
(in $1+3$-spacetime dimensions) is given by the matrix
	\[
	 \Lambda  = \begin{pmatrix} \cosh  \beta & - \sinh  \beta  & 0 & 0 \\
	 - \sinh  \beta & \cosh \beta  & 0 & 0 \\
	 0 & 0  & 1 & 0 \\
	 0 & 0 & 0 & 1\\
	 \end{pmatrix} \; , \quad \cosh \beta = \frac{1}{\sqrt{1-{\it v}^2}} \; , \;  \sinh \beta = \frac{  {\it v} }{\sqrt{1- {\it v}^2}} \; . 
	 \]
},
\begin{align*}
	& {\mathcal W}^{(n)}  \bigl( (t_1 \cosh  \beta, \vec x_1 +  t_1   \sinh \beta), \ldots,
	(t_n \cosh  \beta, \vec x_n +  t_n   \sinh \beta) \bigr) 
	\\
	&= \langle \varphi (0, \vec x_1)\Omega_0 , {\rm e}^{-i (t_1-t_2 )\cosh \beta H_{\vec {\it v}} } 
		\varphi (0, \vec x_2) \cdots  {\rm e}^{-i (t_{n-1}-t_n )\cosh \beta H_{\vec {\it v}} } 	
		\varphi (0, \vec x_n) \Omega_0 \rangle 
\end{align*}
and can be analytically continued (in the relative variables) to imaginary times. For $s_i \geqslant s_{i+1}$, where $i= 1, \ldots, n-1$, define the modified Schwinger functions,
\begin{align*}
	& {\mathcal S}_{\vec {\it v}}^{(n)}  \bigl( t_1 \cosh  \beta, \vec x_1 +  t_1 
	\sinh \beta, \ldots,
	t_n \cosh  \beta, \vec x_n +  t_n  \sinh \beta
	\bigr) 
	\\
	& \doteq\langle \varphi (0, \vec x_1)\Omega_0 , {\rm e}^{- (t_1-t_2 )\cosh \beta H_{\vec {\it v}} } 
		\varphi (0, \vec x_2) \cdots {\rm e}^{- (t_{n-1}-t_n )\cosh \beta H_{\vec {\it v}} } 
		\varphi (0, \vec x_n) \Omega_0 \rangle . 
\end{align*}
Using the Flat Tube Theorem, the possibility to vary $\vec {\it v}$ ensures that the Wightman functions 
(in difference coordinates $x_{j}-x_{j+1}$) extend by analytic continuation to the forward tube
\[
\mathcal{T}^{n-1} = \mathbb{R}^{(n-1)d} - i (V_+)^{n-1} \; , \qquad V_+ = \{ (t, \vec x) \in \mathbb{R}^d \mid |\vec x| < |t| \} \; . 
\]
The novelty of our approach is that the modified Schwinger functions appear as the expectation values of complex classical fields.

\subsection{Compactified Time\label{Compactified Time}}

For the quantum theory reconstructed from a classical theory with periodic time, namely, $\boldsymbol{X}=S^1\times \mathbb{R}^{d-1}$ or $\boldsymbol{X} =S^1\times \mathbb{T}^{d-1}$, the spectrum condition 
no longer holds. In fact, the Osterwalder-Schrader reconstruction provides thermal equilibrium states, and the spectrum of both the 
generators of the time-evolution and of the spatial translations is symmetric around zero. Therefore, one might wonder whether quantization by reflection positivity yields  a  quantum field whose equations of 
motion are invariant under Poincar\'e transformations. In order to answer this question, it is instructive to compare our result with the free thermal neutral scalar Wightman field, whose expectation values 
are specified by the averaged two-point function, 
\begin{align*}
	{\mathcal W}_\beta^{(2)} \bigl( (t, \alpha), (t' \alpha') \bigr) 
	&= \langle  \alpha , \coth (\beta \mu ) {\rm e}^{-i (t-t') \mu} \alpha'  \rangle_{ \mathfrak{H}_{-\frac{1}{2} } } \;, 
	\; \;  \alpha, \alpha' \in \mathfrak{H}_{-\frac{1}{2}} (\mathbb{R}^d) \; , \end{align*}
where $\beta$ denotes the inverse temperature. 
Although this two-point function is {\em not} invariant under boosts, there is no problem to consider 
the expectation values of Lorentz boosted quantum fields (suppressing the rescaling factor $1/ \sqrt{1-\vec v \,^2}$). Since $\mu > \mu_+ / (1+ | \vec {\it v}| )$, the expression
\[
	{\mathcal W}_\beta^{(2)} \bigl( ( t  ,  	\vec x -  t \vec {\it v}) \, , \, ( t' ,
	\vec x' -  t' \vec {\it v} ) \bigr) 
	 = \left( \frac{1}{2\mu}\coth ( \beta \mu  ) 
	 {\rm e}^{-i (t-t')  \mu_+ } \right)
	(\vec x, \vec x')\; 
\]
allows an analytic continuation to imaginary times into the strip 
	\[ 
		\Bigl\{ (t-t') \in \mathbb{C} \mid - \tfrac{\beta} {1+ | \vec {\it v}| } < \Im  (t-t')  < 0 \Bigr\} \; . 
	\]
Thus, the Flat Tube Theorem ensures that ${\mathcal W}_\beta^{(2)}(t, \vec x)$ is analyticity in the tube 
\begin{equation}
\label{tuboid}
{\mathcal T}_\beta = \mathbb{R}^{d} - i \left(V^{+} \cap (\beta \mathbf{e}_1 - V^{+})\right) \; , 
\end{equation}
where $\mathbf{e}_{1} = (1, 0, \ldots, 0)$ is the unit vector in the  time-direction distinguished by the rest-frame. 

For interacting quantum field theories, this question was addressed by Bros and Buchholz, who formulated a {\em relativistic KMS condition} \cite{BB1,BB2}.  The relativistic KMS condition ensures analyticity of the two point-function in the domain \eqref{tuboid}.  They verified that the relativistic KMS condition holds for a large class of quantum field theories satisfying the 
nuclearity condition of Buchholz and Wich\-mann~\cite{BW-Nuclearity}.  Nuclearity, however, has not been established 
for the models considered in Constructive Quantum Field Theory.  Only recently
the {\em relativistic KMS condition}  has  been proved  for the 
${\mathscr P}(\varphi)_{2}$-models using multiple reflection positivity~\cite{JR}. 

For real $\vec {\it v} $, the function appearing in the corollary at the end of \S VII can be expressed in terms of the two-point function,
\begin{align*}
	 &\left( \tfrac{\coth ( \beta \mu_+  )}{2\mu} 
	 {\rm e}^{-i  \frac{(t-t')}{1-  {\it v}^2} \mu_+}\right) (\vec x, \vec x')\\
	 & \qquad \qquad \qquad =  	{\mathcal W}_{\sqrt{1-{\it v}^2} \beta}^{(2)} 
	  \Bigl( \bigl(  \tfrac{ t+ \vec {\it v} \cdot \vec x}{\sqrt{1-  {\it v}^2}} \,  , \, 
	\tfrac{\vec x}{\sqrt{1-  {\it v}^2}}  \bigr) \, , \, \bigl(  \tfrac{ t'+ \vec {\it v} \cdot \vec x'}{\sqrt{1-  {\it v}^2}} \, , \, 
	\tfrac{\vec x'}{\sqrt{1-  {\it v}^2}}  \bigr)\Bigr)  \;,
\end{align*}
which is analytic in the domain 
	\[ 
		| \Im (\vec x- \vec x') |   - \beta (1-{\it v}^2) 
	  	< \Im \bigl( (t-t') + \vec {\it v} \cdot \vec x \bigr) < -  | \Im ( \vec x- \vec x') |  \; , 
	\]
and the boundary values satisfy the KMS  condition for $ \Im ( \vec x- \vec x')=0$.

\end{document}